\documentclass[onecolumn,preprint]{emulateapj}
\usepackage{graphicx,times}
\usepackage{amssymb,amsmath}
\bibpunct{(}{)}{;}{a}{}{,}

\usepackage[T1]{fontenc}
\usepackage{pdftexcmds}
\usepackage{multirow}
\usepackage{rotating}
\usepackage{float}

\def\hi{H{\textsc i} }

\def\co{$^{12}$CO}
\def\co13{$^{13}$CO}
\def\co18{C$^{18}$O}
\def\cm3{cm$^{-3}$}
\def\cm2{cm$^{-2}$}

\def\deg{$^\circ$}
\def\cm2{cm$^{-2}$}

\def\nh3{NH$_3$}
\def\n2h{N$_2$H$^+$}
\def\hc3n{HC$_3$N}
\def\h2{H$_2$}
\def\nh{n(H$_2$)}

 \def\pasa{PASA}
 
\usepackage{array}
\newcolumntype{C}[1]{>{\centering\arraybackslash}p{#1}}
 
\usepackage{color}
\usepackage{cancel,soul,ulem}
\usepackage{multirow,amsmath}

\usepackage{enumerate}   
 \usepackage[flushleft]{threeparttable}  

\shorttitle{The Fundamental Performance of FAST  with 19-beam Receiverat L Band }
\shortauthors{P. Jiang et al. }

\begin{document}

\title{The Fundamental Performance of  FAST  with 19-beam Receiver at L Band }

\author{ Peng Jiang\altaffilmark{1,2},
Ning-Yu Tang\altaffilmark{1,2}, 
Li-Gang Hou\altaffilmark{1,2},
Meng-Ting Liu\altaffilmark{1,2}, 
Marko Kr$\rm\check{c}$o\altaffilmark{1,2}, 
Lei Qian\altaffilmark{1,2}, 
Jing-Hai Sun\altaffilmark{1,2}, 
Tao-Chung Ching\altaffilmark{1,2}, 
Bin Liu\altaffilmark{1,2}, 
Yan Duan\altaffilmark{1,2}, 
You-Ling Yue\altaffilmark{1,2}, 
Heng-Qian Gan\altaffilmark{1,2}, 
Rui Yao\altaffilmark{1,2}, 
Hui Li\altaffilmark{1,2}, 
Gao-Feng Pan\altaffilmark{1,2}, 
Dong-Jun Yu\altaffilmark{1,2}, 
Hong-Fei Liu\altaffilmark{1,2}, 
Di Li\altaffilmark{1,2}, 
Bo Peng\altaffilmark{1,2},
Jun Yan\altaffilmark{1,2}, 
and FAST Collaboration
} 


\altaffiltext{1}{National Astronomical Observatories, CAS, Beijing 100012, China; {\it pjiang@nao.cas.cn,nytang@nao.cas.cn, ghq@nao.cas.cn}} 
\altaffiltext{2}{CAS Key Laboratory of FAST, National Astronomical Observatories, Chinese Academy of Sciences, Beijing 100101, China}

\begin{abstract}
The Five-hundred-meter Aperture Spherical radio Telescope (FAST) passed national acceptance and is taking pilot cycle of `Shared-Risk' observations. The 19-beam receiver covering 1.05-1.45 GHz was used for most of these observations. The electronics gain fluctuation of the system is better than 1\% over 3.5 hours, enabling enough stability for observations. Pointing accuracy, aperture efficiency and system temperature are three key parameters of FAST.  The measured standard deviation of pointing accuracy is 7.9$''$, which satisfies the initial design of FAST. When zenith angle is less than 26.4$^\circ$, the aperture efficiency and system temperature around 1.4 GHz are $\sim$ 0.63 and less than 24 K for central beam, respectively.   The measured value of these two parameters are better than  designed value of 0.6 and 25 K, respectively. The sensitivity and stability of the 19-beam backend are confirmed to satisfy expectation by spectral \hi observations toward N672  and polarization observations toward 3C286. The performance allows FAST to take sensitive observations in various scientific goals,  from studies of pulsar to galaxy evolution. 

\end{abstract}
\keywords{instrumentation: detectors --- line: profiles}

\section{Introduction}
\label{sec:introduction}

The Five-hundred-meter Aperture Spherical radio Telescope (FAST) with a diameter of 500 m passed its national acceptance on January 11, 2020. It operates at frequencies ranging from 70 to 3000 MHz and is the most sensitive single dish telescope in this frequency range.

FAST commissioning began when construction was completed on September 25, 2016. The details of FAST commissioning have been introduced in Jiang et al.\ (2019). A lot of observations, especially pulsar related observations were taken during the commissioning phase. A series of primary results including the first detection of a pulsar by FAST have been published in the FAST issue of Science China Physics, Mechanics \& Astronomy.  

With the effort of the commissioning group, the performance of FAST has been improved significantly during the last two years. One example is the much improved Radio Frequency Interference (RFI) environment. In this paper we will review the current instrumental properties based on measurements with the 19 beam receiver that covers 1.05 to 1.45 GHz. 

In section \ref{sec:19beam_receiver}, we present the properties of the 19 beam receiver system including calibration of the noise dipole, the spatial distribution of all 19 beams, pointing accuracy, aperture efficiency and system temperature. In section \ref{sec:backend}, the performance of the spectral backend including stability and sensitivity, standing wave, and polarization is presented. Observation modes and status of RFI are presented in section \ref{sec:observation}.  Summary is presented in section \ref{sec:summary}. 

\section{Properties of the 19 Beam Receiver}
\label{sec:19beam_receiver}

\subsection{Measurement of the Noise Dipole}
\label{subsec:noise_dipole}
The FAST 19-beam L-band Array contains a temperature stabilized noise 
injection system. The noise is injected between the feed and the low 
noise amplifiers. The noise source is a single diode whose signal is 
split into each beam and polarization. The diode itself is always 
powered, while it can be switched in and out of the signal path using a 
solid state switch. This was done in order to improve stability at very 
high switching speeds. It takes less than 1 microsecond for the noise 
power to stabilize once switched on or off. The noise diode has two adjustable 
power output modes, but are currently kept at approximately 1.1, and 12.5K.

\begin{figure}
\centering
\includegraphics[width=1.0\textwidth]{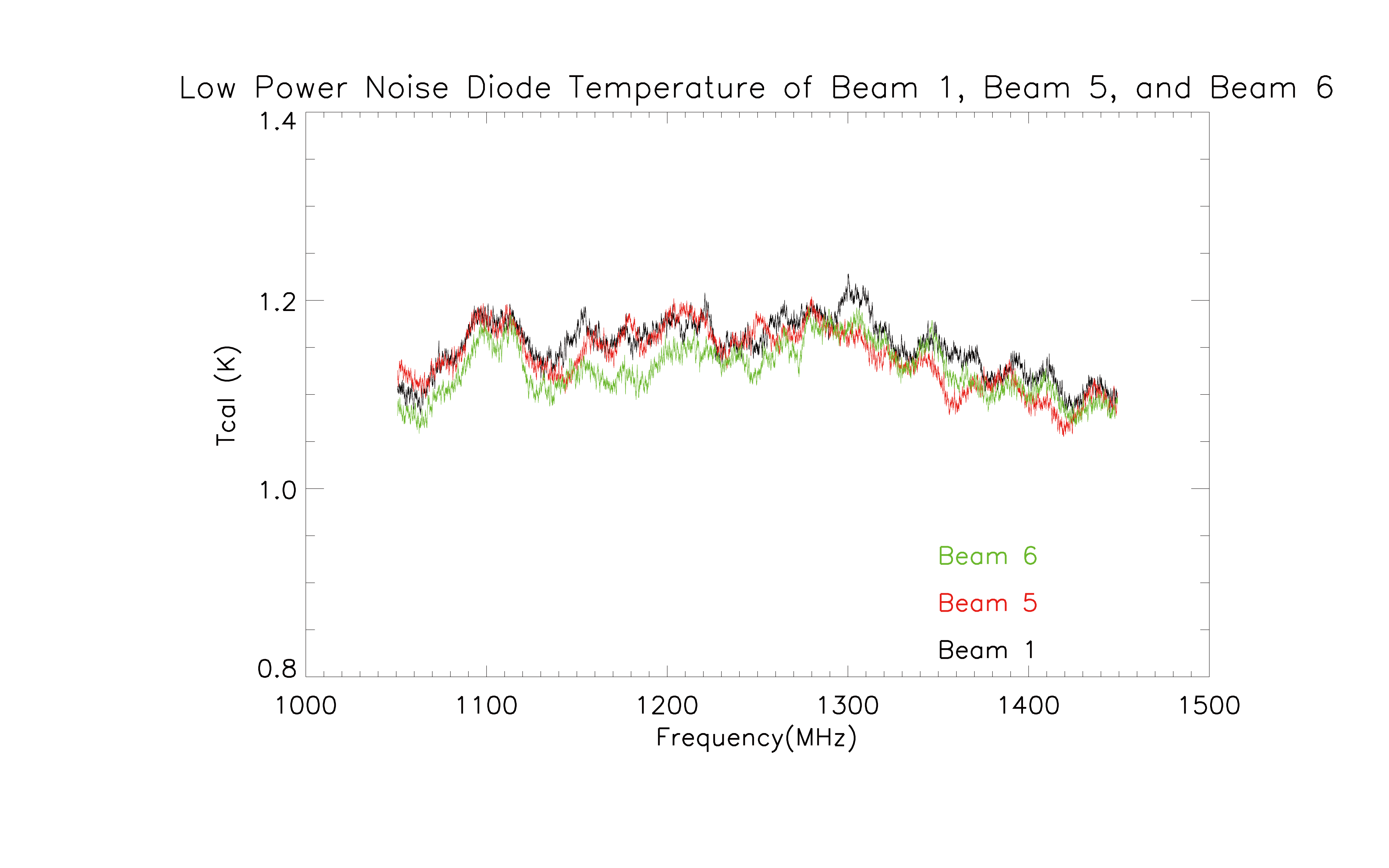}
\caption{The low power noise diode temperatures of Beam 1, Beam 5, and Beam 6.
The 3 beams plotted here are representative of the rest. The data for all 19 beams and polarizations is available and is being used for subsequent calibration.}
\label{lowcal}
\end{figure}

\begin{figure}
\centering
\includegraphics[width=1.0\textwidth]{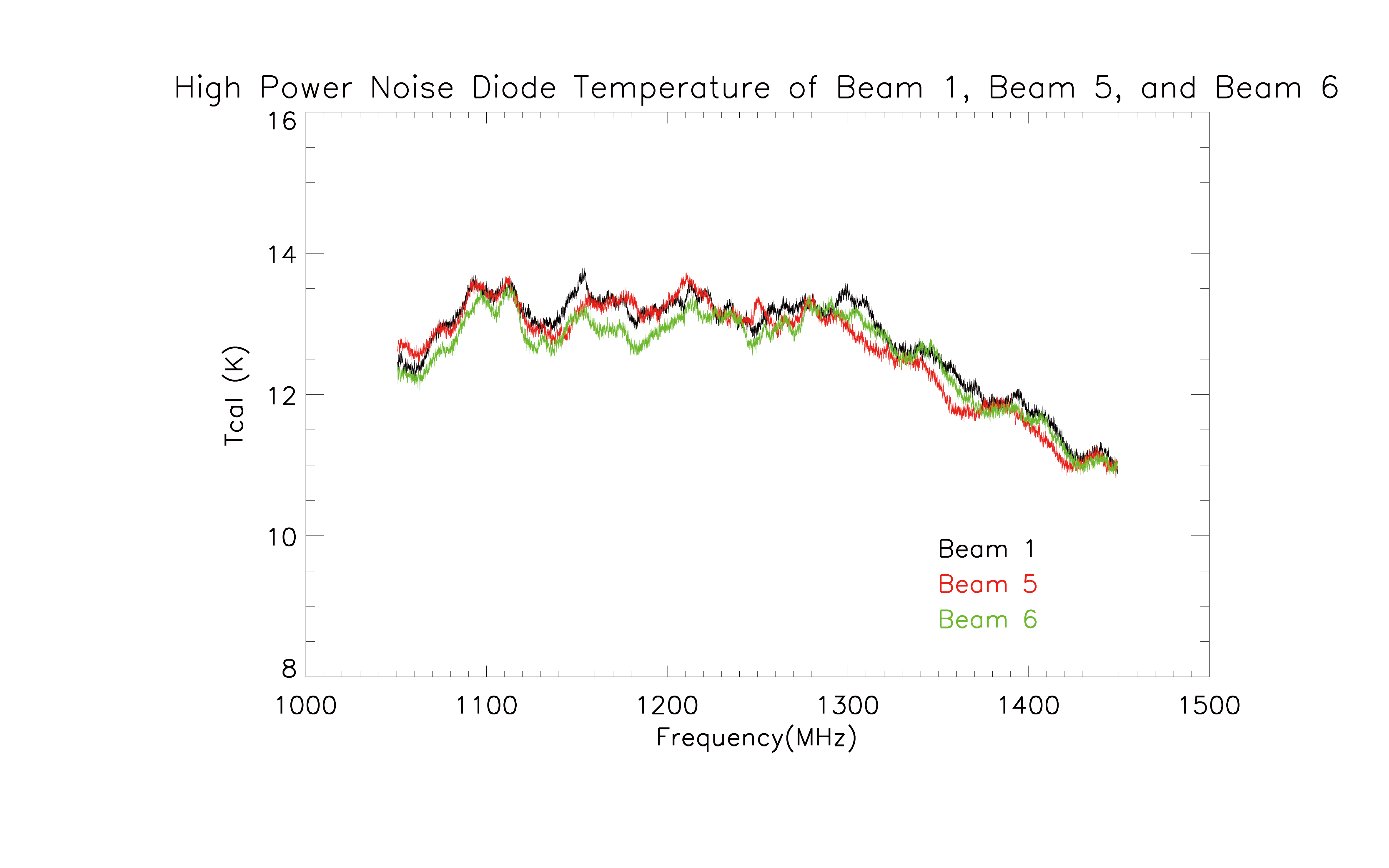}
\caption{The high power noise diode temperatures of Beam 1, Beam 5, and Beam 6.
The 3 beams plotted here are representative of the rest. The data for all 19 beams and polarizations is available and is being used for subsequent calibration.}
\label{highcal}
\end{figure}

In order to test the performance and stability of the noise diode, we 
conducted a series of hot load measurements whereby the feed cabin was 
lowered to the ground and a foam absorber was placed directly under the 
feed so that it completely covered all of the beams. We then 
periodically measured the temperature of the absorber with a 
thermometer, while the noise diode was continuously switched on and off 
(''a winking CAL'') with a period of 1.00663296 seconds. Figure \ref{lowcal} and Figure \ref{highcal} show 
the measured noise diode temperatures (Tcal) for both the low, and high 
power modes with respect to frequency. The diode is tuned such as to 
minimize the diode temperatures near 1420.4 MHz to 
reduce the impact on the system temperature.

\begin{figure}
\centering
\includegraphics[width=1.0\textwidth]{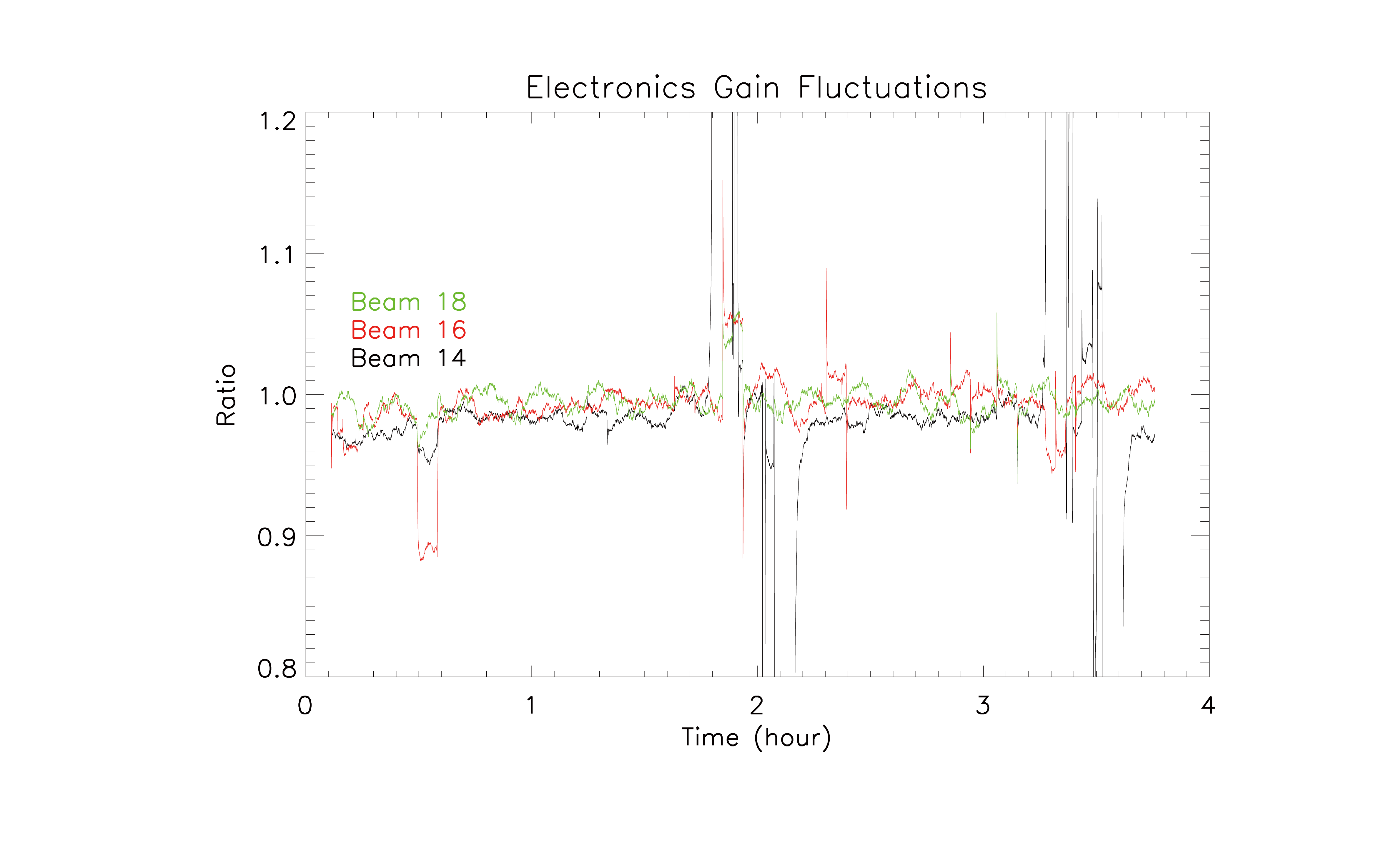}
\caption{The electronics gain fluctuations of the system over several hours. 
The vertical shows the ratio of fluctuations with respect to the mean temperature over time of Beams 14, 16, and 18. The three beams presented here represent the strongest electronics gain fluctuations out of all 19. 
}
\label{plot3}
\end{figure}

Figure ~\ref{plot3} shows the electronics gain fluctuations of the system over several hours. This was done assuming that the noise diode temperature is 
absolutely constant. It shows that the electronics gain fluctuations 
within the FAST 19-beam receiver/backend signal path are typically on 
the order of a few percent over timescales of a few minutes. This 
implies that science projects for whom flux calibration is important 
should fire the noise diode at least every few minutes in order to 
account for typical electronics gain fluctuations. The larger 
fluctuations in Figure 3 are caused by RFI leaking into the 
signal path during the hot load measurements, and instability in the low 
noise amplifiers.  The  electronics gain fluctuations of the 
rest beams are presented in the Figure~\ref{fig:electrongain_restbeams}.

\begin{figure}
\centering
\includegraphics[width=1.0\textwidth]{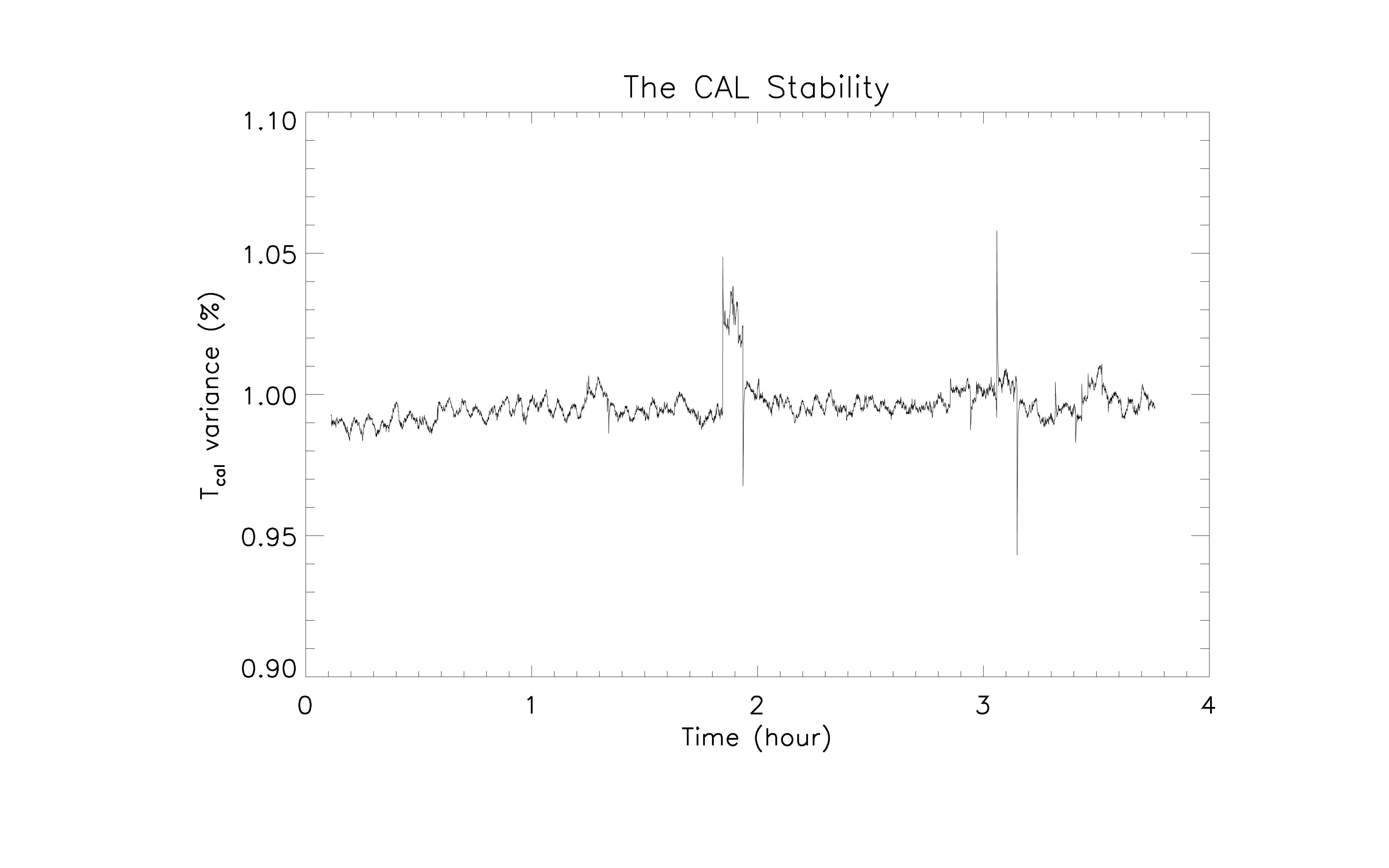}
\caption{The noise diode stability.
An upper limit on the noise diode stability as a function of time measured by monitoring the median value of the electronics gain in all 19 beams during hot load measurements.}
\label{plot4}
\end{figure}

The electronics gain fluctuations in Figure \ref{plot3} are mostly uncorrelated 
between the different beams. By taking a median value of all 19 beams, 
we can produce Figure \ref{plot4} which yields an upper limit on the temperature
fluctuations in the noise diode. While laboratory measurements of the 
noise diode itself yielded temperature fluctuations on the order of 
$\sim$0.1$\%$ over several hours, we see that once placed within the full signal
path we get an upper limit of $\sim$1$\%$, implying that it is likely impossible
to obtain accurate flux calibration better than $\sim$1$\%$. Considering FAST's 
pointing accuracy, and beam size the best flux calibration for a point 
source is approximately $\sim$2$\%$, meaning that the noise diode is certainly 
operating within our tolerance limits. This lower limit is estimated by looking at the shape of the beam response function at 1420MHz within the pointing uncertainty limits. The effect is weaker at longer wavelengths where the pointing uncertainty is smaller compared to the beam size.

\subsection{Beam Properties and Pointing Accuracy}
\label{subsec:beam_pointing}

To measure the properties of FAST's 19-beam receiver, one method is to directly make mapping observations
toward radio sources on the sky. The raster scan (or raster scanning)
observation mode has been realized for FAST, which is used to map radio sources
and measure the 19-beam properties.

\subsubsection{Observations}

\begin{figure}
\centering
\includegraphics[width=0.48\textwidth, angle=0]{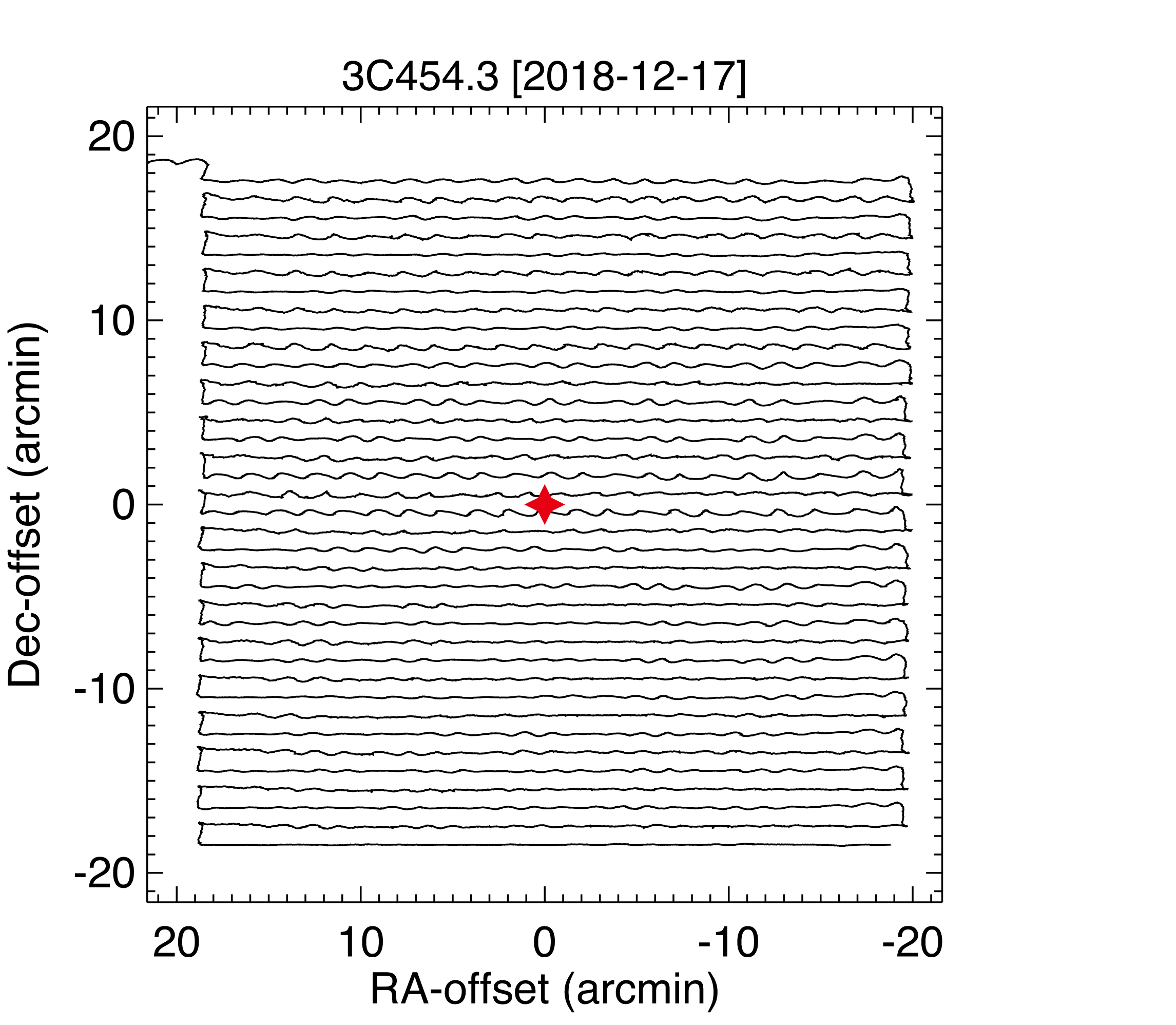}
\includegraphics[width=0.48\textwidth, angle=0]{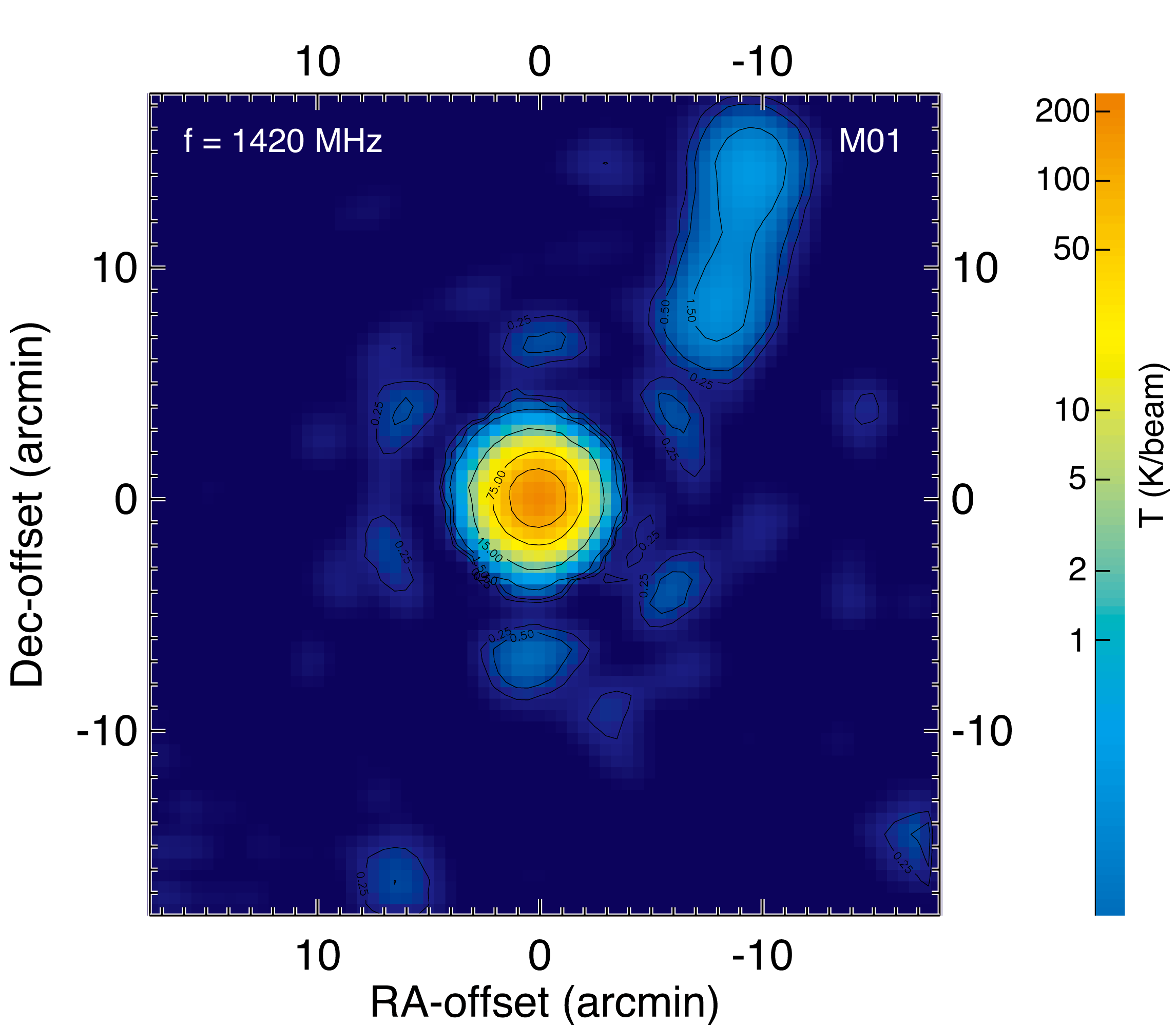}
\caption{{\it Left}: The telescope track for a raster scan along the Dec
  direction. The covered sky area is $\sim$37$^\prime \times
  37^\prime$. The target source (3C454.3) is marked by a red star in the
  center of the plot. The X- and Y-axes indicate the RA- and
  Dec-offsets relative to the equatorial coordinates of 3C454.3 (RA:
  22h53m57.7479s, Dec: +16d08m53.561s, in J2000), respectively. {\it
    Right}: distribution map of total intensity obtained by the
  central beam (M01) near the frequency of 1420\,MHz. The intensity
  distributions in a logarithmic-scale are indicated by different
  colors. The contours show the intensities with levels of
  0.25, 0.5, 1.5, 15, 75 and 150 K.}
\label{track}
\end{figure}

From November 28 to December 29, 2018, mapping observation
have been conducted toward 3C380, 1902+319, 1859+129, 3C454.3, and
2023+318, etc. In the observations, a sky area centering on the target
source and covering $\sim37^\prime (\rm RA)\times 37^\prime(\rm Dec)$
is mapped by a raster scan along the RA or Dec direction with the 19-beam
receiver. An example of the telescope track for a raster scan along
Dec is given in Fig.~\ref{track}. The scan velocity is
15$^{\prime}$/min. The separation between two sub-scans is 1$^\prime$,
which satisfies Nyquist sampling. A total of $\sim$100 minutes is
needed to acquire such a map. Observational data are simultaneously
recorded with the pulsar mode, the narrow-band spectral-line mode, and
also the wide-band spectral-line mode. The number of channels is 4096
for the pulsar backend and 65536 for the spectral-line backend
(wideband mode), corresponding to a frequency coverage from 1000 $-$
1500\,MHz, which is somewhat broader than the designed bandwidth of
the 19-beam receiver (1050 $-$ 1450\,MHz). The sampling time, i.e.,
the integration time for raw data, is set to 196.608$\mu$s for the
pulsar mode and 1.00663296s for the spectral-line modes. The intensity
for each of the 19 beams is calibrated by the periodic injection of a
high-intensity noise signal ($\sim$10\,K, see
Sect.~\ref{subsec:noise_dipole}) with a period of
0.2s.

\subsubsection{Data processing procedure}

A standard pipeline for processing the mapping data of the 19-beam
receiver is still under development. In the following
analysis the mapping data is processed using IDL. The data recorded by
the pulsar backend are adopted, which have dual linear polarizations
(XX and YY) and a relatively high time-resolution. We first compress the
data raw data in the time dimension to have an identical
time-resolution with the periodic injected noise. The antenna
temperature $T_a(\nu)$ for each of the polarization is then converted
from the raw counts by using the noise diode calibrator. The observed
intensity (XX or YY, in K) corresponding to a frequency can be
extracted from the spectra after a removing of the radio frequency
interference (RFI). Together with the data recording time given by the
backend, one series of pairs of $\lbrace t_{obs}, intensity \rbrace$
can be obtained. During the raster scan observation, the phase center
of the feed is measured by a real-time measurement system (see Jiang
et al. 2019 for a detail), which can be converted to the telescope
pointing of the central beam (M01) in the horizontal coordinate system
(alt, az). The telescope pointing in the equatorial coordinates (ra,
dec in J2000) are then calculated from the horizontal coordinates alt
and az after considering the precession, nutation, aberration and also
the refraction effect of the atmosphere. With the time information
given by the measurement system, we can derive another series of pairs
of $\lbrace t_{obs}, coordinates \rbrace$. By cross-matching time
between the two series of flow data, a data-cube with information of
RA, Dec and intensity is derived. Then the data-cube can be regridded
to construct the intensity map.

\begin{figure}
\centering
\includegraphics[width=0.8\textwidth, angle=0]{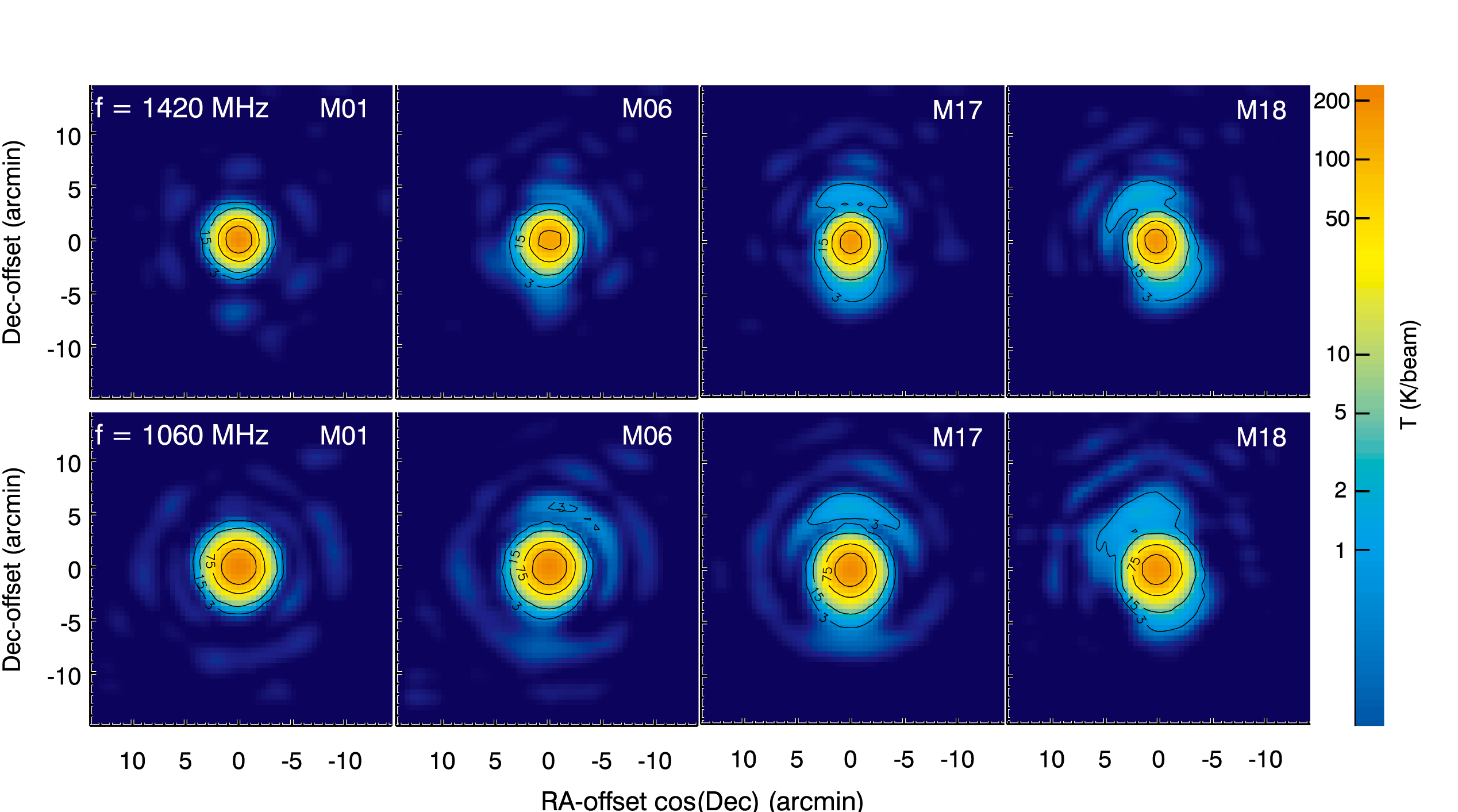}
\caption{{\it Upper panels}: distribution maps of total intensity
  obtained by raster scan observations of beam M01, M06, M17, and M18
  near the frequency of 1420\,MHz. {\it Lower panels}: similar to the
  upper panels, but for the observation frequency near 1060\,MHz. The
  target source is 3C454.3. The observation date is December 17,
  2018. In each map, the intensity distributions in a
  logarithmic-scale are indicated by different colors. Contours (black
  lines) show the intensities with levels of 3, 15, 75 and 150 K. The
  X- and Y-axises indicate the RA- and Dec-offsets relative to the
  equatorial coordinates of 3C454.3. A correction factor cos(Dec) is
  applied for the X-axis in each of the plots.}
\label{beam}
\end{figure}

\begin{figure}
\centering
\includegraphics[width=0.9\textwidth, angle=0]{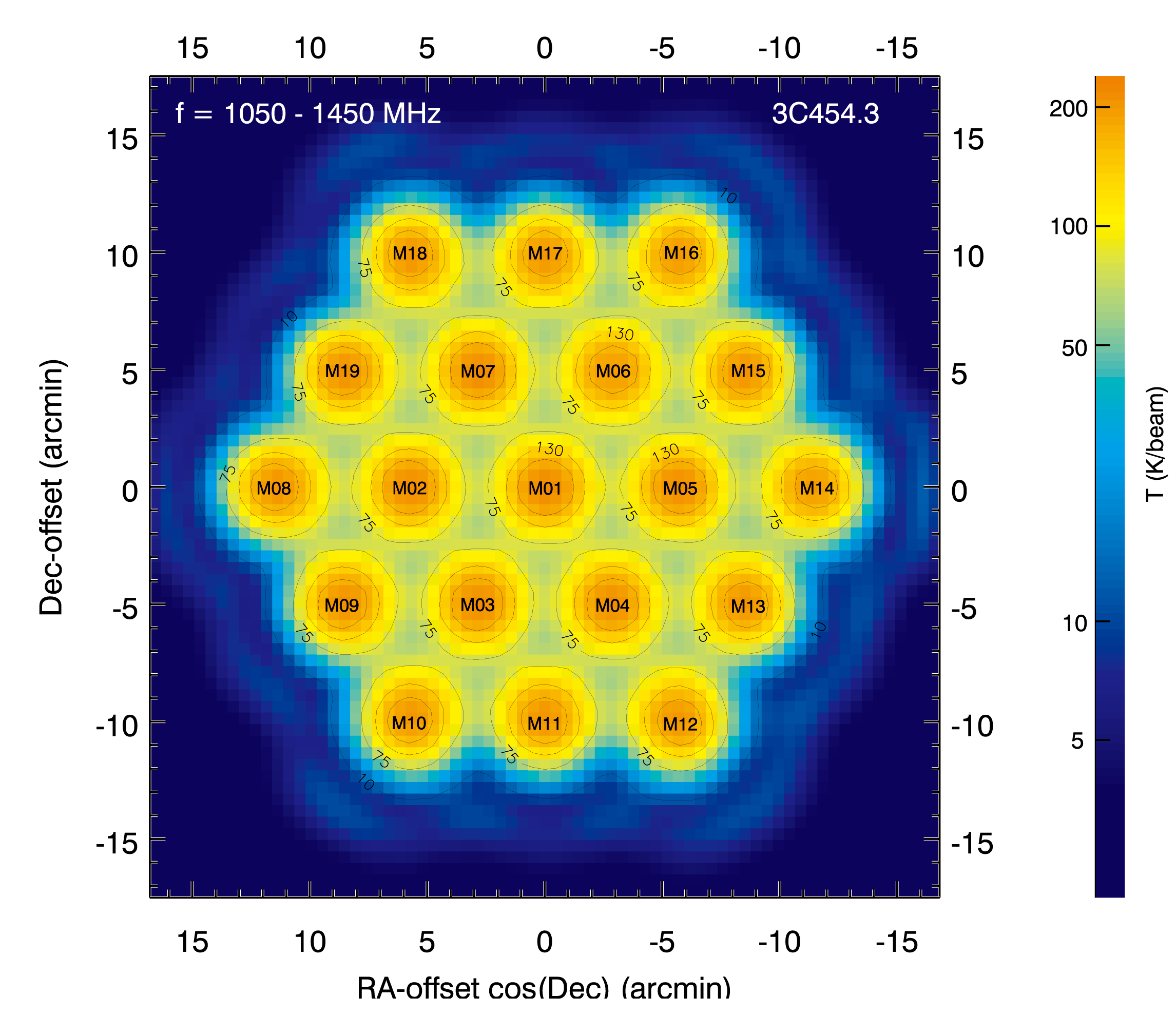}
\caption{Combined response of the 19-beam receiver of FAST. The target
  object is 3C454.3. The frequency coverage used to infer the total
  intensity is from 1050\,MHz to 1450\,MHz. The intensity
  distributions in a logarithmic-scale are indicated by different
  colors. Contours (black lines) show the intensities with levels of
  10, 75, 130 and 175 K. The X- and Y-axes indicate the RA- and
  Dec-offsets relative to the center of beam M01 (see
  Table~\ref{beamcen}). The numbering of the 19 beams (M01-M19) is
  marked in the plot.}
\label{19beam}
\end{figure}

An example of a total intensity map is given in
Fig.~\ref{track}. The continuum background is calculated and removed
to show the sidelobes and the weak field sources more clearly in the
plots of logarithmic-scale. We compare it with the continuum map of
the same sky area obtained by the NRAO/VLA Sky Survey
(NVSS) \footnote{https://www.cv.nrao.edu/nvss/postage.shtml} at
1.4\,GHz, and found that the detected positions of four weak radio
sources near RA- and Dec-offsets of ($-9^\prime$,+15$^\prime$),
($-$8$^\prime$,+8$^\prime$), (6.5$^\prime$, $-$16.5$^\prime$), and
($-$17$^\prime$,$-$15$^\prime$) are consistent. Then, the background
radio sources are fitted by 2-D Gaussian models and removed. Similar
procedure of data processing is made to the mapping data of other
beams, and examples of the results are given in Fig.~\ref{beam}. The
patterns of beam response shown in Fig.~\ref{beam} are smilar to the
radiation patterns given by simulations (Smith et al. 2016). According to the angular distance of the beam center to that of
the central beam M01, $d_{\rm M01}$, the rest 18 elements of the
19-beam receiver are distributed in three concentric ring-shaped area
(see Fig.~\ref{19beam}): 1st-ring) M02-M07, with $d_{\rm M01} \sim
5.8$ arcmin; 2nd-ring) M09, M11, M13, M15, M17, M19, with $d_{\rm M01}
\sim 10.0$ arcmin; 3rd-ring) M08, M10, M12, M14, M16, M18, with
$d_{\rm M01} \sim 11.6$ arcmin. As the increase of $d_{\rm M01}$, the
sidelobes become more and more significant. The morphology of the
sidelobes and their angular distances to the beam center depend on the
observation frequency. Other notable features are the noncircular
morphology and coma of the main beam, which are negligible for the
central beam, but seem to become more and more significant as the
increase of $d_{\rm M01}$ for the outer beams.

\begin{table}
\caption[]{\label{beamcen} Parameters of the beam center ($X_c$,
  $Y_c$) for each of the 19 beams (also see Fig.~\ref{19beam}),
  obtained by fitting the observed total
  intensity maps towards 3C454.3 observed between December 07
  and December 17, 2018. For the central beam M01, $X_c =
  (RA_{measured}-RA_{source})cos(Dec)$ and $Y_c = Dec_{measured} -
  Dec_{source}$, are the pointing errors of FAST in RA and Dec,
  respectively. For other beams, $X_c$ and $Y_c$ are taken as the
  angular offsets relative to the true center of beam M01. The mean
  values of $X_c$ and $Y_c$ for each of the 19 beams are given in the
  last two columns.}
\setlength{\tabcolsep}{3pt} \small
 \begin{tabular}{crrrrrrrrrr}
   \hline\noalign{\smallskip}

   & \multicolumn{2}{c}{(2018-12-07)}  &  \multicolumn{2}{c}{(2018-12-08)}    &  \multicolumn{2}{c}{(2018-12-16)}    &  \multicolumn{2}{c}{(2018-12-17)}   &       \multicolumn{2}{c}{Mean values}      \\
Scan Direction   & \multicolumn{2}{c}{(Along RA)}  &  \multicolumn{2}{c}{(Along Dec)}    &  \multicolumn{2}{c}{(Along RA)}    &  \multicolumn{2}{c}{(Along Dec)}   &       \multicolumn{2}{c}{ }      \\   
\hline
   & $X_c$ &  $Y_c$  & $X_c$ &  $Y_c$ & $X_c$ &  $Y_c$ & $X_c$ &  $Y_c$ & $X_c$ &  $Y_c$ \\
   Beam No.  & ($^\prime$)  & ($^\prime$)   & ($^\prime$)   & ($^\prime$)   & ($^\prime$)   & ($^\prime$)   & ($^\prime$)   & ($^\prime$)   & ($^\prime$)   & ($^\prime$)  \\
%
   \hline\noalign{\smallskip}
\hline\noalign{\smallskip}
M01  &  0.00    &  0.00     & $-$0.02  &   0.01     & $-$0.06  & $-$0.05   & $-$0.06  & $-$0.02  & $-$0.04   & $-$0.02 \\
M02  &  5.81    &  0.03     &  5.81    &   0.03     &  5.69    & $-$0.05   &  5.73    & $-$0.04  &  5.76     & $-$0.01 \\
M03  &  2.89    & $-$4.97   &  2.94    &  $-$4.93   &  2.81    & $-$5.06   &  2.83    & $-$4.99  &  2.86     & $-$4.98 \\  
M04  & $-$2.84  & $-$4.99   & $-$2.81  &  $-$5.00   & $-$2.95  & $-$5.03   & $-$2.95  & $-$5.02  & $-$2.89   & $-$5.01 \\
M05  & $-$5.75  & $-$0.03   & $-$5.74  &  $-$0.01   & $-$5.82  & $-$0.04   & $-$5.80  &  0.01    &  $-$5.78  & $-$0.02 \\
M06  & $-$2.90  &  5.00     & $-$2.91  &   5.00     & $-$2.92  &  4.94     & $-$2.94  &  4.97    &  $-$2.92  &  4.98   \\
M07  &  2.89    &  5.01     &  2.85    &   5.03     &  2.83    &  4.90     &  2.83    &  4.95    &   2.85    &  4.97   \\
M08  & 11.58    &  0.01     & 11.60    &   0.05     & 11.49    & $-$0.06   & 11.53    & $-$0.07  &  11.55    & $-$0.02 \\
M09  &  8.67    & $-$4.96   &  8.74    &  $-$4.93   &  8.56    & $-$5.07   &  8.64    & $-$5.06  &   8.65    & $-$5.01 \\
M10  &  5.81    & $-$9.99   &  5.89    &  $-$9.94   &  5.69    &$-$10.10   &  5.70    &$-$10.04  &   5.78    &$-$10.02 \\
M11  &  0.08    & $-$9.99   &  0.13    &  $-$9.96   & $-$0.09  &$-$10.05   & $-$0.10  &$-$10.06  &   0.00    &$-$10.04 \\
M12  & $-$5.69  &$-$10.06   & $-$5.65  & $-$10.10   & $-$5.90  &$-$10.07   & $-$5.90  &$-$10.07  &  $-$5.78  &$-$10.07 \\
M13  & $-$8.57  & $-$5.07   & $-$8.60  &  $-$5.05   & $-$8.75  & $-$5.05   & $-$8.74  & $-$5.04  &  $-$8.67  & $-$5.05 \\
M14  & ---      & ---       &$-$11.58  &  $-$0.05   & ---      & ---       &$-$11.64  &  0.00    & $-$11.61  & $-$0.02 \\
M15  & $-$8.67  &  4.93     & $-$8.69  &   5.00     & $-$8.67  &  4.96     & $-$8.72  &  4.99    &  $-$8.68  &  4.99   \\
M16  & $-$5.81  & 10.01     & $-$5.82  &  10.05     & $-$5.81  & 10.00     & $-$5.85  & 10.02    &  $-$5.83  & 10.02   \\
M17  & $-$0.02  & 10.01     & $-$0.02  &  10.03     & $-$0.01  &  9.95     & $-$0.07  &  9.99    &  $-$0.03  & 10.00   \\
M18  &  5.79    & 10.04     &  5.77    &  10.06     &  5.74    &  9.94     &  5.73    &  9.96    &   5.76    & 10.00   \\
M19  &  8.67    &  5.01     &  8.66    &   5.07     &  8.58    &  4.92     &  8.60    &  4.94    &   8.63    &  4.98   \\
   \noalign{\smallskip}\hline
\end{tabular}
\tablecomments{For beam M14, the parameters of beam
  center can not be well fitted from the observation data of December
  07 and December 16, 2018, due to a serious scanning effect.}
\end{table}

To fit the main-beam power pattern $P(\theta,\psi)$ with consideration
of elliptcity and coma, and determine the combined response of the
19-element multi-beam receiver, we follow the definition of a 'skew
Gaussian' given by Heiles et al. (2004). Nine parameters for the $i$-th beam
are the peak intensity $A_0$, beam center $X_c$ and $Y_c$, the average
beamwidth $\Theta_{0}$, beam ellipticity $\Theta_{1}$, beam
orientation $\phi_{\rm beam}$, coma $\alpha_{\rm coma}$ with
orientation $\phi_{\rm coma}$, and a factor $\epsilon$. For the
central beam M01, $X_c = (RA_{measured}-RA_{source})cos(Dec)$ and $Y_c
= Dec_{measured} - Dec_{source}$, are the pointing errors of FAST in
RA and Dec, respectively. For other beams, the fitting parameter $X_c$
and $Y_c$ are taken as the angular offsets relative to the true center
of beam M01. For each beam, the main-beam power pattern is
parameterized in the polar coordinates ($\theta$, $\phi$) as:
\begin{equation}
      P(\theta,\phi) = A_0 exp[-\frac{\theta^2(1-min\lbrace\alpha_{\rm coma}\frac{\theta_{coma}}{\Theta_0},\epsilon\rbrace)}{\Theta^2}],
\label{beamshape}      
\end{equation}
where $\theta_{coma} = \theta cos(\phi - \phi_{coma})$, $\Theta =
\Theta_0 + \Theta_1cos2(\phi-\phi_{beam})$. Here, $\theta$ is the
angular distance of the position to the true beam center, $\phi$ is
the position angle in the equatorial coordinates (RA, Dec in J2000),
and defined to be zero along the positive RA-offset axis and increases
towards positive Dec-offset axis. Follow Heiles et al. (2004), we also take
$\epsilon$ as a constant parameter 0.75 during the fit. Together with
the term min$\lbrace\alpha_{\rm
  coma}\frac{\theta_{coma}}{\Theta_0},\epsilon\rbrace$, they are used
to prevent the coma term from unduly distorting the beam far from the
beam center (Heiles et al. 2004). The MPFIT
package\footnote{http://www.physics.wisc.edu/~craigm/idl/fitting.html}
is adopted to make the minimization. In the following, we give a
general discussion about the combined beam response, the beamwidth and
also the pointing errors of FAST, obtained by fitting the observed
total intensity (XX+YY) maps.

\subsubsection{The Convolved Beam Pattern of the 19-beam Receiver}

To measure the beam properties, we analyzed the mapping data toward
3C454.3, which is the strongest source at 1.4~GHz among the observed
$37^\prime\times37^\prime$ targets (1902+319, 1859+129, 3C454.3, and
2023+318, etc) by FAST. Raster scan covering a sky area of
$\sim37^\prime\times37^\prime$ were made four times along RA or Dec
directions. The fitted parameters of the beam center for each of the
19 beams is given in Table~\ref{beamcen}. With the fitted beam center,
we combine the intensity distribution maps derived by each of the 19
beams to give a composite response of the 19-beam receiver as shown in
Fig.~\ref{19beam}. The observed layout of the 19-beam receiver is
consistent with the schematic diagram given by Dunning et al. (2017)
and Li et al. (2018). Around the main beams, the responses of first
sidelobes for the 12 outer beams (M08-M19) are also visible in the
logarithmic-scale plot. Similar results were obtained by analyzing the
observation data toward other sources, e.g., 3C380, 2023+318.

It is necessary to emphasize that the intensity distributions given in
Fig~\ref{beam} and Fig.~\ref{19beam} are the beam response to the
point source 3C454.3 verseus the angular offsets. They are the
convolution of the beam pattern with the brightness distribution on
the sky.

\subsubsection{Beamwidth}

\begin{table}
\caption[]{\label{beamw} Half-power beamwidth of the 19-beam receiver
  of FAST for different observation frequency, which are used for the
  plot given in Fig.~\ref{beamwidth}. }
\setlength{\tabcolsep}{5pt} \small
 \begin{tabular}{cccccccccccccc}
   \hline\noalign{\smallskip}
   \hline
   Beam No.          &  \multicolumn{13}{c}{Frequency (MHz)}  \\  
                  & 1060 & 1080 & 1100 & 1120 & 1140 & 1300 & 1320 & 1340 & 1360 & 1380 & 1400 & 1420 & 1440 \\  
   \hline\noalign{\smallskip}
\hline\noalign{\smallskip}
M01  & 3.44$^\prime$ & 3.37$^\prime$ & 3.31$^\prime$ & 3.29$^\prime$ & 3.29$^\prime$ & 3.01$^\prime$ & 2.98$^\prime$ & 2.93$^\prime$ & 2.89$^\prime$ & 2.85$^\prime$ & 2.82$^\prime$ & 2.82$^\prime$ & 2.82$^\prime$ \\
M02  & 3.44$^\prime$ & 3.37$^\prime$ & 3.31$^\prime$ & 3.33$^\prime$ & 3.33$^\prime$ & 3.04$^\prime$ & 3.00$^\prime$ & 2.96$^\prime$ & 2.92$^\prime$ & 2.88$^\prime$ & 2.85$^\prime$ & 2.85$^\prime$ & 2.84$^\prime$ \\
M03  & 3.44$^\prime$ & 3.38$^\prime$ & 3.30$^\prime$ & 3.33$^\prime$ & 3.32$^\prime$ & 3.03$^\prime$ & 3.00$^\prime$ & 2.95$^\prime$ & 2.91$^\prime$ & 2.87$^\prime$ & 2.84$^\prime$ & 2.84$^\prime$ & 2.83$^\prime$ \\
M04  & 3.44$^\prime$ & 3.38$^\prime$ & 3.31$^\prime$ & 3.34$^\prime$ & 3.32$^\prime$ & 3.05$^\prime$ & 3.01$^\prime$ & 2.97$^\prime$ & 2.93$^\prime$ & 2.90$^\prime$ & 2.86$^\prime$ & 2.86$^\prime$ & 2.85$^\prime$ \\
M05  & 3.43$^\prime$ & 3.36$^\prime$ & 3.29$^\prime$ & 3.31$^\prime$ & 3.32$^\prime$ & 3.03$^\prime$ & 2.99$^\prime$ & 2.95$^\prime$ & 2.91$^\prime$ & 2.87$^\prime$ & 2.83$^\prime$ & 2.83$^\prime$ & 2.82$^\prime$ \\
M06  & 3.44$^\prime$ & 3.39$^\prime$ & 3.31$^\prime$ & 3.34$^\prime$ & 3.34$^\prime$ & 3.07$^\prime$ & 3.03$^\prime$ & 3.01$^\prime$ & 3.01$^\prime$ & 2.99$^\prime$ & 2.95$^\prime$ & 2.94$^\prime$ & 2.99$^\prime$ \\
M07  & 3.45$^\prime$ & 3.39$^\prime$ & 3.31$^\prime$ & 3.33$^\prime$ & 3.34$^\prime$ & 3.02$^\prime$ & 2.99$^\prime$ & 2.95$^\prime$ & 2.92$^\prime$ & 2.88$^\prime$ & 2.86$^\prime$ & 2.86$^\prime$ & 2.86$^\prime$ \\
M08  & 3.46$^\prime$ & 3.42$^\prime$ & 3.38$^\prime$ & 3.35$^\prime$ & 3.34$^\prime$ & 3.04$^\prime$ & 3.02$^\prime$ & 2.98$^\prime$ & 2.95$^\prime$ & 2.92$^\prime$ & 2.90$^\prime$ & 2.88$^\prime$ & 2.86$^\prime$ \\
M09  & 3.48$^\prime$ & 3.41$^\prime$ & 3.36$^\prime$ & 3.33$^\prime$ & 3.30$^\prime$ & 3.04$^\prime$ & 3.01$^\prime$ & 2.97$^\prime$ & 2.94$^\prime$ & 2.91$^\prime$ & 2.88$^\prime$ & 2.88$^\prime$ & 2.86$^\prime$ \\
M10  & 3.52$^\prime$ & 3.46$^\prime$ & 3.40$^\prime$ & 3.38$^\prime$ & 3.35$^\prime$ & 3.08$^\prime$ & 3.05$^\prime$ & 3.03$^\prime$ & 3.00$^\prime$ & 2.97$^\prime$ & 2.95$^\prime$ & 2.94$^\prime$ & 2.92$^\prime$ \\
M11  & 3.49$^\prime$ & 3.43$^\prime$ & 3.38$^\prime$ & 3.37$^\prime$ & 3.34$^\prime$ & 3.05$^\prime$ & 3.02$^\prime$ & 3.00$^\prime$ & 2.97$^\prime$ & 2.94$^\prime$ & 2.92$^\prime$ & 2.91$^\prime$ & 2.90$^\prime$ \\
M12  & 3.49$^\prime$ & 3.44$^\prime$ & 3.39$^\prime$ & 3.37$^\prime$ & 3.34$^\prime$ & 3.07$^\prime$ & 3.05$^\prime$ & 3.03$^\prime$ & 3.00$^\prime$ & 2.97$^\prime$ & 2.95$^\prime$ & 2.94$^\prime$ & 2.90$^\prime$ \\
M13  & 3.48$^\prime$ & 3.41$^\prime$ & 3.34$^\prime$ & 3.33$^\prime$ & 3.32$^\prime$ & 3.04$^\prime$ & 3.01$^\prime$ & 2.98$^\prime$ & 2.95$^\prime$ & 2.93$^\prime$ & 2.91$^\prime$ & 2.91$^\prime$ & 2.89$^\prime$ \\
M14  & 3.53$^\prime$ & 3.48$^\prime$ & 3.43$^\prime$ & 3.41$^\prime$ & 3.38$^\prime$ & 3.10$^\prime$ & 3.07$^\prime$ & 3.04$^\prime$ & 3.01$^\prime$ & 2.98$^\prime$ & 2.95$^\prime$ & 2.94$^\prime$ & 2.92$^\prime$ \\
M15  & 3.49$^\prime$ & 3.42$^\prime$ & 3.36$^\prime$ & 3.36$^\prime$ & 3.36$^\prime$ & 3.06$^\prime$ & 3.01$^\prime$ & 2.98$^\prime$ & 2.94$^\prime$ & 2.91$^\prime$ & 2.89$^\prime$ & 2.88$^\prime$ & 2.86$^\prime$ \\
M16  & 3.51$^\prime$ & 3.46$^\prime$ & 3.41$^\prime$ & 3.39$^\prime$ & 3.34$^\prime$ & 3.10$^\prime$ & 3.07$^\prime$ & 3.05$^\prime$ & 3.02$^\prime$ & 2.99$^\prime$ & 2.97$^\prime$ & 2.95$^\prime$ & 2.93$^\prime$ \\
M17  & 3.49$^\prime$ & 3.42$^\prime$ & 3.35$^\prime$ & 3.34$^\prime$ & 3.33$^\prime$ & 3.02$^\prime$ & 3.00$^\prime$ & 2.99$^\prime$ & 2.97$^\prime$ & 2.94$^\prime$ & 2.92$^\prime$ & 2.92$^\prime$ & 2.91$^\prime$ \\
M18  & 3.55$^\prime$ & 3.51$^\prime$ & 3.37$^\prime$ & 3.33$^\prime$ & 3.29$^\prime$ & 3.05$^\prime$ & 3.04$^\prime$ & 3.00$^\prime$ & 2.98$^\prime$ & 2.96$^\prime$ & 2.95$^\prime$ & 2.94$^\prime$ & 2.93$^\prime$ \\
M19  & 3.42$^\prime$ & 3.41$^\prime$ & 3.36$^\prime$ & 3.33$^\prime$ & 3.34$^\prime$ & 3.03$^\prime$ & 3.00$^\prime$ & 2.98$^\prime$ & 2.95$^\prime$ & 2.93$^\prime$ & 2.91$^\prime$ & 2.90$^\prime$ & 2.89$^\prime$ \\

   \noalign{\smallskip}\hline
\end{tabular}
\end{table}

\begin{figure}
\centering
\includegraphics[width=0.5\textwidth, angle=90]{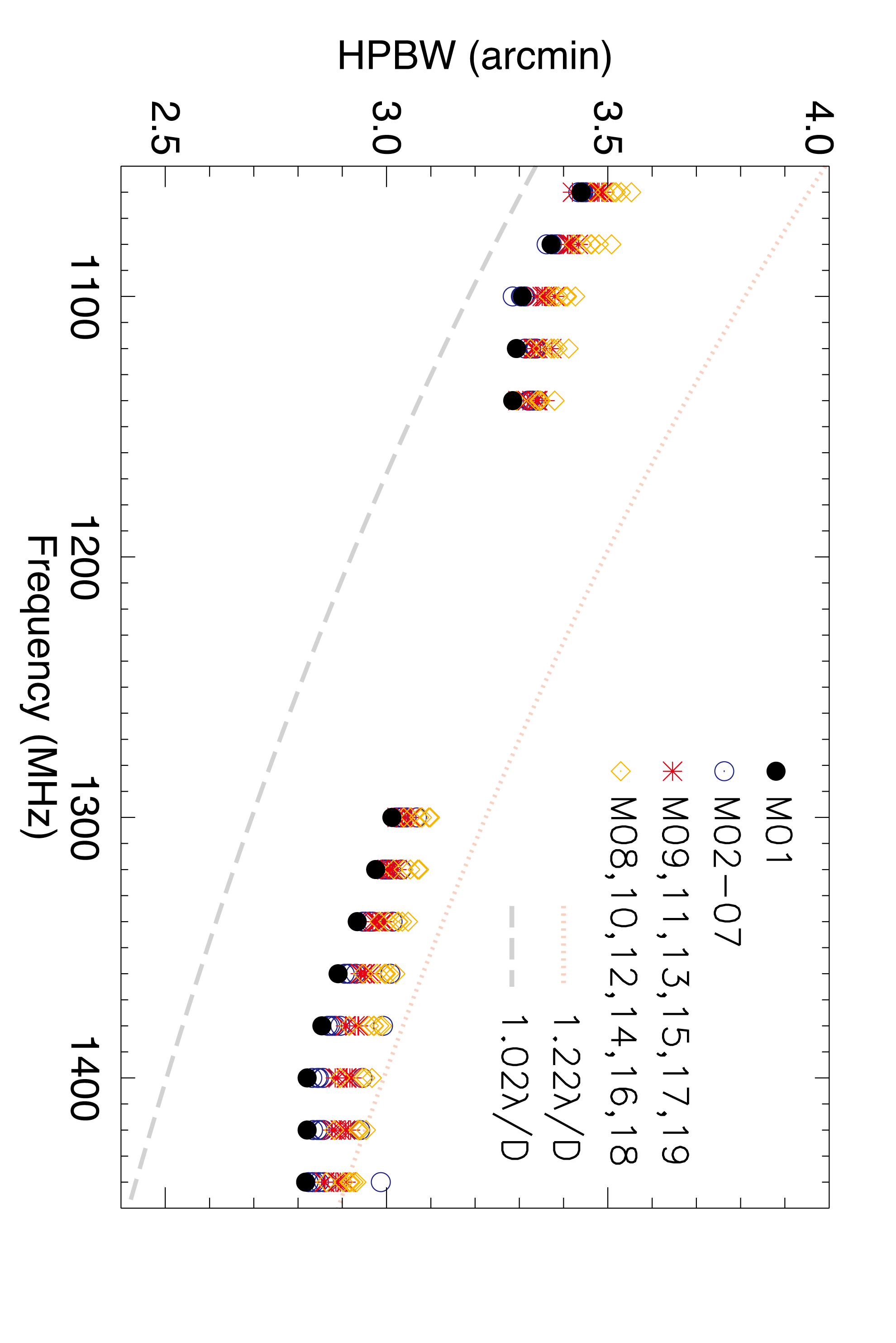}
\caption{Half-power beamwidth (HPBW) verseus observation frequency for
  the 19-beam receiver of FAST. The target source is 3C454.3. The
  observation date is December 17, 2018. The beamwidths corresponding
  to the frequency range of 1160$-$1280\,MHz can not be well fitted,
  as the strong RFI appear during the observation.}
\label{beamwidth}
\end{figure}

The half-power beamwidth (HPBW) is related to the average beamwidth
$\Theta_0$ by HPBW $= 2(ln2)^{1/2}\Theta_0$, here, $\Theta_0$ is
derived by fitting the total intensity distribution maps with a 'skew
Gaussian' model as shown in Eq.~\ref{beamshape}. The HPBW as a
function of the observation frequency for the 19 beams are given in
Fig.~\ref{beamwidth}. In comparison to the central beam, the outer
beams tend to have larger beamwidths, with a difference no more than
0.2~arcmin. The HPBW decreases with the increase of frequency. As
discussed in Jiang et al. (2019), the theoretical HPBW of a telescope
with diameter $D$ is HPBW $= 1.02\lambda/D$ for a uniformly
illuminated aperture and HPBW $= 1.22\lambda/D$ for a cosine-tapered
illuminated aperture. Here, a diameter of $D=300$m is assumed. As
shown in Fig.~\ref{beamwidth}, the fitted beamwidths of the 19-beam
receiver fall within the two theoretical curves of HPBW verseus
frequency, but with a less steep slope than that of a standard
$\lambda/D$ proportionality.

\begin{figure}
\centering
\includegraphics[width=0.8\textwidth, angle=0]{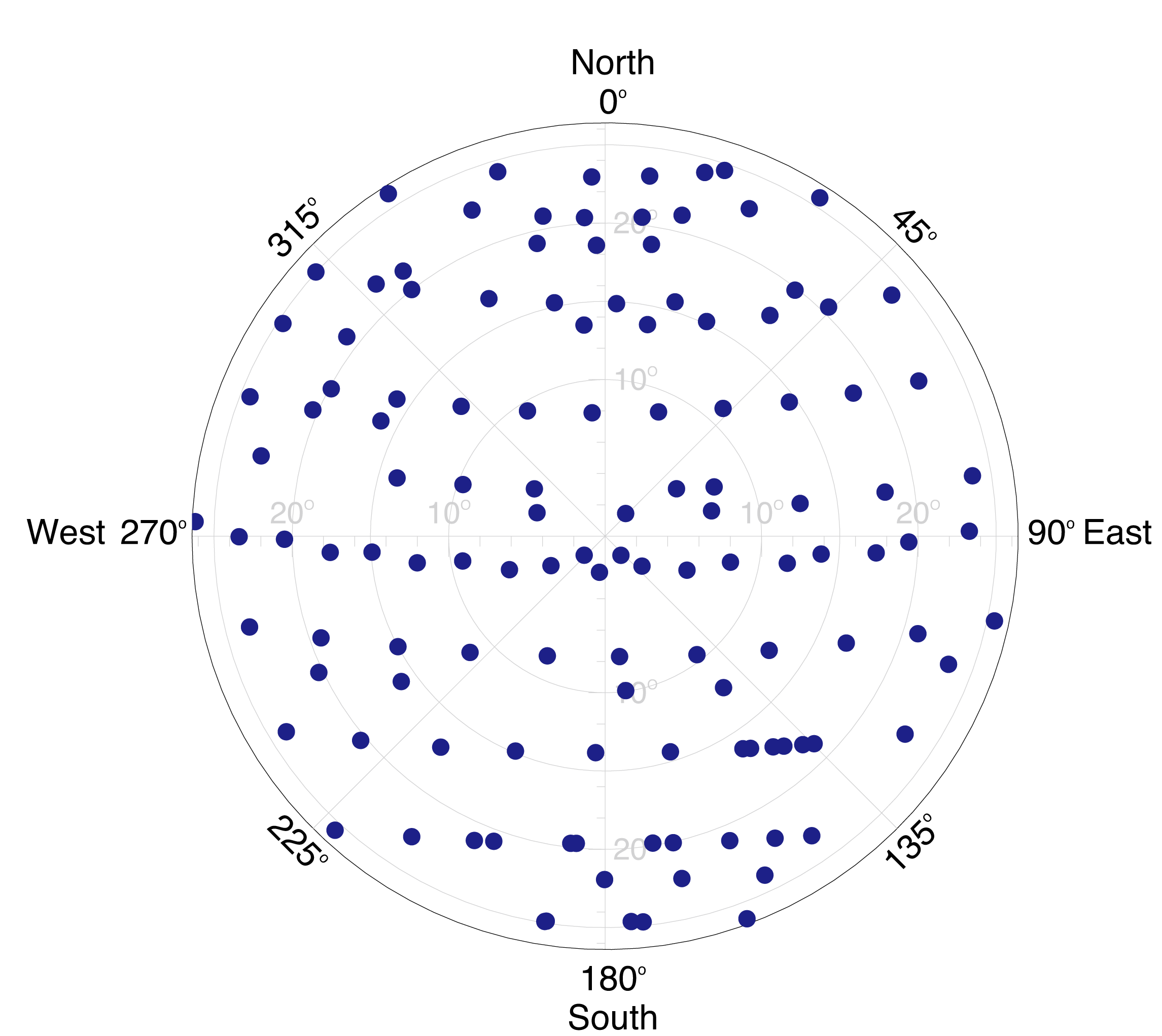}
\caption{Distributions of the observed pointing calibrators in the sky
  coverage of FAST with full gain (zenith angle $\lesssim
  26.4^\circ$).  The observation period is from February 16 to March
  15, 2019. In this plot, the zenith angle is from $0^\circ$ to
  26.4$^\circ$ as indicated by grey circles, the azimuth angle is from
  $0^\circ$ to 360$^\circ$. The directions marked in the periphery of
  the plot are the geographic north, east, south and west at the FAST
  site.}
\label{pointingdis}
\end{figure}

\subsubsection{Pointing errors}

As discussed in Jiang et al. (2019), the analysis of pointing errors
can be used to evaluate the pointing accuracy, improve the pointing of
the telescope, and also guide the error analysis of
observations. Systematic observations have been made to measure the
pointing errors of FAST by raster scan observations. As we only care
about the pointing errors of the central beam M01, the mapping area
toward a target source is set to $\sim7^\prime\times7^\prime$. About 7
minutes is needed to acquire such a map with subscan separation of
1$^\prime$. The target sources are selected from the catalogue of
pointing calibrators given by Condon \& Yin (2001), which contains
3399 strong and compact radio sources with accurate positions from the
NVSS, and uniformly covering the sky north of $\delta = - 40^\circ$
(J2000). The dataset used in the analysis are observed from February
16 to March 15, 2019, which include 126 raster scan observations along
RA or Dec directions. As shown in Fig.~\ref{pointingdis}, the observed
pointing calibrators distribute widely in the sky coverage of FAST
with full gain (zenith angle $\lesssim 26.4^\circ$).

\begin{figure}
\centering
\includegraphics[width=0.5\textwidth, angle=90]{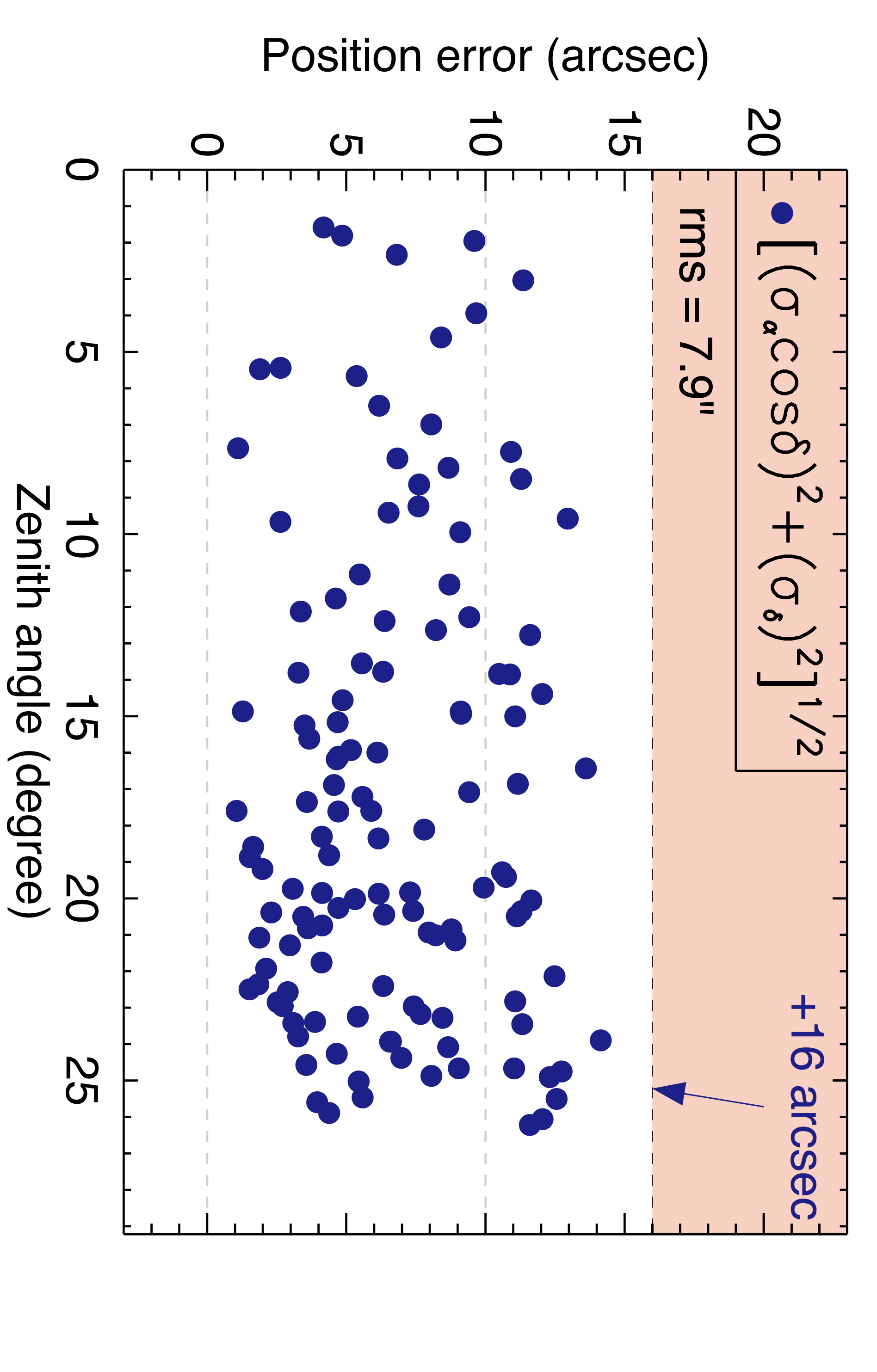}
\caption{Pointing error
  $[(\sigma_{\alpha}cos\delta)^2+(\sigma_\delta)^2]^{1/2}$ verseus
  zenith angle for the 19-beam receiver of FAST. }
\label{pointingerror}
\end{figure}

The observation data are processed in a similar procedure as described
in Sect.2.2.2. To fit the pointing errors in RA and Dec directions, a
2-D Gaussian model is adopted, which is good enough as the central
beam do not present significant beam ellipticity and coma feature
(Fig.~\ref{track} and Fig.~\ref{beam}). The pointing errors
$[(\sigma_{\alpha}cos\delta)^2+(\sigma_\delta)^2]^{1/2}$ are typically
smaller than 16$^{\prime\prime}$ as shown in
Fig.~\ref{pointingerror}. The rms of pointing errors is
7.9$^{\prime\prime}$, which is less than one-tenth of the beamwidth
near 1450\,MHz (beamwidth$_{1450}$/10 $\sim$ 2.8$^\prime$/10 =
16.8$^{\prime\prime}$).

\subsection{Aperture  Efficiency and System Temperature}
\label{subsec:tsys_eff}

Aperture  efficiency ($\eta$) and system temperature (T$\rm_{sys}$)  are two basic parameters of FAST. Both of them are expected to vary as a function of zenith angle ($\rm\theta_{ZA}$) and frequency.   The 19 beams are  divided into 4 categories  according to  separation to the central beam.  The 4 categories are : 1)  Beam 1;  2)  Beam 2,  3, 4, 5, 6, and 7;  3)  Beam 8, 10, 12, 14, 16,  and 18; 4) Beam 9, 11, 13, 15, 17 and 19.  Though the performance of aperture efficiency and system temperature in each category is expected to be same,  asymmetry of the receiver platform may lead to deviation. Thus we obtained the measurements toward all 19 beams. 

The aperture efficiency curve was obtained  by repeating observations of  the calibrator 3C286 and its off position at different zenith angles. The position switch mode that is introduced in section \ref{sec:observation} was adopted. This mode provides quick switch  between ON and OFF source position.  The separation between ON and OFF position was  selected to allow for measuring aperture efficiency curve of two beams (e.g, Beam 1 and 19) simultaneously, which increases measurement efficiency. 

Observations of Beam 1, 2, 8, and 19 were taken during August 7, 2019 and  August 24, 2019.   To make it clear, we take the aperture efficiency measurement of Beam 1 and 19 as an example.  At first, the telescope tracked the calibrator 3C286 with Beam 1 (source ON  for Beam 1, source OFF position for Beam 19) for 90 s, and then switched to the sky position with offset of  (-8.85$^{\prime}$, -5.11$^{\prime}$).  At this position, the calibrator 3C286 was OFF for Beam 1 but  lay in Beam 19.  After tracking  for 90s, the telescope switched back to track 3C286 with Beam 1.  A cycle consists of one ON and one OFF phase for each beam. A total of 90 cycles were taken during available tracking time of  $\sim$ 6 hours of 3C286. Supplementary measurements of the aperture efficiency curve for the rest 15 beams were taken by tracking 3C286 during December 19, 2019 and December 25, 2019.  During these observations, the total tracking time in a cycle was reduced into 60 s, including 30 s for both source on and source off. A total of 85 cycles and $\sim $ 3 hours were taken for measurement of both two beams.  

The system temperature curve of 19 beams was obtained by continuously tracking a clean sky position ($\alpha_{2000}$=23$^{h}$30$^m$0$^s$.0, $\delta_{2000}$=25$^{\circ}$39$^{\prime}$10.6") from rise to set on September 14, 2019.  

In the above measurements, spectral backend with sampling time of 1.00663296 s and channel width of 7.62939 kHz  was adopted to record the data.   The high noise signal ($\sim 11$ K) was injected at a synchronized period of 2.01326592 s.  The duration time of cal on and cal off is 1.00663296 s. During calculation, channel width of the data were smoothed to 1 MHz. The aperture efficiency was derived for each observation cycle while the system temperature was derived for each sampling time. 

\subsubsection{Aperture efficiency}
\label{subsubsec:eta}

Based on absolute measurement of noise dipole in Section \ref{subsec:noise_dipole}, the observed data were calibrated into antenna temperature $T_A$ in K .  The antenna temperature of 3C286,  $T_A^{3C286}$ was derived by the difference between source ON and source OFF data for each beam.  

3C286 is a stable flux calibrator. The frequency dependency of the spectral flux density of 3C286 could be fitted with a polynomial function (Perley \&  Butler 2017),

\begin{equation}
log(S) = a_0 + a_1 log(\nu_G)+a_2 [log(\nu_G)]^2+a_3 [log(\nu_G)]^3, 
\label{eq:polyfit}
\end{equation}
where S and $\nu_G$ are the spectral flux density in Jy and the frequency respectively.  We adopted the value of $a_0=1.2481$, $a_1=-0.4507$, $a_2=-0.1798$ and $a_3=0.0357$ that is valid for 3C286 at frequency range of [0.05-50] GHz (Perley \&  Butler 2017).  

Measured gain ($G$) of FAST  is expressed as $G_0=A_\textrm{eff}/2k$ , in which $A_\textrm{eff}$ is  effective illumination area. In observation, $G$ is calculated by ratio between antenna temperature and flux density, $G = T_A^{3C286}/S_A^{3C286}$.  The prefect gain $G_0$ of FAST is $G_0=A_\textrm{geo}/2k$ for single polarization, in which  $A_\textrm{geo}$ and $k$ are geometric  illumination area and Boltzmann constant respectively.  The geometric illumination area with diameter of 300 m leads to  $G_0$= 25.6 K/Jy for FAST.  

The aperture efficiency $\eta$ of FAST  is derived by, 

\begin{equation}
\eta = G/G_0 = A_\textrm{eff}/A_\textrm{geo} =  \frac{2k T_A^{3C286}}{A_\textrm{geo} S_A^{3C286}}
\end{equation}

For the FAST system, $\eta$ is contributed by 6 main components:

\begin{enumerate} 
\item $\eta\rm_{sf}$, the reflection efficiency of main reflector. It is described by $Ruze$ equation, $\eta_{sf}=e^{-(\frac{4\pi\varepsilon}{\lambda})^2}$, in which $\varepsilon$ and $\lambda$ are root mean square (RMS) of the surface error and observational wavelength, respectively.  The RMS of the surface error is contributed by controlling accuracy of panels, which is  $\sim 5$ mm at 21 cm (Jiang et al.\ 2019). This  results in $\eta_{sf}$ of  $\sim$ 91\%.

\item  $\eta\rm_{bl}$, the efficiency when the shielding of the feed cabin is considered. The diameter of feed cabin of $\sim 10$ m leads to  $\eta\rm_{bl}$ value of 99.9\%. 

\item $\eta\rm_{s}$, spillover efficiency. 

\item $\eta\rm_t$, illumination efficiency of the feed. It equals 76\% when a -13 dB Gaussian illumination is adopted (Jiang et al. 2019). 

\item $\eta\rm_{misc}$, efficiency of other aspects including the offset of the feed phase, matching loss of feed. 

\item  $\eta\rm_{sloss}$, the percent of effective surface area compared to a 300-m diameter paraboloid.  The value is 1 when $\rm\theta_{ZA}$ is less than 26.4$^\circ$.  When $\rm\theta_{ZA}$ is greater than 26.4$^\circ$, the reflection area decreases. $\eta\rm_{sloss}$ would decrease and  reach $\sim$ 2/3 when $\rm\theta_{ZA}= 40^\circ$. 

\end{enumerate}

The aperture efficiency is a synthetical result of above effects and is described by the following equation,

\begin{equation}
\eta = \rm \eta_{sf}\cdot \eta_{bl} \cdot \eta_{s}\cdot \eta_t \cdot \eta_{misc} \cdot \eta_{sloss}
\label{eq:efficiency}
\end{equation}

\begin{figure}
\centering
  \includegraphics[width=0.45\textwidth]{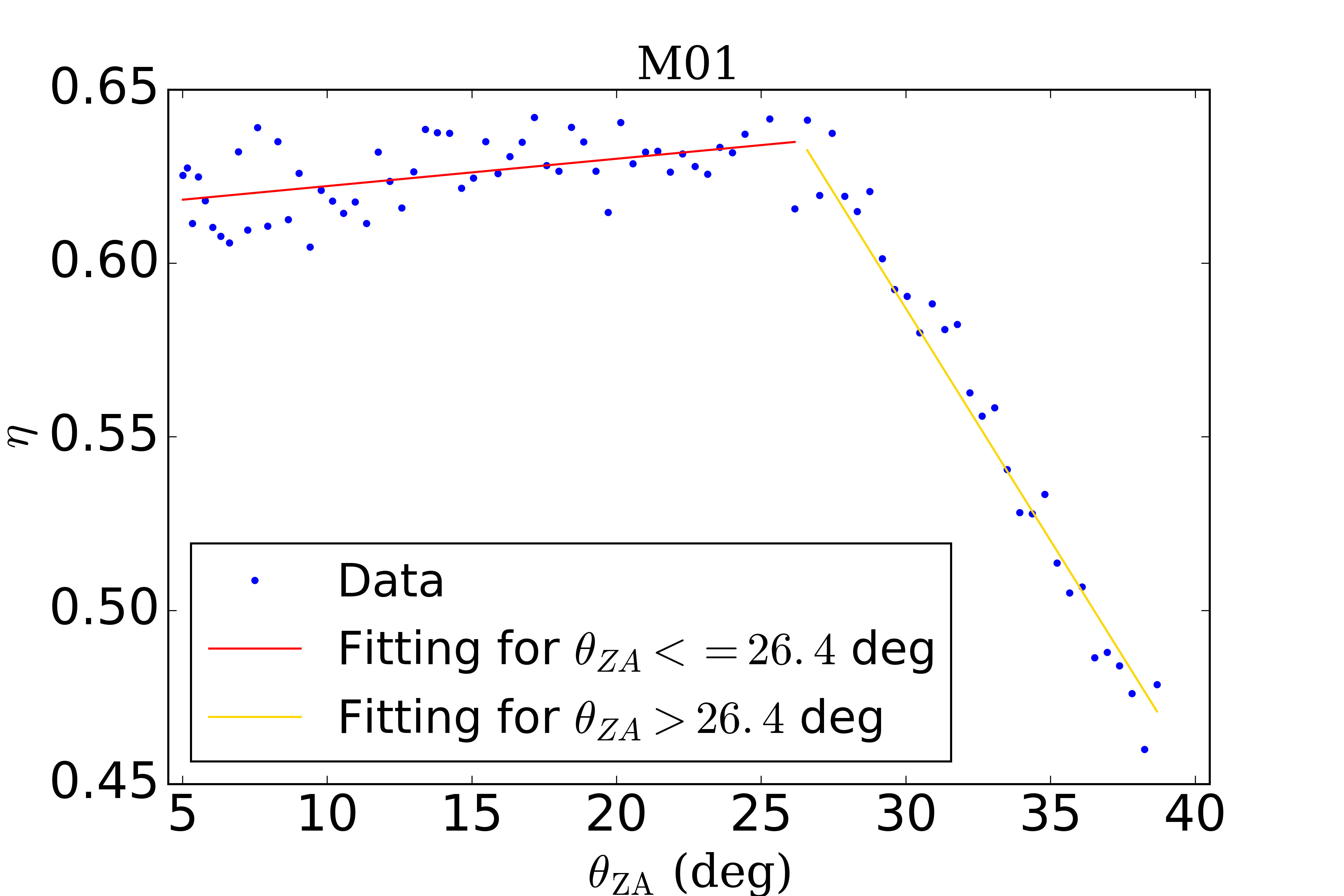}
  \includegraphics[width=0.45\textwidth]{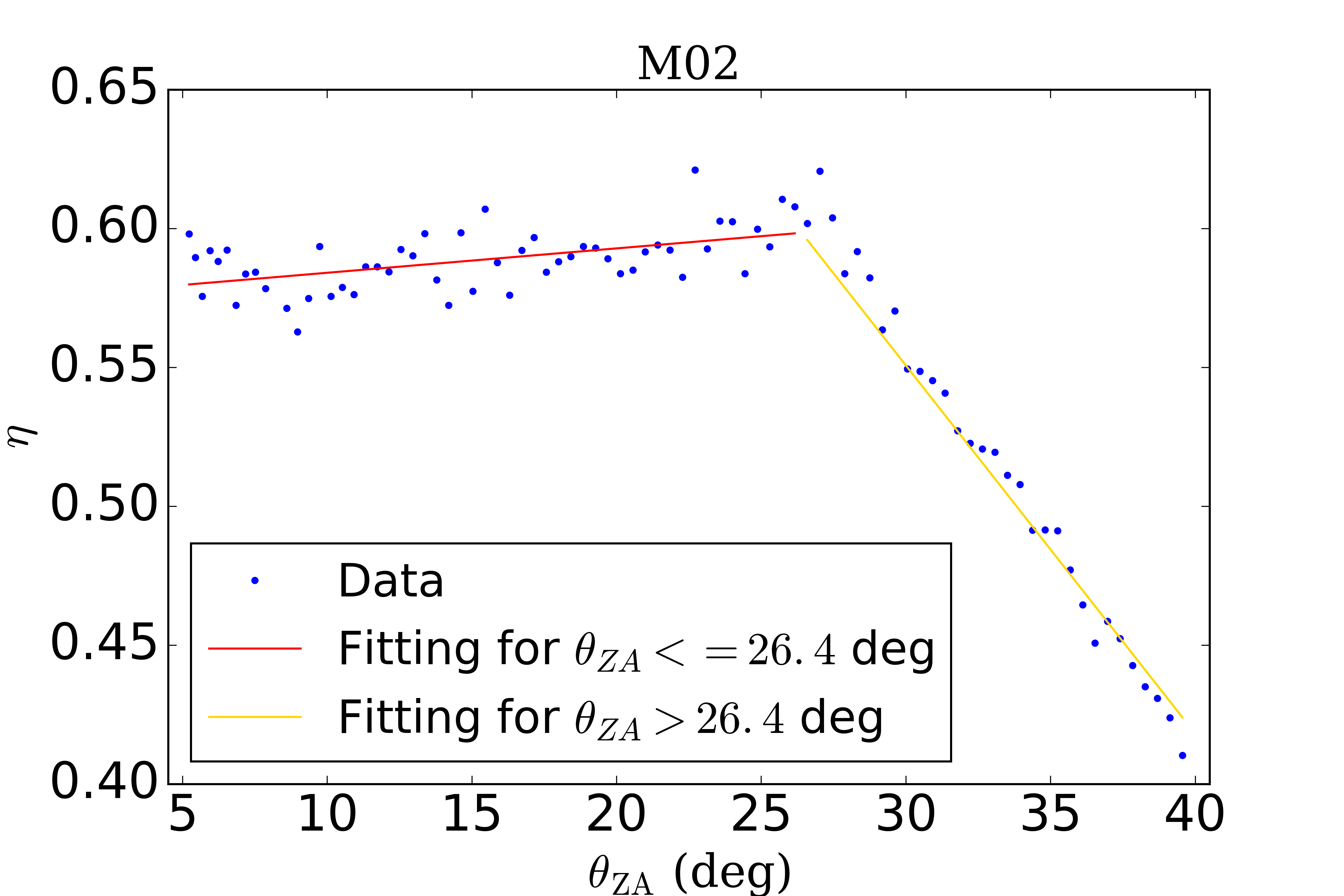}
  \includegraphics[width=0.45\textwidth]{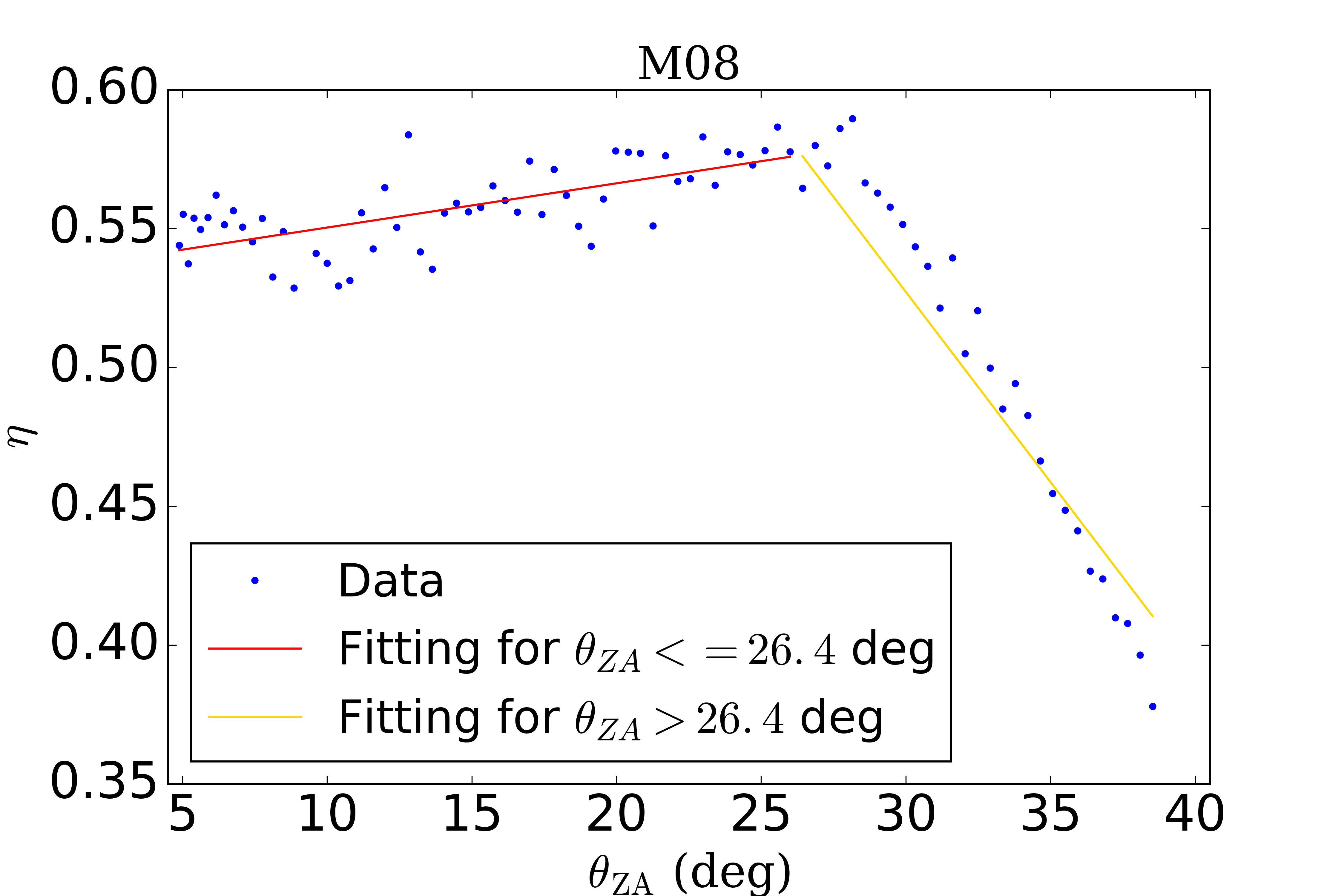}
  \includegraphics[width=0.45\textwidth]{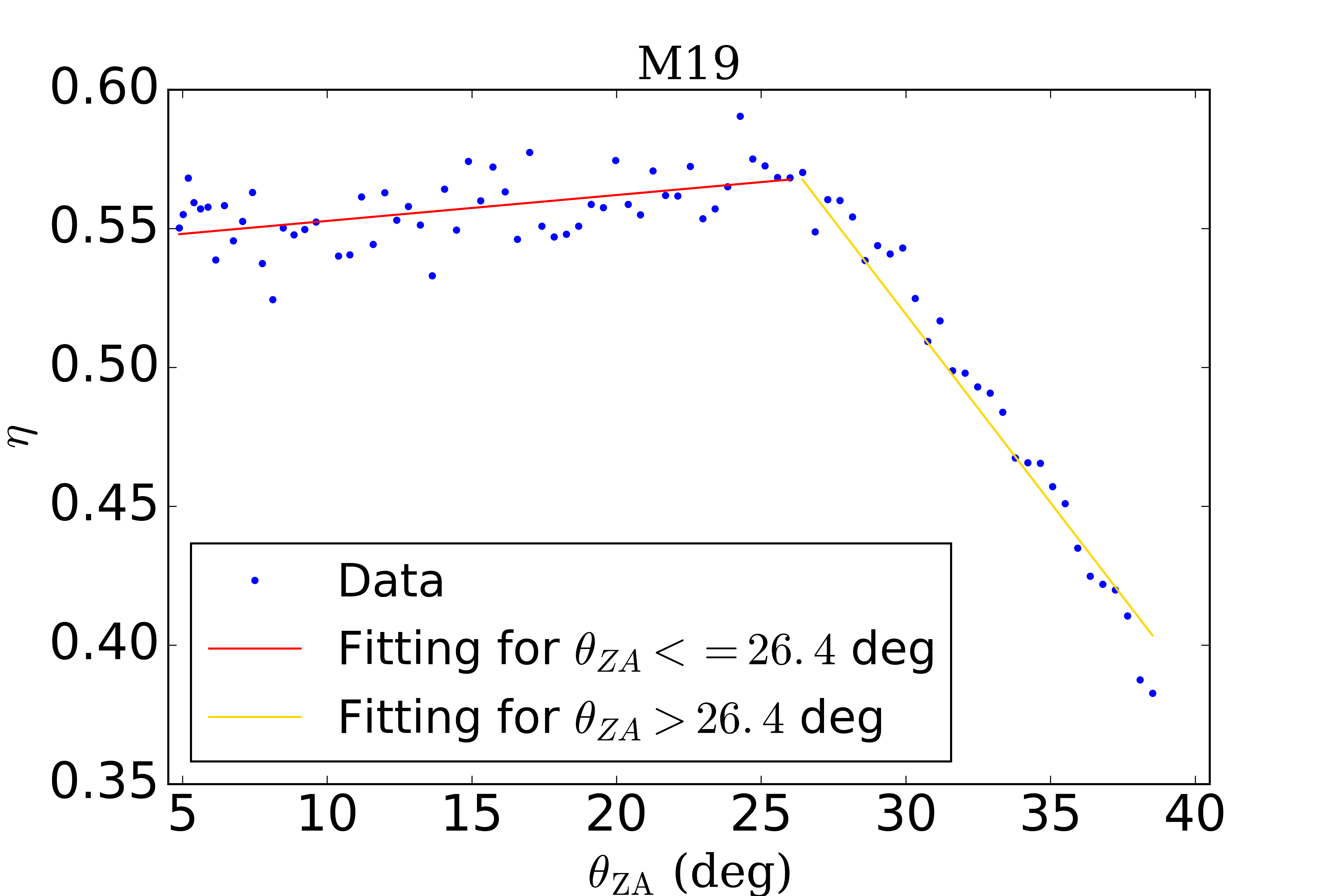}
  \caption{Measured $\eta$ curve as a function of zenith angle $\theta\rm_{ZA}$ at 1400 MHz for Beam 1, 2, 8 and 19.   Fitting results when $\theta\rm_{ZA}$ $\leq$ 26.4$^\circ$ and $\theta\rm_{ZA}$ > 26.4$^\circ$ are represented with red and gold solid line, respectively.}
\label{fig:etafit}
\end{figure}

The value of $\eta\rm_{sf}$,$\eta\rm_{bl}$,$\eta\rm_{s}$,$\eta\rm_t$,and $\eta\rm_{misc}$  are almost constant. The value of  $\eta$ is mainly determined  by $\eta\rm_{sloss}$. As shown in Fig. \ref{fig:etafit},  the value $\eta$ at all frequency would keep almost constant when $\rm\theta_{ZA}$ is less than 26.4$^\circ$ and decreases linearly when $\rm\theta_{ZA}$>26.4$^\circ$. This is consistent with the fact that FAST would lose part of reflection panels when $\rm\theta_{ZA}$>26.4$^\circ$. 

We fitted variation of $\eta$ as a function of $\rm\theta_{ZA}$ at specific frequency with two linear equations. The formula is shown as follows, 

\begin{equation} 
\eta = \left\{
                        \begin{array}{lr}
                          a\rm\ \theta_{ZA}+b,\ 0^\circ \leq \theta_{ZA} \leq 26.4^\circ  \\
                          c\rm\ \theta_{ZA}+d,\ 26.4^\circ < \theta_{ZA} \leq 40^\circ.
                        \end{array}
\right.
\end{equation}
In which $d$ satisfies the equation $d=b+26.4(a-c)$. The fitting result of parameter $a$,$b$ and $c$ for different frequencies are shown in Table \ref{table:etapara}. As an example, the fitting results of $\eta$ for Beam 1, 2, 8 and 19 are shown in Fig. \ref{fig:etafit}. The fitting results of rest 15 beams are shown in Fig. \ref{fig:etafit_restbeams}.

The gain of FAST could be expressed with $G= \eta G_0$, in which $G_0$ = 25.6 K/Jy. Flux density (in unit of Jy) of point source is converted from antenna temperature by dividing the gain value.  The averaged gain values of 19 beams within ZA of 26.4 $^{\circ}$  are shown in Table \ref{table:gainvalue}. The Beam 16 has smallest gain value among 19 beams. The gain ratio of Beam 16 compared to Beam 1 reaches  $0.83\pm 0.02$ at 1400 and 1450 MHz. 

\subsubsection{System Temperature}
\label{subsubsec:tsys}

Observation of sky position ($\alpha_{2000}$=23$^{h}$30$^m$0$^s$.0, $\delta_{2000}$=25$^{\circ}$39$^{\prime}$10.6") allows us to measure system temperature from zenith angle of 40$^{\circ}$ to nearly $0^{\circ}$.  The data were calibrated into antenna temperature in K with noise curve. For a single dish telescope at L band, T$\rm_{sys}$ consists four main sources (Campbell et al.\ 2002):

\begin{enumerate}
    \item  Noise contribution from receiver, $T\rm_{rec}$. For 19 beam receiver of FAST, it is measured as 7-9 K, including $\sim$ 4 K from low noise amplifier (LAN) (see section 4.1.1 in Jiang et al. 2019 for details). 
    
    \item  Continuum brightness temperature of the sky, $T\rm_{sky}$. This includes 2.73 K from cosmic microwave background (CMB) and non-thermal emission from the Milky Way.  The value of  $T\rm_{sky}$ at 1.4 GHz is $\sim  3.48$ K  toward position ($\alpha_{2000}$=23$^{h}$30$^m$0$^s$.0, $\delta_{2000}$=25$^{\circ}$39$^{\prime}$10.6")  (CHIPASS survey; Calabretta et al. 2014).
    
    \item Emission from the Earth's atmosphere $T\rm_{atm}$. This value should be in a few K. 
    
    \item Radiation from the surrounding terrain, $T_{scat}$. This contribution originates from  side lobe of FAST. Unlike other single dish with fixed surface and horn, the illumination area of horn varies for different zenith angle, leading to different value of $T_{scat}$. 
\end{enumerate}

System temperature are synthetical result of the above components. 
\begin{equation}
    \rm T_{sys}=T_{rec}+T_{sky}+T_{atm}+T_{scat}
\label{eq:tsys}
\end{equation}

\begin{figure}
\centering
  \includegraphics[width=0.45\textwidth]{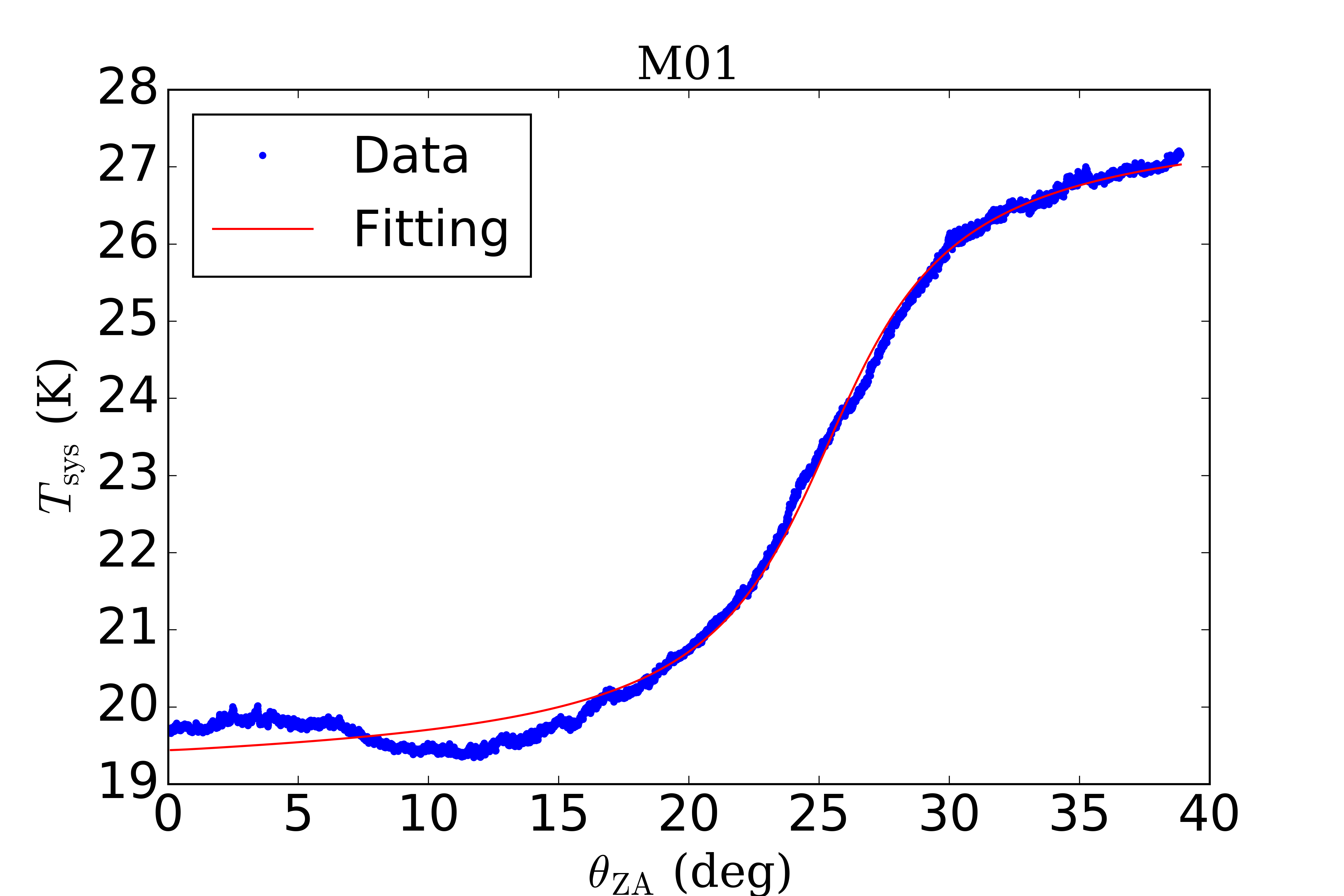}
  \includegraphics[width=0.45\textwidth]{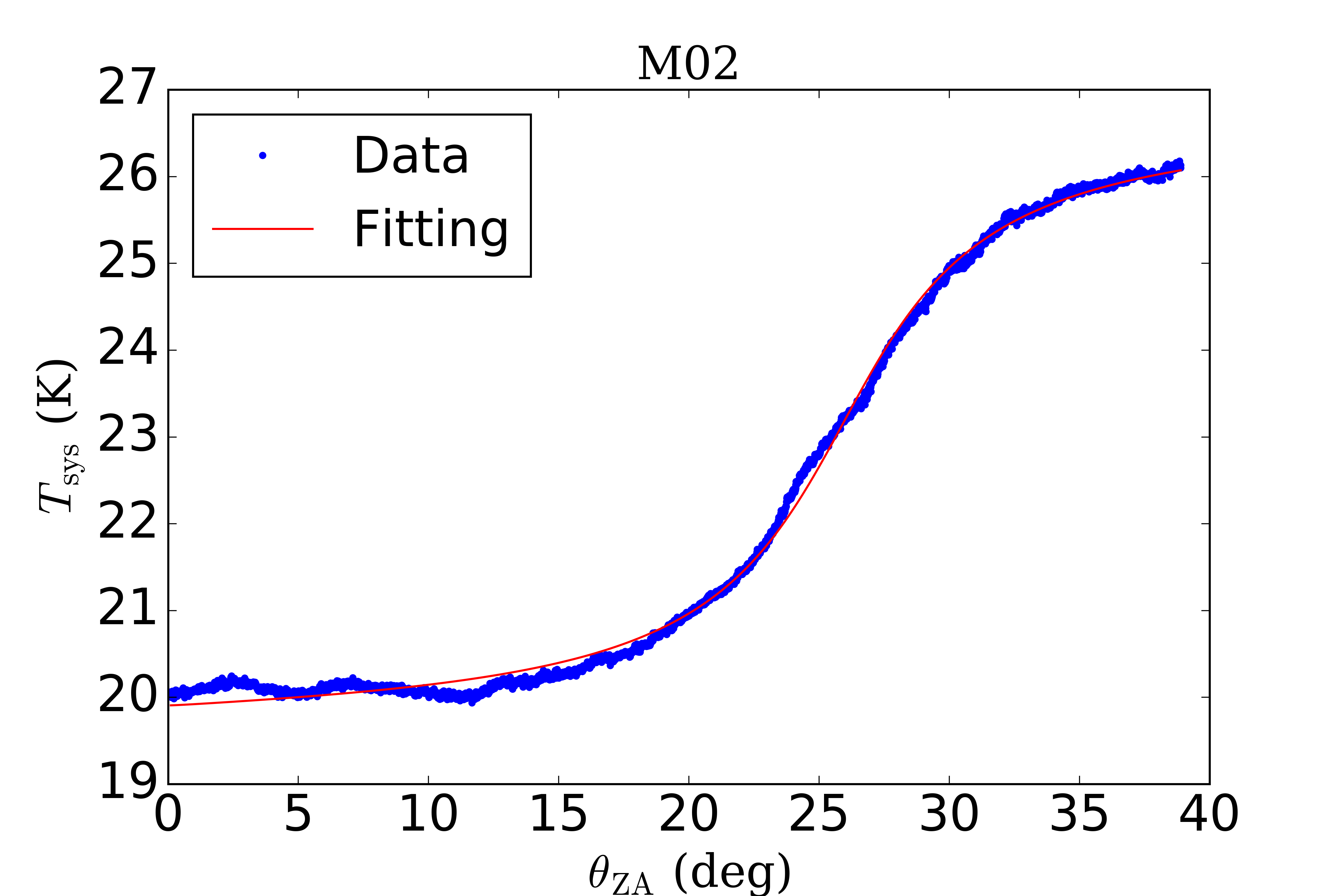}
  \includegraphics[width=0.45\textwidth]{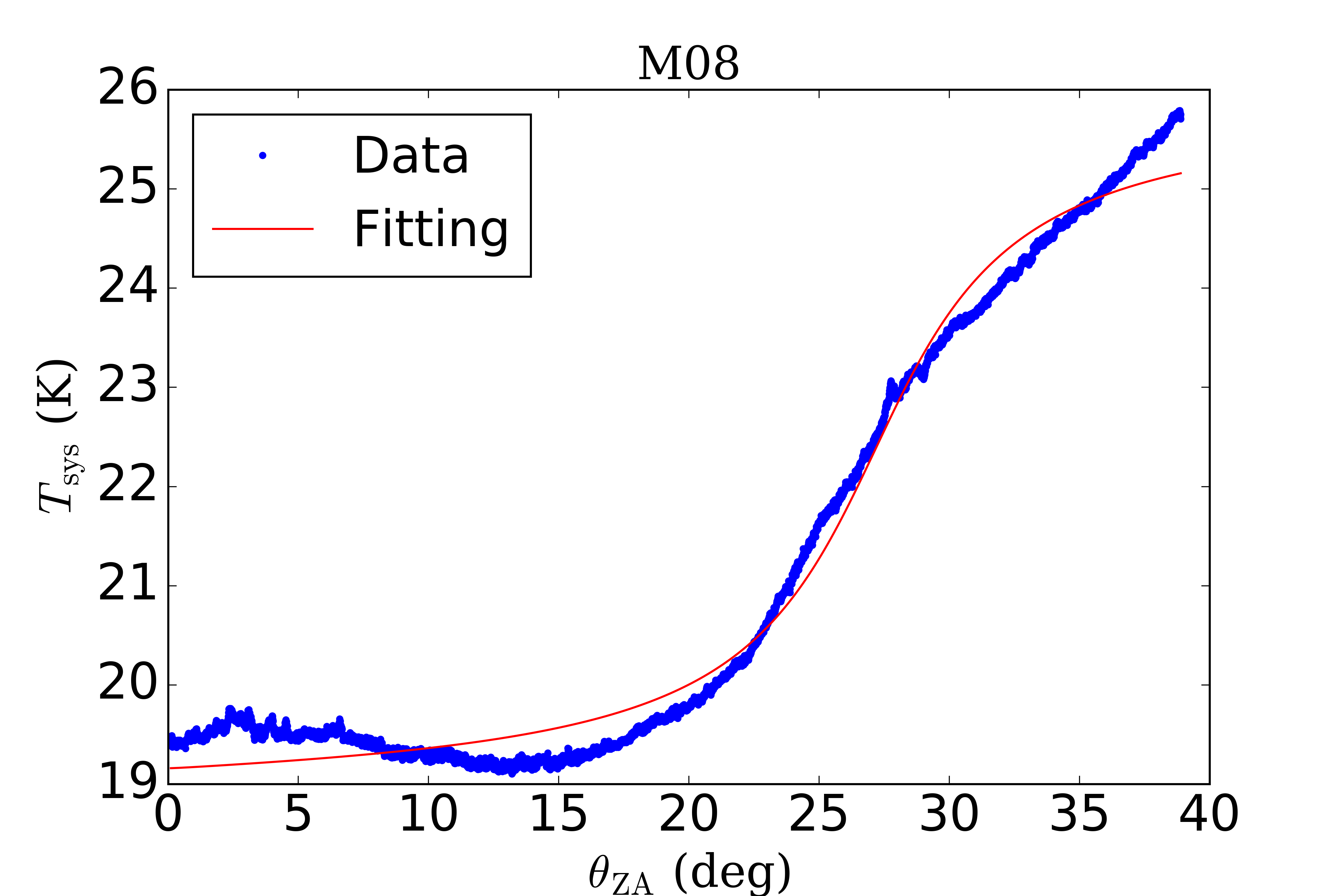}
  \includegraphics[width=0.45\textwidth]{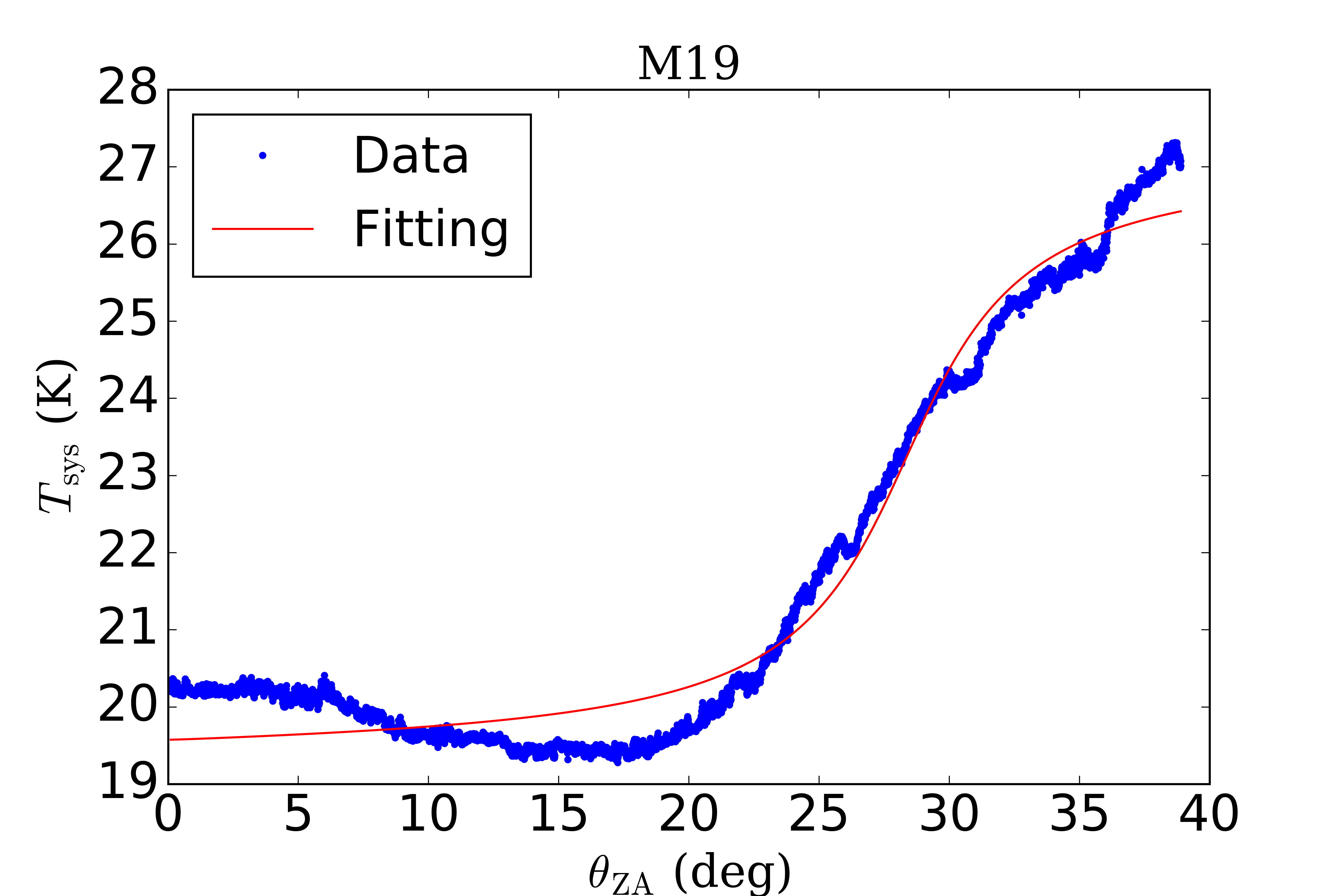}
  \caption{Measured $T\rm_{sys}$ curve as a function of zenith angle $\theta\rm_{ZA}$ at 1400 MHz for Beam 1, 2, 8 and 19.  The fitting result is represented with red solid line.}
\label{fig:tsysfit}
\end{figure}

The variation of system temperature $T\rm_{sys}$ as a function of zenith angle could be fitted with the following formula,

\begin{equation} 
T\rm_{sys} = P_0\arctan(\sqrt{1+{\theta_{ZA}}^n}-P_1)+P_2,
\label{eq:tsysfit}
\end{equation}
which is valid for $0^\circ \leq \theta\rm_{ZA} <= 40^\circ$. The value of $P_0$, $P_1$, $P_2$ and $n$ in different frequencies are shown in Table \ref{table:tsys}. An example of fitting $T\rm_{sys}$  curve is shown in Fig. \ref{fig:tsysfit}.  The results for the rest beams are shown in Fig. \ref{fig:tsysfit_restbeams}. 

A significant feature in the $T\rm_{sys}$ profile is that the $T\rm_{sys}$ value of most beams reaches its minimum at ZA  range of [10$^{\circ}$,15$^{\circ}$]. When ZA is smaller than 10$^{\circ}$, the $T\rm_{sys}$ value would increase by a maximum value of 1 K. Though central hole of spherical surface was shielded with metal mesh during observations, the leakage of ground  emission ($\sim 300$ K) from the central hole goes into the main beam and contributes $T\rm_{sys}$ increase when ZA is less than 10$^{\circ}$. When ZA gets larger than 15$^{\circ}$, the $T\rm_{sys}$ value increases by 5-7 K, which arises from  emission of nearby mountain through sidelobe. 

\subsubsection{Discussion}
\label{subsubsec:discuss}

The observations lie in $\theta\rm_{ZA}$ range of [4.9,40] deg for $\eta$ curve. The fitting results are expected to be valid for $\theta\rm_{ZA}$ range of [0,40] degree for the following reasons. When $\theta\rm_{ZA}<4.9^{\circ}$, there is no loss of panel. In this case, $\eta$  should keep almost constant as shown in the fitting curves.    

\begin{table}[!ht]
\centering
\fontsize{6}{4}\selectfont
\renewcommand{\arraystretch}{1.15}
\caption{Details of fitting parameters of aperture efficiency for all 19 beams.}
 \label{table:etapara}
\begin{tabular}{ccccccccccc}
 \hline\noalign{\smallskip}
 \hline\noalign{\smallskip}
\multicolumn{1}{c}{ \multirow{2}{*}{Beam}} & \multicolumn{1}{c}{ \multirow{2}{*}{Paras}} & \multicolumn{9}{c}{Freq (MHz)} \\
 \cline{3-11}
& & \multicolumn{1}{c}{1050} &\multicolumn{1}{c}{1100} & \multicolumn{1}{c}{1150}& \multicolumn{1}{c}{1200}& \multicolumn{1}{c}{1250} & \multicolumn{1}{c}{1300} & \multicolumn{1}{c}{1350} & \multicolumn{1}{c}{1400} & \multicolumn{1}{c}{1450} \\
 \noalign{\smallskip}\hline
M01 & a/1e-4 & 3.31 $\pm$ 2.23 & 2.54 $\pm$ 1.85 & 0.34 $\pm$ 1.83 & 5.03 $\pm$ 2.03 & 4.94 $\pm$ 2.02 & 9.83 $\pm$ 1.85 & 8.50 $\pm$ 1.93 & 7.87 $\pm$ 1.95 & 10.64 $\pm$ 1.90  \\
M01 & b/1e-1 & 6.19 $\pm$ 0.04 & 6.32 $\pm$ 0.03 & 6.43 $\pm$ 0.03 & 6.22 $\pm$ 0.03 & 6.22 $\pm$ 0.03 & 6.08 $\pm$ 0.03 & 6.12 $\pm$ 0.03 & 6.14 $\pm$ 0.03 & 5.98 $\pm$ 0.03  \\
M01 & c/1e-2 & -1.58 $\pm$ 0.02 & -1.61 $\pm$ 0.03 & -1.37 $\pm$ 0.04 & -1.40 $\pm$ 0.04 & -1.42 $\pm$ 0.03 & -1.38 $\pm$ 0.03 & -1.40 $\pm$ 0.02 & -1.34 $\pm$ 0.02 & -1.26 $\pm$ 0.02  \\
\hline
M02 & a/1e-4 & 9.46 $\pm$ 3.53 & 6.28 $\pm$ 3.09 & 9.01 $\pm$ 3.14 & 12.13 $\pm$ 3.12 & 10.98 $\pm$ 3.06 & 11.31 $\pm$ 2.16 & 8.71 $\pm$ 2.09 & 8.78 $\pm$ 2.06 & 13.07 $\pm$ 1.98  \\
M02 & b/1e-1 & 5.97 $\pm$ 0.06 & 5.97 $\pm$ 0.05 & 5.99 $\pm$ 0.05 & 5.81 $\pm$ 0.05 & 5.81 $\pm$ 0.05 & 5.65 $\pm$ 0.04 & 5.75 $\pm$ 0.03 & 5.75 $\pm$ 0.03 & 5.62 $\pm$ 0.03  \\
M02 & c/1e-2 & -1.54 $\pm$ 0.03 & -1.62 $\pm$ 0.03 & -1.46 $\pm$ 0.03 & -1.44 $\pm$ 0.02 & -1.42 $\pm$ 0.02 & -1.29 $\pm$ 0.02 & -1.32 $\pm$ 0.02 & -1.33 $\pm$ 0.02 & -1.21 $\pm$ 0.02  \\
\hline
M03 & a/1e-4 & 1.79 $\pm$ 4.21 & 4.39 $\pm$ 3.55 & 2.41 $\pm$ 4.46 & 5.95 $\pm$ 4.69 & 6.90 $\pm$ 3.21 & 7.41 $\pm$ 2.60 & 9.99 $\pm$ 2.63 & 7.66 $\pm$ 2.60 & 10.30 $\pm$ 2.72  \\
M03 & b/1e-1 & 6.41 $\pm$ 0.07 & 6.24 $\pm$ 0.06 & 6.42 $\pm$ 0.07 & 6.21 $\pm$ 0.07 & 6.17 $\pm$ 0.05 & 6.07 $\pm$ 0.04 & 5.91 $\pm$ 0.04 & 5.99 $\pm$ 0.04 & 5.86 $\pm$ 0.04  \\
M03 & c/1e-2 & -1.70 $\pm$ 0.03 & -1.80 $\pm$ 0.03 & -1.56 $\pm$ 0.03 & -1.63 $\pm$ 0.02 & -1.59 $\pm$ 0.02 & -1.56 $\pm$ 0.02 & -1.53 $\pm$ 0.02 & -1.47 $\pm$ 0.02 & -1.35 $\pm$ 0.03  \\
\hline
M04 & a/1e-4 & 3.57 $\pm$ 4.76 & 1.87 $\pm$ 4.08 & 3.82 $\pm$ 3.93 & 3.90 $\pm$ 4.60 & 8.46 $\pm$ 4.65 & 4.29 $\pm$ 3.22 & 2.75 $\pm$ 3.28 & 1.85 $\pm$ 3.26 & 5.10 $\pm$ 3.20  \\
M04 & b/1e-1 & 6.24 $\pm$ 0.08 & 6.15 $\pm$ 0.07 & 6.25 $\pm$ 0.06 & 6.05 $\pm$ 0.08 & 5.98 $\pm$ 0.08 & 5.92 $\pm$ 0.05 & 5.89 $\pm$ 0.05 & 5.89 $\pm$ 0.05 & 5.75 $\pm$ 0.05  \\
M04 & c/1e-2 & -1.80 $\pm$ 0.03 & -1.82 $\pm$ 0.03 & -1.65 $\pm$ 0.02 & -1.57 $\pm$ 0.06 & -1.58 $\pm$ 0.02 & -1.42 $\pm$ 0.02 & -1.45 $\pm$ 0.02 & -1.36 $\pm$ 0.02 & -1.24 $\pm$ 0.02  \\
\hline
M05 & a/1e-4 & -3.97 $\pm$ 2.87 & -4.25 $\pm$ 2.72 & -4.44 $\pm$ 2.72 & -7.54 $\pm$ 3.48 & -4.09 $\pm$ 4.19 & -5.41 $\pm$ 2.35 & -3.33 $\pm$ 2.26 & -4.51 $\pm$ 2.42 & -4.58 $\pm$ 2.53  \\
M05 & b/1e-1 & 6.45 $\pm$ 0.05 & 6.48 $\pm$ 0.04 & 6.68 $\pm$ 0.04 & 6.42 $\pm$ 0.06 & 6.39 $\pm$ 0.07 & 6.26 $\pm$ 0.04 & 6.10 $\pm$ 0.04 & 6.17 $\pm$ 0.04 & 6.11 $\pm$ 0.04  \\
M05 & c/1e-2 & -1.82 $\pm$ 0.03 & -1.86 $\pm$ 0.03 & -1.72 $\pm$ 0.03 & -1.56 $\pm$ 0.03 & -1.60 $\pm$ 0.03 & -1.53 $\pm$ 0.02 & -1.45 $\pm$ 0.03 & -1.44 $\pm$ 0.03 & -1.30 $\pm$ 0.03  \\
\hline
M06 & a/1e-4 & 1.69 $\pm$ 2.84 & 0.74 $\pm$ 2.59 & -0.00 $\pm$ 4.17 & 1.06 $\pm$ 5.18 & 0.97 $\pm$ 3.49 & 0.33 $\pm$ 1.82 & 3.11 $\pm$ 1.80 & -0.03 $\pm$ 1.76 & 1.37 $\pm$ 1.76  \\
M06 & b/1e-1 & 6.16 $\pm$ 0.04 & 6.22 $\pm$ 0.04 & 6.33 $\pm$ 0.07 & 6.01 $\pm$ 0.08 & 6.04 $\pm$ 0.06 & 5.92 $\pm$ 0.03 & 5.83 $\pm$ 0.03 & 5.89 $\pm$ 0.03 & 5.75 $\pm$ 0.03  \\
M06 & c/1e-2 & -1.74 $\pm$ 0.03 & -1.80 $\pm$ 0.03 & -1.60 $\pm$ 0.03 & -1.50 $\pm$ 0.04 & -1.44 $\pm$ 0.03 & -1.33 $\pm$ 0.02 & -1.38 $\pm$ 0.02 & -1.35 $\pm$ 0.02 & -1.22 $\pm$ 0.02  \\
\hline
M07 & a/1e-4 & 1.71 $\pm$ 2.43 & 0.08 $\pm$ 2.18 & 0.77 $\pm$ 2.41 & 2.03 $\pm$ 3.25 & 2.24 $\pm$ 3.64 & 3.46 $\pm$ 1.82 & 5.46 $\pm$ 1.60 & 6.70 $\pm$ 1.65 & 6.91 $\pm$ 1.49  \\
M07 & b/1e-1 & 6.10 $\pm$ 0.04 & 6.04 $\pm$ 0.04 & 6.18 $\pm$ 0.04 & 5.90 $\pm$ 0.05 & 6.02 $\pm$ 0.06 & 5.86 $\pm$ 0.03 & 5.77 $\pm$ 0.03 & 5.73 $\pm$ 0.03 & 5.62 $\pm$ 0.02  \\
M07 & c/1e-2 & -1.68 $\pm$ 0.02 & -1.64 $\pm$ 0.02 & -1.52 $\pm$ 0.02 & -1.39 $\pm$ 0.02 & -1.43 $\pm$ 0.02 & -1.32 $\pm$ 0.02 & -1.34 $\pm$ 0.02 & -1.36 $\pm$ 0.02 & -1.22 $\pm$ 0.02  \\
\hline
M08 & a/1e-4 & 13.91 $\pm$ 3.13 & 17.00 $\pm$ 2.98 & 14.04 $\pm$ 3.81 & 16.81 $\pm$ 4.98 & 17.71 $\pm$ 2.95 & 17.29 $\pm$ 2.75 & 16.95 $\pm$ 2.29 & 15.92 $\pm$ 2.31 & 17.76 $\pm$ 2.43  \\
M08 & b/1e-1 & 5.76 $\pm$ 0.05 & 5.75 $\pm$ 0.05 & 5.81 $\pm$ 0.06 & 5.62 $\pm$ 0.08 & 5.54 $\pm$ 0.05 & 5.46 $\pm$ 0.04 & 5.41 $\pm$ 0.04 & 5.34 $\pm$ 0.04 & 5.22 $\pm$ 0.04  \\
M08 & c/1e-2 & -1.60 $\pm$ 0.06 & -1.70 $\pm$ 0.05 & -1.52 $\pm$ 0.06 & -1.50 $\pm$ 0.06 & -1.50 $\pm$ 0.05 & -1.47 $\pm$ 0.05 & -1.45 $\pm$ 0.05 & -1.37 $\pm$ 0.05 & -1.28 $\pm$ 0.05  \\
\hline
M09 & a/1e-4 & -1.26 $\pm$ 2.03 & -0.43 $\pm$ 1.99 & -2.68 $\pm$ 1.94 & 0.12 $\pm$ 1.67 & 0.96 $\pm$ 1.60 & 1.53 $\pm$ 1.53 & 1.09 $\pm$ 1.36 & 1.00 $\pm$ 1.11 & 0.27 $\pm$ 0.91  \\
M09 & b/1e-1 & 5.96 $\pm$ 0.03 & 5.95 $\pm$ 0.03 & 5.98 $\pm$ 0.03 & 5.75 $\pm$ 0.03 & 5.77 $\pm$ 0.03 & 5.74 $\pm$ 0.02 & 5.64 $\pm$ 0.02 & 5.57 $\pm$ 0.02 & 5.53 $\pm$ 0.02  \\
M09 & c/1e-2 & -1.39 $\pm$ 0.04 & -1.43 $\pm$ 0.03 & -1.28 $\pm$ 0.04 & -1.24 $\pm$ 0.04 & -1.24 $\pm$ 0.03 & -1.18 $\pm$ 0.04 & -1.09 $\pm$ 0.04 & -1.06 $\pm$ 0.03 & -1.01 $\pm$ 0.03  \\
\hline
M10 & a/1e-4 & 14.36 $\pm$ 3.80 & 15.37 $\pm$ 3.39 & 15.65 $\pm$ 3.78 & 15.20 $\pm$ 4.85 & 18.06 $\pm$ 2.92 & 17.97 $\pm$ 2.67 & 19.34 $\pm$ 2.49 & 16.72 $\pm$ 2.52 & 18.57 $\pm$ 2.65  \\
M10 & b/1e-1 & 6.01 $\pm$ 0.06 & 5.98 $\pm$ 0.05 & 6.04 $\pm$ 0.06 & 5.81 $\pm$ 0.08 & 5.86 $\pm$ 0.05 & 5.75 $\pm$ 0.04 & 5.72 $\pm$ 0.04 & 5.69 $\pm$ 0.04 & 5.60 $\pm$ 0.04  \\
M10 & c/1e-2 & -1.79 $\pm$ 0.04 & -1.83 $\pm$ 0.04 & -1.74 $\pm$ 0.03 & -1.73 $\pm$ 0.03 & -1.78 $\pm$ 0.03 & -1.74 $\pm$ 0.03 & -1.72 $\pm$ 0.03 & -1.66 $\pm$ 0.03 & -1.58 $\pm$ 0.03  \\
\hline
M11 & a/1e-4 & 7.68 $\pm$ 5.51 & 2.98 $\pm$ 4.73 & 4.16 $\pm$ 4.57 & 3.38 $\pm$ 6.44 & 2.45 $\pm$ 4.86 & 4.27 $\pm$ 3.39 & 5.99 $\pm$ 3.36 & 5.03 $\pm$ 3.28 & 8.23 $\pm$ 3.05  \\
M11 & b/1e-1 & 5.90 $\pm$ 0.09 & 5.94 $\pm$ 0.08 & 6.00 $\pm$ 0.07 & 5.81 $\pm$ 0.10 & 5.80 $\pm$ 0.08 & 5.67 $\pm$ 0.06 & 5.66 $\pm$ 0.05 & 5.61 $\pm$ 0.05 & 5.45 $\pm$ 0.05  \\
M11 & c/1e-2 & -1.59 $\pm$ 0.04 & -1.59 $\pm$ 0.03 & -1.55 $\pm$ 0.03 & -1.35 $\pm$ 0.05 & -1.36 $\pm$ 0.03 & -1.31 $\pm$ 0.03 & -1.35 $\pm$ 0.02 & -1.27 $\pm$ 0.02 & -1.18 $\pm$ 0.02  \\
\hline
M12 & a/1e-4 & -3.00 $\pm$ 5.82 & -2.08 $\pm$ 4.75 & 1.33 $\pm$ 4.29 & 4.98 $\pm$ 5.10 & 8.50 $\pm$ 4.02 & -0.98 $\pm$ 3.17 & 2.94 $\pm$ 3.06 & 2.49 $\pm$ 2.71 & 2.77 $\pm$ 2.60  \\
M12 & b/1e-1 & 6.06 $\pm$ 0.09 & 5.96 $\pm$ 0.08 & 5.91 $\pm$ 0.07 & 5.82 $\pm$ 0.08 & 5.76 $\pm$ 0.06 & 5.70 $\pm$ 0.05 & 5.58 $\pm$ 0.05 & 5.53 $\pm$ 0.04 & 5.48 $\pm$ 0.04  \\
M12 & c/1e-2 & -1.56 $\pm$ 0.05 & -1.49 $\pm$ 0.04 & -1.45 $\pm$ 0.05 & -1.56 $\pm$ 0.05 & -1.69 $\pm$ 0.03 & -1.35 $\pm$ 0.04 & -1.40 $\pm$ 0.03 & -1.34 $\pm$ 0.03 & -1.27 $\pm$ 0.03  \\
\hline
M13 & a/1e-4 & -1.33 $\pm$ 4.57 & 0.41 $\pm$ 4.05 & 2.41 $\pm$ 4.07 & -1.35 $\pm$ 4.89 & 2.44 $\pm$ 3.87 & 2.60 $\pm$ 3.22 & 4.14 $\pm$ 3.11 & 5.50 $\pm$ 2.95 & 5.83 $\pm$ 2.77  \\
M13 & b/1e-1 & 6.18 $\pm$ 0.08 & 6.12 $\pm$ 0.07 & 6.22 $\pm$ 0.07 & 6.23 $\pm$ 0.08 & 6.04 $\pm$ 0.06 & 6.00 $\pm$ 0.05 & 5.94 $\pm$ 0.05 & 5.85 $\pm$ 0.05 & 5.71 $\pm$ 0.05  \\
M13 & c/1e-2 & -1.49 $\pm$ 0.04 & -1.59 $\pm$ 0.03 & -1.53 $\pm$ 0.04 & -1.54 $\pm$ 0.07 & -1.54 $\pm$ 0.03 & -1.57 $\pm$ 0.02 & -1.59 $\pm$ 0.03 & -1.53 $\pm$ 0.03 & -1.44 $\pm$ 0.03  \\
\hline
M14 & a/1e-4 & 0.35 $\pm$ 3.81 & -0.57 $\pm$ 3.77 & 4.27 $\pm$ 3.61 & 3.01 $\pm$ 4.84 & 7.37 $\pm$ 4.78 & 2.58 $\pm$ 2.61 & 4.43 $\pm$ 2.19 & 5.43 $\pm$ 2.22 & 5.53 $\pm$ 2.19  \\
M14 & b/1e-1 & 5.91 $\pm$ 0.06 & 5.98 $\pm$ 0.06 & 5.91 $\pm$ 0.06 & 5.76 $\pm$ 0.08 & 5.73 $\pm$ 0.08 & 5.78 $\pm$ 0.04 & 5.64 $\pm$ 0.04 & 5.59 $\pm$ 0.04 & 5.53 $\pm$ 0.04  \\
M14 & c/1e-2 & -1.56 $\pm$ 0.04 & -1.67 $\pm$ 0.04 & -1.65 $\pm$ 0.04 & -1.57 $\pm$ 0.04 & -1.64 $\pm$ 0.03 & -1.68 $\pm$ 0.02 & -1.61 $\pm$ 0.03 & -1.60 $\pm$ 0.04 & -1.51 $\pm$ 0.04  \\
\hline
M15 & a/1e-4 & 1.33 $\pm$ 2.89 & -0.59 $\pm$ 2.55 & -3.57 $\pm$ 3.01 & 3.71 $\pm$ 5.82 & -1.46 $\pm$ 2.36 & -2.34 $\pm$ 1.73 & -2.80 $\pm$ 1.71 & -3.55 $\pm$ 1.68 & -2.20 $\pm$ 1.64  \\
M15 & b/1e-1 & 5.62 $\pm$ 0.05 & 5.78 $\pm$ 0.04 & 5.83 $\pm$ 0.05 & 5.59 $\pm$ 0.09 & 5.68 $\pm$ 0.04 & 5.61 $\pm$ 0.03 & 5.53 $\pm$ 0.03 & 5.50 $\pm$ 0.03 & 5.46 $\pm$ 0.03  \\
M15 & c/1e-2 & -1.42 $\pm$ 0.04 & -1.53 $\pm$ 0.03 & -1.34 $\pm$ 0.04 & -1.53 $\pm$ 0.03 & -1.42 $\pm$ 0.03 & -1.32 $\pm$ 0.03 & -1.27 $\pm$ 0.03 & -1.21 $\pm$ 0.03 & -1.13 $\pm$ 0.03  \\
\hline
M16 & a/1e-4 & 9.81 $\pm$ 3.39 & 3.14 $\pm$ 3.06 & 0.38 $\pm$ 4.28 & 3.85 $\pm$ 3.48 & 6.30 $\pm$ 2.65 & 5.25 $\pm$ 2.13 & 2.32 $\pm$ 1.95 & 0.49 $\pm$ 1.90 & 1.14 $\pm$ 1.83  \\
M16 & b/1e-1 & 5.44 $\pm$ 0.06 & 5.52 $\pm$ 0.05 & 5.55 $\pm$ 0.07 & 5.47 $\pm$ 0.06 & 5.29 $\pm$ 0.04 & 5.28 $\pm$ 0.04 & 5.19 $\pm$ 0.03 & 5.17 $\pm$ 0.03 & 5.09 $\pm$ 0.03  \\
M16 & c/1e-2 & -1.58 $\pm$ 0.04 & -1.52 $\pm$ 0.03 & -1.46 $\pm$ 0.03 & -1.51 $\pm$ 0.03 & -1.41 $\pm$ 0.02 & -1.36 $\pm$ 0.02 & -1.28 $\pm$ 0.02 & -1.21 $\pm$ 0.02 & -1.14 $\pm$ 0.02  \\
\hline
M17 & a/1e-4 & -0.89 $\pm$ 3.92 & -4.49 $\pm$ 3.39 & -3.98 $\pm$ 3.71 & -7.57 $\pm$ 4.51 & -3.25 $\pm$ 2.74 & -0.30 $\pm$ 2.47 & -0.80 $\pm$ 2.36 & -1.44 $\pm$ 2.17 & -0.64 $\pm$ 2.08  \\
M17 & b/1e-1 & 5.81 $\pm$ 0.06 & 5.83 $\pm$ 0.06 & 5.85 $\pm$ 0.06 & 5.71 $\pm$ 0.07 & 5.65 $\pm$ 0.05 & 5.47 $\pm$ 0.04 & 5.48 $\pm$ 0.04 & 5.41 $\pm$ 0.04 & 5.28 $\pm$ 0.03  \\
M17 & c/1e-2 & -1.54 $\pm$ 0.02 & -1.52 $\pm$ 0.02 & -1.41 $\pm$ 0.02 & -1.25 $\pm$ 0.03 & -1.33 $\pm$ 0.02 & -1.22 $\pm$ 0.01 & -1.21 $\pm$ 0.02 & -1.16 $\pm$ 0.01 & -1.08 $\pm$ 0.02  \\
\hline
M18 & a/1e-4 & -1.88 $\pm$ 2.27 & -5.55 $\pm$ 2.09 & -5.94 $\pm$ 2.11 & -1.53 $\pm$ 2.53 & -2.21 $\pm$ 3.35 & -0.75 $\pm$ 1.52 & -1.19 $\pm$ 1.52 & 0.24 $\pm$ 1.51 & 0.93 $\pm$ 1.53  \\
M18 & b/1e-1 & 5.64 $\pm$ 0.04 & 5.68 $\pm$ 0.03 & 5.81 $\pm$ 0.03 & 5.60 $\pm$ 0.04 & 5.57 $\pm$ 0.05 & 5.46 $\pm$ 0.02 & 5.38 $\pm$ 0.02 & 5.38 $\pm$ 0.02 & 5.33 $\pm$ 0.02  \\
M18 & c/1e-2 & -1.37 $\pm$ 0.02 & -1.34 $\pm$ 0.02 & -1.34 $\pm$ 0.02 & -1.29 $\pm$ 0.02 & -1.20 $\pm$ 0.02 & -1.13 $\pm$ 0.02 & -1.11 $\pm$ 0.02 & -1.11 $\pm$ 0.02 & -1.05 $\pm$ 0.02  \\
\hline
M19 & a/1e-4 & 11.82 $\pm$ 2.94 & 6.55 $\pm$ 2.78 & 5.01 $\pm$ 3.51 & 10.73 $\pm$ 3.34 & 9.32 $\pm$ 2.79 & 8.64 $\pm$ 2.51 & 10.22 $\pm$ 2.15 & 9.32 $\pm$ 2.21 & 11.03 $\pm$ 2.24  \\
M19 & b/1e-1 & 5.70 $\pm$ 0.05 & 5.88 $\pm$ 0.04 & 5.92 $\pm$ 0.06 & 5.67 $\pm$ 0.05 & 5.69 $\pm$ 0.04 & 5.61 $\pm$ 0.04 & 5.48 $\pm$ 0.03 & 5.43 $\pm$ 0.04 & 5.30 $\pm$ 0.04  \\
M19 & c/1e-2 & -1.76 $\pm$ 0.03 & -1.75 $\pm$ 0.03 & -1.59 $\pm$ 0.03 & -1.60 $\pm$ 0.04 & -1.57 $\pm$ 0.03 & -1.50 $\pm$ 0.03 & -1.44 $\pm$ 0.03 & -1.36 $\pm$ 0.03 & -1.25 $\pm$ 0.03  \\
 \hline\noalign{\smallskip}
\end{tabular}
\end{table}

\begin{table}[!ht]
\centering
\fontsize{6}{4}\selectfont
\renewcommand{\arraystretch}{1.15}
\caption{Details of fitting parameters of system temperature  for all 19 beams.}
 \label{table:tsys}
 
\begin{tabular}{ccccccccccc}
\hline\hline
\multicolumn{1}{c}{ \multirow{2}{*}{Beam}} & \multicolumn{1}{c}{ \multirow{2}{*}{Paras}} & \multicolumn{9}{c}{Freq (MHz)} \\
 \cline{3-11} 
& & \multicolumn{1}{c}{1050} &\multicolumn{1}{c}{1100} & \multicolumn{1}{c}{1150}& \multicolumn{1}{c}{1200}& \multicolumn{1}{c}{1250} & \multicolumn{1}{c}{1300} & \multicolumn{1}{c}{1350} & \multicolumn{1}{c}{1400} & \multicolumn{1}{c}{1450} \\  
 \noalign{\smallskip}\hline

M01 & P$_0$ & 4.94 $\pm$ 0.01 & 5.83 $\pm$ 0.01 & 4.36 $\pm$ 0.01 & 4.24 $\pm$ 0.03 & 4.35 $\pm$ 0.02 & 3.60 $\pm$ 0.01 & 3.10 $\pm$ 0.01 & 2.76 $\pm$ 0.01 & 2.00 $\pm$ 0.01  \\
M01 & P$_1$ & 9.12 $\pm$ 0.02 & 8.75 $\pm$ 0.03 & 7.53 $\pm$ 0.03 & 7.75 $\pm$ 0.08 & 6.88 $\pm$ 0.04 & 9.88 $\pm$ 0.03 & 10.03 $\pm$ 0.04 & 10.00 $\pm$ 0.04 & 11.93 $\pm$ 0.07  \\
M01 & P$_2$ & 26.31 $\pm$ 0.01 & 28.14 $\pm$ 0.01 & 26.89 $\pm$ 0.01 & 27.19 $\pm$ 0.02 & 25.54 $\pm$ 0.01 & 24.59 $\pm$ 0.01 & 23.91 $\pm$ 0.00 & 23.47 $\pm$ 0.00 & 22.23 $\pm$ 0.00  \\
M01 & n & 1.37 $\pm$ 0.00 & 1.34 $\pm$ 0.00 & 1.26 $\pm$ 0.00 & 1.27 $\pm$ 0.01 & 1.21 $\pm$ 0.00 & 1.41 $\pm$ 0.00 & 1.43 $\pm$ 0.00 & 1.42 $\pm$ 0.00 & 1.53 $\pm$ 0.00  \\
\hline
M02 & P$_0$ & 4.10 $\pm$ 0.00 & 4.53 $\pm$ 0.01 & 4.23 $\pm$ 0.02 & 4.17 $\pm$ 0.05 & 4.08 $\pm$ 0.02 & 3.20 $\pm$ 0.00 & 2.47 $\pm$ 0.00 & 2.29 $\pm$ 0.00 & 1.49 $\pm$ 0.00  \\
M02 & P$_1$ & 7.68 $\pm$ 0.01 & 7.89 $\pm$ 0.03 & 6.19 $\pm$ 0.03 & 6.32 $\pm$ 0.07 & 5.73 $\pm$ 0.03 & 8.83 $\pm$ 0.02 & 9.36 $\pm$ 0.02 & 9.32 $\pm$ 0.02 & 11.41 $\pm$ 0.06  \\
M02 & P$_2$ & 25.65 $\pm$ 0.00 & 26.74 $\pm$ 0.01 & 26.63 $\pm$ 0.02 & 27.53 $\pm$ 0.04 & 25.56 $\pm$ 0.02 & 24.14 $\pm$ 0.00 & 23.29 $\pm$ 0.00 & 23.23 $\pm$ 0.00 & 21.87 $\pm$ 0.00  \\
M02 & n & 1.25 $\pm$ 0.00 & 1.28 $\pm$ 0.00 & 1.10 $\pm$ 0.00 & 1.12 $\pm$ 0.01 & 1.07 $\pm$ 0.00 & 1.33 $\pm$ 0.00 & 1.37 $\pm$ 0.00 & 1.37 $\pm$ 0.00 & 1.48 $\pm$ 0.00  \\
\hline
M03 & P$_0$ & 4.89 $\pm$ 0.01 & 5.22 $\pm$ 0.01 & 4.92 $\pm$ 0.02 & 4.31 $\pm$ 0.02 & 4.70 $\pm$ 0.02 & 3.69 $\pm$ 0.01 & 3.10 $\pm$ 0.00 & 2.80 $\pm$ 0.01 & 2.01 $\pm$ 0.01  \\
M03 & P$_1$ & 8.16 $\pm$ 0.02 & 8.12 $\pm$ 0.03 & 6.53 $\pm$ 0.02 & 7.59 $\pm$ 0.06 & 6.29 $\pm$ 0.03 & 8.83 $\pm$ 0.03 & 9.46 $\pm$ 0.03 & 9.32 $\pm$ 0.04 & 10.91 $\pm$ 0.06  \\
M03 & P$_2$ & 27.87 $\pm$ 0.01 & 28.30 $\pm$ 0.01 & 27.66 $\pm$ 0.01 & 27.93 $\pm$ 0.02 & 26.41 $\pm$ 0.02 & 24.67 $\pm$ 0.01 & 23.34 $\pm$ 0.00 & 23.22 $\pm$ 0.01 & 21.86 $\pm$ 0.00  \\
M03 & n & 1.29 $\pm$ 0.00 & 1.29 $\pm$ 0.00 & 1.15 $\pm$ 0.00 & 1.25 $\pm$ 0.01 & 1.12 $\pm$ 0.00 & 1.33 $\pm$ 0.00 & 1.39 $\pm$ 0.00 & 1.37 $\pm$ 0.00 & 1.46 $\pm$ 0.00  \\
\hline
M04 & P$_0$ & 4.23 $\pm$ 0.00 & 5.00 $\pm$ 0.01 & 3.52 $\pm$ 0.01 & 3.24 $\pm$ 0.02 & 3.60 $\pm$ 0.01 & 2.70 $\pm$ 0.00 & 2.42 $\pm$ 0.00 & 2.08 $\pm$ 0.00 & 1.31 $\pm$ 0.00  \\
M04 & P$_1$ & 8.05 $\pm$ 0.01 & 8.10 $\pm$ 0.03 & 7.31 $\pm$ 0.03 & 7.31 $\pm$ 0.08 & 6.20 $\pm$ 0.03 & 8.88 $\pm$ 0.02 & 9.47 $\pm$ 0.02 & 9.35 $\pm$ 0.02 & 12.18 $\pm$ 0.05  \\
M04 & P$_2$ & 26.10 $\pm$ 0.00 & 27.88 $\pm$ 0.01 & 26.48 $\pm$ 0.01 & 27.47 $\pm$ 0.02 & 26.43 $\pm$ 0.01 & 25.13 $\pm$ 0.00 & 24.59 $\pm$ 0.00 & 23.92 $\pm$ 0.00 & 22.13 $\pm$ 0.00  \\
M04 & n & 1.30 $\pm$ 0.00 & 1.31 $\pm$ 0.00 & 1.26 $\pm$ 0.00 & 1.25 $\pm$ 0.01 & 1.15 $\pm$ 0.00 & 1.36 $\pm$ 0.00 & 1.40 $\pm$ 0.00 & 1.40 $\pm$ 0.00 & 1.56 $\pm$ 0.00  \\
\hline
M05 & P$_0$ & 4.38 $\pm$ 0.00 & 4.75 $\pm$ 0.01 & 3.43 $\pm$ 0.01 & 3.19 $\pm$ 0.03 & 3.50 $\pm$ 0.01 & 2.82 $\pm$ 0.00 & 2.47 $\pm$ 0.00 & 2.09 $\pm$ 0.00 & 1.41 $\pm$ 0.00  \\
M05 & P$_1$ & 8.18 $\pm$ 0.01 & 8.44 $\pm$ 0.03 & 6.79 $\pm$ 0.03 & 6.69 $\pm$ 0.09 & 6.42 $\pm$ 0.04 & 8.59 $\pm$ 0.02 & 9.29 $\pm$ 0.02 & 9.49 $\pm$ 0.02 & 11.49 $\pm$ 0.05  \\
M05 & P$_2$ & 26.13 $\pm$ 0.00 & 27.32 $\pm$ 0.01 & 26.06 $\pm$ 0.01 & 26.75 $\pm$ 0.02 & 25.96 $\pm$ 0.01 & 24.51 $\pm$ 0.00 & 23.88 $\pm$ 0.00 & 23.76 $\pm$ 0.00 & 22.65 $\pm$ 0.00  \\
M05 & n & 1.32 $\pm$ 0.00 & 1.34 $\pm$ 0.00 & 1.21 $\pm$ 0.00 & 1.20 $\pm$ 0.01 & 1.19 $\pm$ 0.00 & 1.34 $\pm$ 0.00 & 1.40 $\pm$ 0.00 & 1.41 $\pm$ 0.00 & 1.52 $\pm$ 0.00  \\
\hline
M06 & P$_0$ & 5.28 $\pm$ 0.01 & 6.22 $\pm$ 0.02 & 4.54 $\pm$ 0.02 & 4.29 $\pm$ 0.03 & 4.65 $\pm$ 0.02 & 3.77 $\pm$ 0.01 & 3.28 $\pm$ 0.01 & 2.99 $\pm$ 0.01 & 2.23 $\pm$ 0.01  \\
M06 & P$_1$ & 9.01 $\pm$ 0.03 & 8.49 $\pm$ 0.04 & 7.29 $\pm$ 0.05 & 7.96 $\pm$ 0.08 & 7.25 $\pm$ 0.04 & 9.39 $\pm$ 0.04 & 10.41 $\pm$ 0.05 & 10.16 $\pm$ 0.06 & 11.65 $\pm$ 0.09  \\
M06 & P$_2$ & 27.57 $\pm$ 0.01 & 29.40 $\pm$ 0.02 & 27.50 $\pm$ 0.02 & 28.62 $\pm$ 0.03 & 26.86 $\pm$ 0.02 & 25.10 $\pm$ 0.01 & 24.27 $\pm$ 0.01 & 23.75 $\pm$ 0.01 & 22.46 $\pm$ 0.01  \\
M06 & n & 1.35 $\pm$ 0.00 & 1.32 $\pm$ 0.00 & 1.23 $\pm$ 0.00 & 1.27 $\pm$ 0.01 & 1.22 $\pm$ 0.00 & 1.37 $\pm$ 0.00 & 1.45 $\pm$ 0.00 & 1.42 $\pm$ 0.00 & 1.50 $\pm$ 0.00  \\
\hline
M07 & P$_0$ & 3.72 $\pm$ 0.00 & 4.43 $\pm$ 0.02 & 3.44 $\pm$ 0.02 & 3.34 $\pm$ 0.03 & 3.38 $\pm$ 0.01 & 2.88 $\pm$ 0.00 & 2.24 $\pm$ 0.00 & 1.95 $\pm$ 0.00 & 1.25 $\pm$ 0.00  \\
M07 & P$_1$ & 8.78 $\pm$ 0.02 & 8.15 $\pm$ 0.05 & 6.38 $\pm$ 0.04 & 6.94 $\pm$ 0.08 & 6.41 $\pm$ 0.03 & 9.30 $\pm$ 0.02 & 10.59 $\pm$ 0.03 & 10.72 $\pm$ 0.04 & 13.50 $\pm$ 0.10  \\
M07 & P$_2$ & 26.36 $\pm$ 0.00 & 27.90 $\pm$ 0.01 & 26.27 $\pm$ 0.01 & 27.01 $\pm$ 0.03 & 25.08 $\pm$ 0.01 & 24.22 $\pm$ 0.00 & 23.02 $\pm$ 0.00 & 22.70 $\pm$ 0.00 & 21.36 $\pm$ 0.00  \\
M07 & n & 1.35 $\pm$ 0.00 & 1.30 $\pm$ 0.00 & 1.16 $\pm$ 0.00 & 1.19 $\pm$ 0.01 & 1.16 $\pm$ 0.00 & 1.38 $\pm$ 0.00 & 1.46 $\pm$ 0.00 & 1.47 $\pm$ 0.00 & 1.61 $\pm$ 0.00  \\
\hline
M08 & P$_0$ & 4.57 $\pm$ 0.01 & 3.93 $\pm$ 0.01 & 4.85 $\pm$ 0.04 & 4.33 $\pm$ 0.04 & 3.96 $\pm$ 0.02 & 3.20 $\pm$ 0.01 & 2.63 $\pm$ 0.01 & 2.24 $\pm$ 0.01 & 1.77 $\pm$ 0.01  \\
M08 & P$_1$ & 7.17 $\pm$ 0.02 & 7.87 $\pm$ 0.03 & 5.78 $\pm$ 0.02 & 6.80 $\pm$ 0.06 & 6.40 $\pm$ 0.03 & 8.69 $\pm$ 0.03 & 9.06 $\pm$ 0.05 & 9.87 $\pm$ 0.06 & 11.21 $\pm$ 0.09  \\
M08 & P$_2$ & 25.78 $\pm$ 0.01 & 25.29 $\pm$ 0.01 & 27.01 $\pm$ 0.04 & 27.23 $\pm$ 0.04 & 24.97 $\pm$ 0.02 & 23.97 $\pm$ 0.01 & 22.68 $\pm$ 0.01 & 22.43 $\pm$ 0.01 & 21.43 $\pm$ 0.01  \\
M08 & n & 1.17 $\pm$ 0.00 & 1.25 $\pm$ 0.00 & 1.02 $\pm$ 0.00 & 1.14 $\pm$ 0.01 & 1.10 $\pm$ 0.00 & 1.30 $\pm$ 0.00 & 1.34 $\pm$ 0.00 & 1.38 $\pm$ 0.00 & 1.46 $\pm$ 0.00  \\
\hline
M09 & P$_0$ & 3.47 $\pm$ 0.01 & 3.53 $\pm$ 0.01 & 3.28 $\pm$ 0.02 & 3.28 $\pm$ 0.02 & 3.18 $\pm$ 0.02 & 2.61 $\pm$ 0.01 & 2.07 $\pm$ 0.01 & 1.72 $\pm$ 0.00 & 1.26 $\pm$ 0.00  \\
M09 & P$_1$ & 8.88 $\pm$ 0.02 & 9.07 $\pm$ 0.05 & 7.01 $\pm$ 0.05 & 8.56 $\pm$ 0.10 & 7.15 $\pm$ 0.06 & 11.58 $\pm$ 0.07 & 12.20 $\pm$ 0.09 & 13.30 $\pm$ 0.11 & 17.47 $\pm$ 0.22  \\
M09 & P$_2$ & 25.34 $\pm$ 0.00 & 25.91 $\pm$ 0.01 & 25.78 $\pm$ 0.01 & 26.73 $\pm$ 0.02 & 25.03 $\pm$ 0.02 & 24.10 $\pm$ 0.01 & 23.22 $\pm$ 0.01 & 22.99 $\pm$ 0.00 & 21.81 $\pm$ 0.00  \\
M09 & n & 1.34 $\pm$ 0.00 & 1.36 $\pm$ 0.00 & 1.20 $\pm$ 0.00 & 1.33 $\pm$ 0.01 & 1.22 $\pm$ 0.01 & 1.51 $\pm$ 0.00 & 1.55 $\pm$ 0.00 & 1.60 $\pm$ 0.01 & 1.76 $\pm$ 0.01  \\
\hline
M10 & P$_0$ & 3.96 $\pm$ 0.00 & 3.79 $\pm$ 0.01 & 3.68 $\pm$ 0.01 & 3.49 $\pm$ 0.02 & 3.42 $\pm$ 0.01 & 2.80 $\pm$ 0.00 & 2.33 $\pm$ 0.00 & 1.93 $\pm$ 0.00 & 1.41 $\pm$ 0.00  \\
M10 & P$_1$ & 6.83 $\pm$ 0.01 & 7.30 $\pm$ 0.01 & 5.43 $\pm$ 0.01 & 6.64 $\pm$ 0.04 & 5.78 $\pm$ 0.02 & 7.75 $\pm$ 0.01 & 8.41 $\pm$ 0.02 & 8.59 $\pm$ 0.02 & 10.30 $\pm$ 0.03  \\
M10 & P$_2$ & 25.98 $\pm$ 0.00 & 26.39 $\pm$ 0.00 & 26.23 $\pm$ 0.01 & 26.51 $\pm$ 0.01 & 24.95 $\pm$ 0.01 & 23.77 $\pm$ 0.00 & 23.21 $\pm$ 0.00 & 22.64 $\pm$ 0.00 & 21.54 $\pm$ 0.00  \\
M10 & n & 1.18 $\pm$ 0.00 & 1.22 $\pm$ 0.00 & 1.04 $\pm$ 0.00 & 1.16 $\pm$ 0.00 & 1.08 $\pm$ 0.00 & 1.25 $\pm$ 0.00 & 1.30 $\pm$ 0.00 & 1.32 $\pm$ 0.00 & 1.43 $\pm$ 0.00  \\
\hline
M11 & P$_0$ & 4.91 $\pm$ 0.01 & 5.14 $\pm$ 0.01 & 4.41 $\pm$ 0.01 & 4.03 $\pm$ 0.02 & 4.02 $\pm$ 0.01 & 3.39 $\pm$ 0.01 & 3.02 $\pm$ 0.01 & 2.70 $\pm$ 0.01 & 1.95 $\pm$ 0.01  \\
M11 & P$_1$ & 8.77 $\pm$ 0.03 & 8.48 $\pm$ 0.04 & 7.85 $\pm$ 0.03 & 9.20 $\pm$ 0.08 & 7.58 $\pm$ 0.03 & 9.63 $\pm$ 0.04 & 10.56 $\pm$ 0.05 & 10.49 $\pm$ 0.05 & 12.10 $\pm$ 0.08  \\
M11 & P$_2$ & 27.18 $\pm$ 0.01 & 28.28 $\pm$ 0.01 & 27.61 $\pm$ 0.01 & 27.91 $\pm$ 0.02 & 25.57 $\pm$ 0.01 & 24.46 $\pm$ 0.01 & 24.18 $\pm$ 0.01 & 24.07 $\pm$ 0.01 & 23.14 $\pm$ 0.00  \\
M11 & n & 1.33 $\pm$ 0.00 & 1.31 $\pm$ 0.00 & 1.26 $\pm$ 0.00 & 1.35 $\pm$ 0.01 & 1.24 $\pm$ 0.00 & 1.38 $\pm$ 0.00 & 1.45 $\pm$ 0.00 & 1.44 $\pm$ 0.00 & 1.52 $\pm$ 0.00  \\
\hline
M12 & P$_0$ & 4.74 $\pm$ 0.01 & 4.62 $\pm$ 0.01 & 3.94 $\pm$ 0.01 & 3.51 $\pm$ 0.02 & 3.44 $\pm$ 0.01 & 2.90 $\pm$ 0.01 & 2.65 $\pm$ 0.00 & 2.35 $\pm$ 0.01 & 1.66 $\pm$ 0.01  \\
M12 & P$_1$ & 7.44 $\pm$ 0.02 & 7.82 $\pm$ 0.02 & 7.00 $\pm$ 0.02 & 8.36 $\pm$ 0.09 & 6.68 $\pm$ 0.03 & 8.30 $\pm$ 0.03 & 8.85 $\pm$ 0.03 & 8.39 $\pm$ 0.03 & 9.72 $\pm$ 0.06  \\
M12 & P$_2$ & 28.58 $\pm$ 0.01 & 28.71 $\pm$ 0.01 & 28.31 $\pm$ 0.01 & 28.76 $\pm$ 0.02 & 25.83 $\pm$ 0.01 & 24.49 $\pm$ 0.01 & 24.16 $\pm$ 0.00 & 23.77 $\pm$ 0.00 & 22.47 $\pm$ 0.00  \\
M12 & n & 1.24 $\pm$ 0.00 & 1.28 $\pm$ 0.00 & 1.20 $\pm$ 0.00 & 1.30 $\pm$ 0.01 & 1.18 $\pm$ 0.00 & 1.30 $\pm$ 0.00 & 1.35 $\pm$ 0.00 & 1.31 $\pm$ 0.00 & 1.39 $\pm$ 0.00  \\
\hline
M13 & P$_0$ & 5.46 $\pm$ 0.01 & 6.30 $\pm$ 0.02 & 4.78 $\pm$ 0.02 & 4.27 $\pm$ 0.02 & 4.31 $\pm$ 0.02 & 3.69 $\pm$ 0.01 & 3.38 $\pm$ 0.01 & 3.14 $\pm$ 0.01 & 2.33 $\pm$ 0.01  \\
M13 & P$_1$ & 8.62 $\pm$ 0.04 & 7.76 $\pm$ 0.04 & 7.53 $\pm$ 0.03 & 8.90 $\pm$ 0.08 & 7.41 $\pm$ 0.04 & 9.08 $\pm$ 0.05 & 8.92 $\pm$ 0.05 & 9.00 $\pm$ 0.05 & 10.57 $\pm$ 0.08  \\
M13 & P$_2$ & 27.97 $\pm$ 0.01 & 29.57 $\pm$ 0.02 & 28.34 $\pm$ 0.01 & 28.26 $\pm$ 0.02 & 25.94 $\pm$ 0.02 & 24.82 $\pm$ 0.01 & 24.14 $\pm$ 0.01 & 23.82 $\pm$ 0.01 & 22.29 $\pm$ 0.01  \\
M13 & n & 1.32 $\pm$ 0.00 & 1.26 $\pm$ 0.00 & 1.23 $\pm$ 0.00 & 1.34 $\pm$ 0.01 & 1.23 $\pm$ 0.00 & 1.34 $\pm$ 0.00 & 1.34 $\pm$ 0.00 & 1.34 $\pm$ 0.00 & 1.44 $\pm$ 0.01  \\
\hline
M14 & P$_0$ & 4.76 $\pm$ 0.01 & 5.18 $\pm$ 0.02 & 4.32 $\pm$ 0.03 & 3.33 $\pm$ 0.03 & 3.68 $\pm$ 0.02 & 2.92 $\pm$ 0.01 & 2.56 $\pm$ 0.01 & 2.39 $\pm$ 0.01 & 1.73 $\pm$ 0.01  \\
M14 & P$_1$ & 8.69 $\pm$ 0.03 & 8.11 $\pm$ 0.04 & 7.10 $\pm$ 0.05 & 9.13 $\pm$ 0.11 & 7.40 $\pm$ 0.05 & 9.65 $\pm$ 0.06 & 10.67 $\pm$ 0.08 & 10.62 $\pm$ 0.08 & 13.43 $\pm$ 0.18  \\
M14 & P$_2$ & 28.24 $\pm$ 0.01 & 29.19 $\pm$ 0.02 & 27.85 $\pm$ 0.02 & 27.87 $\pm$ 0.02 & 27.31 $\pm$ 0.02 & 25.67 $\pm$ 0.01 & 24.39 $\pm$ 0.01 & 23.82 $\pm$ 0.01 & 22.24 $\pm$ 0.01  \\
M14 & n & 1.32 $\pm$ 0.00 & 1.27 $\pm$ 0.00 & 1.19 $\pm$ 0.00 & 1.34 $\pm$ 0.01 & 1.20 $\pm$ 0.00 & 1.35 $\pm$ 0.00 & 1.44 $\pm$ 0.00 & 1.42 $\pm$ 0.00 & 1.55 $\pm$ 0.01  \\
\hline
M15 & P$_0$ & 4.13 $\pm$ 0.00 & 4.73 $\pm$ 0.01 & 4.33 $\pm$ 0.01 & 3.48 $\pm$ 0.02 & 3.26 $\pm$ 0.01 & 2.56 $\pm$ 0.00 & 2.32 $\pm$ 0.00 & 2.15 $\pm$ 0.00 & 1.39 $\pm$ 0.00  \\
M15 & P$_1$ & 8.49 $\pm$ 0.02 & 7.52 $\pm$ 0.02 & 5.52 $\pm$ 0.02 & 6.02 $\pm$ 0.04 & 6.46 $\pm$ 0.03 & 7.99 $\pm$ 0.02 & 7.92 $\pm$ 0.02 & 8.02 $\pm$ 0.02 & 11.21 $\pm$ 0.03  \\
M15 & P$_2$ & 25.79 $\pm$ 0.00 & 27.53 $\pm$ 0.01 & 27.69 $\pm$ 0.01 & 27.34 $\pm$ 0.02 & 24.98 $\pm$ 0.01 & 23.72 $\pm$ 0.00 & 23.08 $\pm$ 0.00 & 22.96 $\pm$ 0.00 & 21.65 $\pm$ 0.00  \\
M15 & n & 1.35 $\pm$ 0.00 & 1.27 $\pm$ 0.00 & 1.09 $\pm$ 0.00 & 1.12 $\pm$ 0.00 & 1.20 $\pm$ 0.00 & 1.29 $\pm$ 0.00 & 1.29 $\pm$ 0.00 & 1.30 $\pm$ 0.00 & 1.51 $\pm$ 0.00  \\
\hline
M16 & P$_0$ & 6.83 $\pm$ 0.03 & 7.12 $\pm$ 0.03 & 6.59 $\pm$ 0.03 & 5.73 $\pm$ 0.05 & 5.65 $\pm$ 0.03 & 4.64 $\pm$ 0.02 & 4.17 $\pm$ 0.02 & 4.00 $\pm$ 0.02 & 3.23 $\pm$ 0.02  \\
M16 & P$_1$ & 8.69 $\pm$ 0.05 & 8.50 $\pm$ 0.05 & 7.54 $\pm$ 0.04 & 8.36 $\pm$ 0.08 & 7.57 $\pm$ 0.05 & 9.72 $\pm$ 0.07 & 9.71 $\pm$ 0.08 & 9.69 $\pm$ 0.08 & 10.65 $\pm$ 0.11  \\
M16 & P$_2$ & 29.72 $\pm$ 0.03 & 30.43 $\pm$ 0.03 & 29.61 $\pm$ 0.03 & 29.14 $\pm$ 0.05 & 27.19 $\pm$ 0.03 & 25.36 $\pm$ 0.02 & 24.23 $\pm$ 0.02 & 24.15 $\pm$ 0.02 & 23.00 $\pm$ 0.02  \\
M16 & n & 1.30 $\pm$ 0.00 & 1.29 $\pm$ 0.00 & 1.21 $\pm$ 0.00 & 1.26 $\pm$ 0.01 & 1.20 $\pm$ 0.00 & 1.35 $\pm$ 0.00 & 1.36 $\pm$ 0.01 & 1.35 $\pm$ 0.01 & 1.40 $\pm$ 0.01  \\
\hline
M17 & P$_0$ & 4.67 $\pm$ 0.01 & 4.73 $\pm$ 0.01 & 4.10 $\pm$ 0.02 & 3.44 $\pm$ 0.03 & 3.90 $\pm$ 0.02 & 3.27 $\pm$ 0.01 & 2.67 $\pm$ 0.01 & 2.39 $\pm$ 0.00 & 1.74 $\pm$ 0.01  \\
M17 & P$_1$ & 7.77 $\pm$ 0.02 & 7.62 $\pm$ 0.03 & 5.93 $\pm$ 0.03 & 6.74 $\pm$ 0.07 & 6.02 $\pm$ 0.03 & 8.20 $\pm$ 0.02 & 8.84 $\pm$ 0.04 & 8.68 $\pm$ 0.03 & 9.45 $\pm$ 0.05  \\
M17 & P$_2$ & 27.07 $\pm$ 0.01 & 27.82 $\pm$ 0.01 & 26.74 $\pm$ 0.02 & 27.70 $\pm$ 0.03 & 26.60 $\pm$ 0.02 & 25.23 $\pm$ 0.01 & 23.98 $\pm$ 0.01 & 23.48 $\pm$ 0.00 & 22.20 $\pm$ 0.01  \\
M17 & n & 1.26 $\pm$ 0.00 & 1.25 $\pm$ 0.00 & 1.10 $\pm$ 0.00 & 1.15 $\pm$ 0.01 & 1.10 $\pm$ 0.00 & 1.27 $\pm$ 0.00 & 1.33 $\pm$ 0.00 & 1.32 $\pm$ 0.00 & 1.36 $\pm$ 0.00  \\
\hline
M18 & P$_0$ & 5.58 $\pm$ 0.01 & 5.76 $\pm$ 0.02 & 5.85 $\pm$ 0.05 & 4.83 $\pm$ 0.04 & 5.35 $\pm$ 0.03 & 4.28 $\pm$ 0.01 & 3.78 $\pm$ 0.01 & 3.31 $\pm$ 0.01 & 2.66 $\pm$ 0.01  \\
M18 & P$_1$ & 7.37 $\pm$ 0.02 & 7.36 $\pm$ 0.02 & 5.66 $\pm$ 0.03 & 7.19 $\pm$ 0.06 & 6.14 $\pm$ 0.02 & 7.95 $\pm$ 0.02 & 8.05 $\pm$ 0.03 & 8.52 $\pm$ 0.03 & 9.23 $\pm$ 0.04  \\
M18 & P$_2$ & 28.15 $\pm$ 0.01 & 29.09 $\pm$ 0.02 & 29.67 $\pm$ 0.06 & 29.79 $\pm$ 0.04 & 27.99 $\pm$ 0.03 & 25.97 $\pm$ 0.01 & 24.88 $\pm$ 0.01 & 24.46 $\pm$ 0.01 & 23.24 $\pm$ 0.01  \\
M18 & n & 1.19 $\pm$ 0.00 & 1.19 $\pm$ 0.00 & 1.01 $\pm$ 0.00 & 1.18 $\pm$ 0.01 & 1.07 $\pm$ 0.00 & 1.24 $\pm$ 0.00 & 1.23 $\pm$ 0.00 & 1.27 $\pm$ 0.00 & 1.31 $\pm$ 0.00  \\
\hline
M19 & P$_0$ & 4.56 $\pm$ 0.02 & 4.36 $\pm$ 0.02 & 4.70 $\pm$ 0.03 & 4.10 $\pm$ 0.03 & 4.13 $\pm$ 0.02 & 3.38 $\pm$ 0.01 & 2.82 $\pm$ 0.01 & 2.52 $\pm$ 0.01 & 1.96 $\pm$ 0.01  \\
M19 & P$_1$ & 8.61 $\pm$ 0.06 & 9.10 $\pm$ 0.05 & 7.01 $\pm$ 0.04 & 8.58 $\pm$ 0.09 & 7.50 $\pm$ 0.04 & 11.19 $\pm$ 0.07 & 11.27 $\pm$ 0.09 & 11.59 $\pm$ 0.11 & 13.00 $\pm$ 0.18  \\
M19 & P$_2$ & 27.90 $\pm$ 0.02 & 27.57 $\pm$ 0.02 & 28.21 $\pm$ 0.03 & 28.10 $\pm$ 0.03 & 25.91 $\pm$ 0.02 & 24.49 $\pm$ 0.01 & 23.51 $\pm$ 0.01 & 23.29 $\pm$ 0.01 & 21.95 $\pm$ 0.01  \\
M19 & n & 1.29 $\pm$ 0.00 & 1.32 $\pm$ 0.00 & 1.14 $\pm$ 0.00 & 1.28 $\pm$ 0.01 & 1.19 $\pm$ 0.00 & 1.44 $\pm$ 0.00 & 1.44 $\pm$ 0.01 & 1.46 $\pm$ 0.01 & 1.53 $\pm$ 0.01  \\
\hline
\end{tabular}
\end{table}

\begin{table}
\caption[]{The  gain information  for 19 beams within zenith angle of 26.4$^{\circ}$. Mean value of the gain of Beam 1 is expressed in unit of K/Jy.  For the rest 18 beams,  the ratio compared to the gain value of beam 1 in same frequency is presented.  
 \label{table:gainvalue}}
\setlength{\tabcolsep}{1pt}
\footnotesize
\begin{tabular}{cccccccccc}
 \hline\noalign{\smallskip}
 \hline\noalign{\smallskip}
\multicolumn{1}{c}{ \multirow{2}{*}{Beam}}  & \multicolumn{9}{c}{ Gain (K/Jy) / Ratio} \\
 \cline{2-10}
&  \multicolumn{1}{c}{1050} &\multicolumn{1}{c}{1100} & \multicolumn{1}{c}{1150}& \multicolumn{1}{c}{1200}& \multicolumn{1}{c}{1250} & \multicolumn{1}{c}{1300} & \multicolumn{1}{c}{1350} & \multicolumn{1}{c}{1400} & \multicolumn{1}{c}{1450} \\
 \noalign{\smallskip}\hline
M01 & 15.98 $\pm$ 0.27 & 16.27 $\pm$ 0.21 & 16.48 $\pm$ 0.14 & 16.10 $\pm$ 0.25 & 16.12 $\pm$ 0.25 & 15.94 $\pm$ 0.28 & 15.98 $\pm$ 0.27 & 16.02 $\pm$ 0.26 & 15.71 $\pm$ 0.29  \\
\hline
M02 & 0.98 $\pm$ 0.03 & 0.95 $\pm$ 0.03 & 0.95 $\pm$ 0.03 & 0.95 $\pm$ 0.03 & 0.95 $\pm$ 0.03 & 0.93 $\pm$ 0.03 & 0.94 $\pm$ 0.02 & 0.94 $\pm$ 0.02 & 0.95 $\pm$ 0.03  \\
M03 & 1.03 $\pm$ 0.03 & 0.99 $\pm$ 0.03 & 1.00 $\pm$ 0.03 & 1.00 $\pm$ 0.04 & 1.00 $\pm$ 0.03 & 0.99 $\pm$ 0.03 & 0.97 $\pm$ 0.03 & 0.97 $\pm$ 0.03 & 0.98 $\pm$ 0.03  \\
M04 & 1.01 $\pm$ 0.04 & 0.97 $\pm$ 0.03 & 0.98 $\pm$ 0.03 & 0.97 $\pm$ 0.04 & 0.97 $\pm$ 0.04 & 0.96 $\pm$ 0.03 & 0.95 $\pm$ 0.03 & 0.95 $\pm$ 0.03 & 0.95 $\pm$ 0.03  \\
M05 & 1.02 $\pm$ 0.03 & 1.01 $\pm$ 0.02 & 1.03 $\pm$ 0.02 & 1.00 $\pm$ 0.03 & 1.00 $\pm$ 0.03 & 0.99 $\pm$ 0.02 & 0.97 $\pm$ 0.02 & 0.98 $\pm$ 0.02 & 0.98 $\pm$ 0.03  \\
M06 & 0.99 $\pm$ 0.03 & 0.98 $\pm$ 0.02 & 0.98 $\pm$ 0.03 & 0.96 $\pm$ 0.04 & 0.96 $\pm$ 0.03 & 0.95 $\pm$ 0.02 & 0.94 $\pm$ 0.02 & 0.94 $\pm$ 0.02 & 0.94 $\pm$ 0.02  \\
M07 & 0.98 $\pm$ 0.02 & 0.95 $\pm$ 0.02 & 0.96 $\pm$ 0.02 & 0.94 $\pm$ 0.03 & 0.96 $\pm$ 0.03 & 0.95 $\pm$ 0.02 & 0.94 $\pm$ 0.02 & 0.93 $\pm$ 0.02 & 0.93 $\pm$ 0.02  \\
M08 & 0.96 $\pm$ 0.03 & 0.94 $\pm$ 0.03 & 0.93 $\pm$ 0.03 & 0.93 $\pm$ 0.04 & 0.92 $\pm$ 0.03 & 0.92 $\pm$ 0.03 & 0.91 $\pm$ 0.03 & 0.89 $\pm$ 0.03 & 0.89 $\pm$ 0.03  \\
M09 & 0.95 $\pm$ 0.02 & 0.94 $\pm$ 0.02 & 0.92 $\pm$ 0.02 & 0.91 $\pm$ 0.02 & 0.92 $\pm$ 0.02 & 0.93 $\pm$ 0.02 & 0.91 $\pm$ 0.02 & 0.89 $\pm$ 0.02 & 0.90 $\pm$ 0.02  \\
M10 & 0.99 $\pm$ 0.03 & 0.97 $\pm$ 0.03 & 0.97 $\pm$ 0.03 & 0.96 $\pm$ 0.04 & 0.97 $\pm$ 0.03 & 0.96 $\pm$ 0.03 & 0.96 $\pm$ 0.03 & 0.95 $\pm$ 0.03 & 0.95 $\pm$ 0.03  \\
M11 & 0.96 $\pm$ 0.04 & 0.94 $\pm$ 0.04 & 0.94 $\pm$ 0.03 & 0.93 $\pm$ 0.05 & 0.93 $\pm$ 0.04 & 0.92 $\pm$ 0.03 & 0.92 $\pm$ 0.03 & 0.91 $\pm$ 0.03 & 0.91 $\pm$ 0.03  \\
M12 & 0.96 $\pm$ 0.05 & 0.93 $\pm$ 0.04 & 0.92 $\pm$ 0.03 & 0.94 $\pm$ 0.04 & 0.94 $\pm$ 0.03 & 0.91 $\pm$ 0.03 & 0.90 $\pm$ 0.03 & 0.89 $\pm$ 0.02 & 0.90 $\pm$ 0.03  \\
M13 & 0.99 $\pm$ 0.04 & 0.96 $\pm$ 0.03 & 0.97 $\pm$ 0.03 & 0.99 $\pm$ 0.04 & 0.96 $\pm$ 0.03 & 0.97 $\pm$ 0.03 & 0.96 $\pm$ 0.03 & 0.95 $\pm$ 0.03 & 0.95 $\pm$ 0.03  \\
M14 & 0.95 $\pm$ 0.03 & 0.94 $\pm$ 0.03 & 0.93 $\pm$ 0.03 & 0.92 $\pm$ 0.04 & 0.93 $\pm$ 0.04 & 0.93 $\pm$ 0.03 & 0.92 $\pm$ 0.02 & 0.91 $\pm$ 0.02 & 0.91 $\pm$ 0.02  \\
M15 & 0.90 $\pm$ 0.03 & 0.91 $\pm$ 0.02 & 0.90 $\pm$ 0.02 & 0.90 $\pm$ 0.05 & 0.90 $\pm$ 0.02 & 0.90 $\pm$ 0.02 & 0.88 $\pm$ 0.02 & 0.87 $\pm$ 0.02 & 0.88 $\pm$ 0.02  \\
M16 & 0.90 $\pm$ 0.03 & 0.88 $\pm$ 0.02 & 0.86 $\pm$ 0.03 & 0.88 $\pm$ 0.03 & 0.86 $\pm$ 0.02 & 0.86 $\pm$ 0.02 & 0.84 $\pm$ 0.02 & 0.83 $\pm$ 0.02 & 0.83 $\pm$ 0.02  \\
M17 & 0.93 $\pm$ 0.03 & 0.91 $\pm$ 0.03 & 0.90 $\pm$ 0.03 & 0.89 $\pm$ 0.04 & 0.89 $\pm$ 0.02 & 0.88 $\pm$ 0.02 & 0.88 $\pm$ 0.02 & 0.86 $\pm$ 0.02 & 0.86 $\pm$ 0.02  \\
M18 & 0.90 $\pm$ 0.02 & 0.88 $\pm$ 0.02 & 0.89 $\pm$ 0.02 & 0.89 $\pm$ 0.02 & 0.88 $\pm$ 0.03 & 0.87 $\pm$ 0.02 & 0.86 $\pm$ 0.02 & 0.86 $\pm$ 0.02 & 0.87 $\pm$ 0.02  \\
M19 & 0.94 $\pm$ 0.03 & 0.94 $\pm$ 0.02 & 0.93 $\pm$ 0.03 & 0.93 $\pm$ 0.03 & 0.93 $\pm$ 0.03 & 0.92 $\pm$ 0.03 & 0.90 $\pm$ 0.02 & 0.89 $\pm$ 0.02 & 0.89 $\pm$ 0.03  \\
\hline\noalign{\smallskip}
\end{tabular}
\end{table}

\section{Backend}
\label{sec:backend}

\subsection{Stability and Sensitivity of Spectral Baseline}
\label{subsec:spectal_sensitivity}

Stability and sensitivity are two fundamental properties associated with spectral baseline. Stability represents fluctuation level of baseline in a time range. Sensitivity represents response capability of spectrometer. We estimated stability and sensitivity by taking observations  toward  the  H{\sc i} galaxy N672 in April 20, 2019.   

\begin{figure}
\centering
  \includegraphics[width=0.8\textwidth]{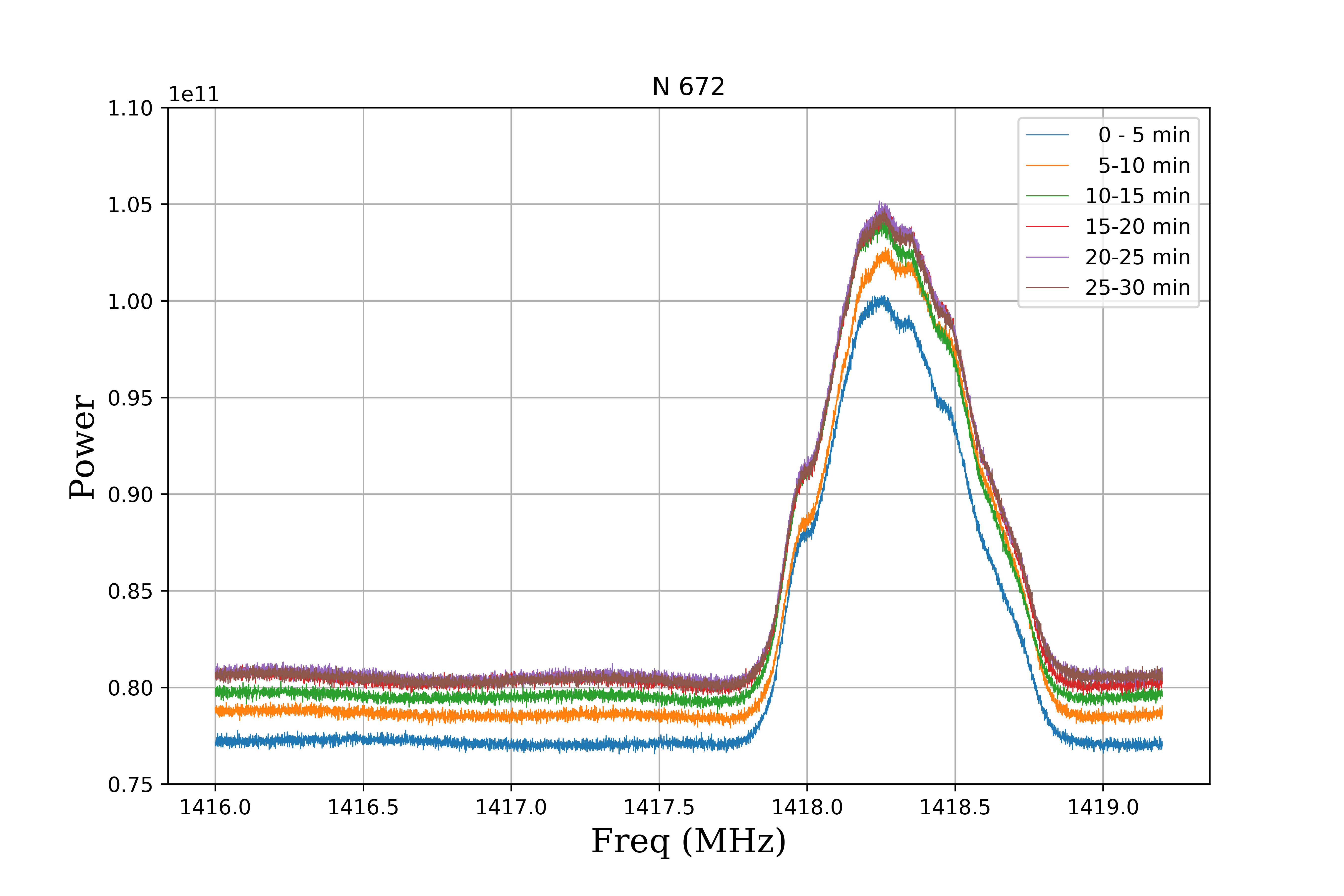}
  \caption{Variance of spectral baseline toward N672 in 30 minutes. Each spectrum is averaged in a time bin of 5 minutes.}
\label{fig:spec_stability}
\end{figure}

A set of observation of 30 minutes without injection of noise signal  was taken to test stability of the baseline.  Averaged bandpass spectra per 5 minutes are shown in Fig. \ref{fig:spec_stability}. The variation of bandpass is $\sim$4\% in 30 minutes. 

\begin{figure}
\centering
  \includegraphics[width=0.8\textwidth]{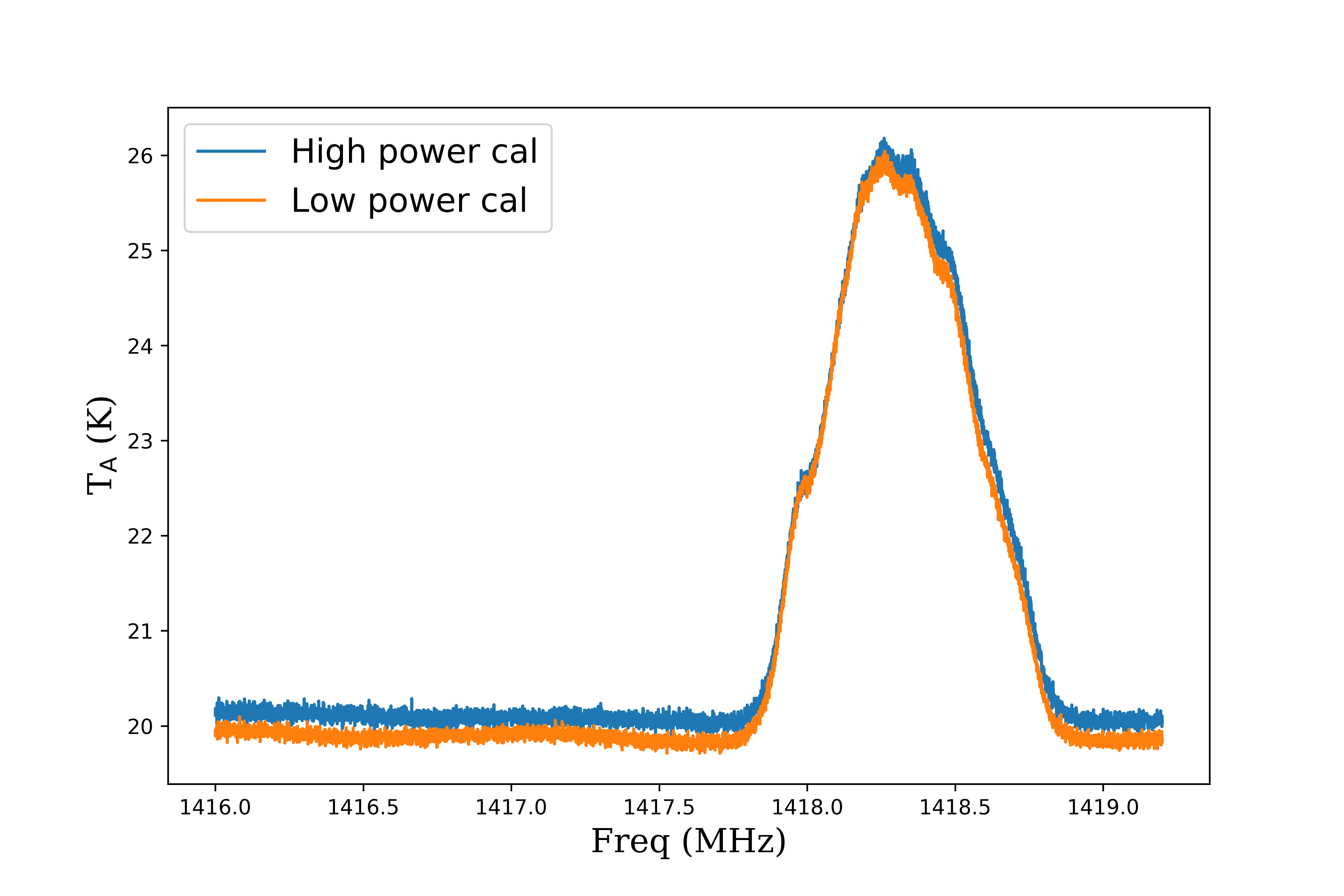}
  \caption{Calibrated H{\sc i} spectrum under high power cal (blue) and low power cal (orange).}
\label{fig:spec_high_low}
\end{figure}

In order to measure sensitivity performance of the backend, noise signal as described in Section \ref{subsec:noise_dipole} was injected  to calibrate observed spectra. Both noise signal with high and low intensity were injected for 5 minutes. The unit of spectrum  was then calibrated into kelvin through the following transformation, 

\begin{equation}
    \rm T_A = T_{cal}\frac{P_{off}^{cal}}{P_{on}^{cal}-P_{off}^{cal}},
\end{equation}
where $T\rm_A$ is calibrated antenna temperature. $T\rm_{cal}$ is noise diode temperature as shown in Fig.\ref{lowcal} and Fig.\ref{highcal}. $P\rm_{on}$ and $P\rm_{off}$ are power value when noise diode is on and off, respectively. As shown in Fig. \ref{fig:spec_high_low}, the derived continuum level under high and low noise injection are consistent within 1\%.  

We estimated sensitivity with root mean square (rms) of a spectrum. RMS is calculated with line-free frequency range. In the total on mode, the expected rms of an averaged spectrum of two polarizations $\sigma\rm_T$ is connected with system temperature $T\rm_{sys}$, channel resolution $\beta$ and integration time $\tau$ with $\sigma_T=T_{sys}/\sqrt{2\beta\tau}$. Periodic injection of high and low noise would lead to increase of $T_{sys}$ by 5.4 K and 0.5 K during an observation cycle (half time for cal on and half time for cal off), respectively.  Comparison between obtained and theoretical rms is shown in Table \ref{table:sensitivity}. It is obvious that  performance of the 19 beam backend is well consistent with that of theoretical value for both low and high intensity noise injection. 

\begin{table}
\caption[]{RMS value of 5 minutes observations toward N672.  Channel resolution is 0.476 kHz during calculation. \label{table:sensitivity}}
\setlength{\tabcolsep}{1pt}
\small
\begin{tabular}{C{3cm}  C{3cm} C{3cm} }
  \hline\noalign{\smallskip}
  \hline\noalign{\smallskip}
Cal Intensity  &  Obtained value & Theoretical value \\
 &  (mK) & (mK) \\
  \hline\noalign{\smallskip}
Low    &  40.1  &  38.4 \\
High   &  48.0  &  47.7  \\
 \noalign{\smallskip}\hline
\end{tabular}
\end{table}

\subsection{Standing Waves}
\label{subsec:standingwave}

The antenna structure and radio frequency (RF) devices of a radio telescope can cause reflection of electromagnetic (EM) wave. The coherent superposition of the received signal and its reflection wave results a periodic fluctuations in the frequency band-pass, which is called "standing wave". This is a common phenomenon seen in the spectroscopic observations with radio telescopes \citep[i.e.][]{Briggs1997, Popping2008}. 
The major contribution to the standing wave for FAST is caused by the reflection between the dish and the receiver cabin. The relation between the standing wave frequency and the reflecting distance is given by Equation  \ref{equ:sdw}:
\begin{equation}
  D=\frac{c}{2\,\Delta f}, \label{equ:sdw}
\end{equation}
where $D$ is the distance between the receiver cabin and the dish, $c$ is the speed of light, and $\Delta f$ is the width of the frequency ripple. For FAST, $D = 138$\,m relates $\Delta f \sim 1$\,MHz. At L-band, the corresponding velocity width is $\sim 200$\,km\,s$^{-1}$, which locates within the range of interest for extra-galactic \hi observations.

\begin{figure}
\centering
\includegraphics[width=0.49\textwidth, angle=0]{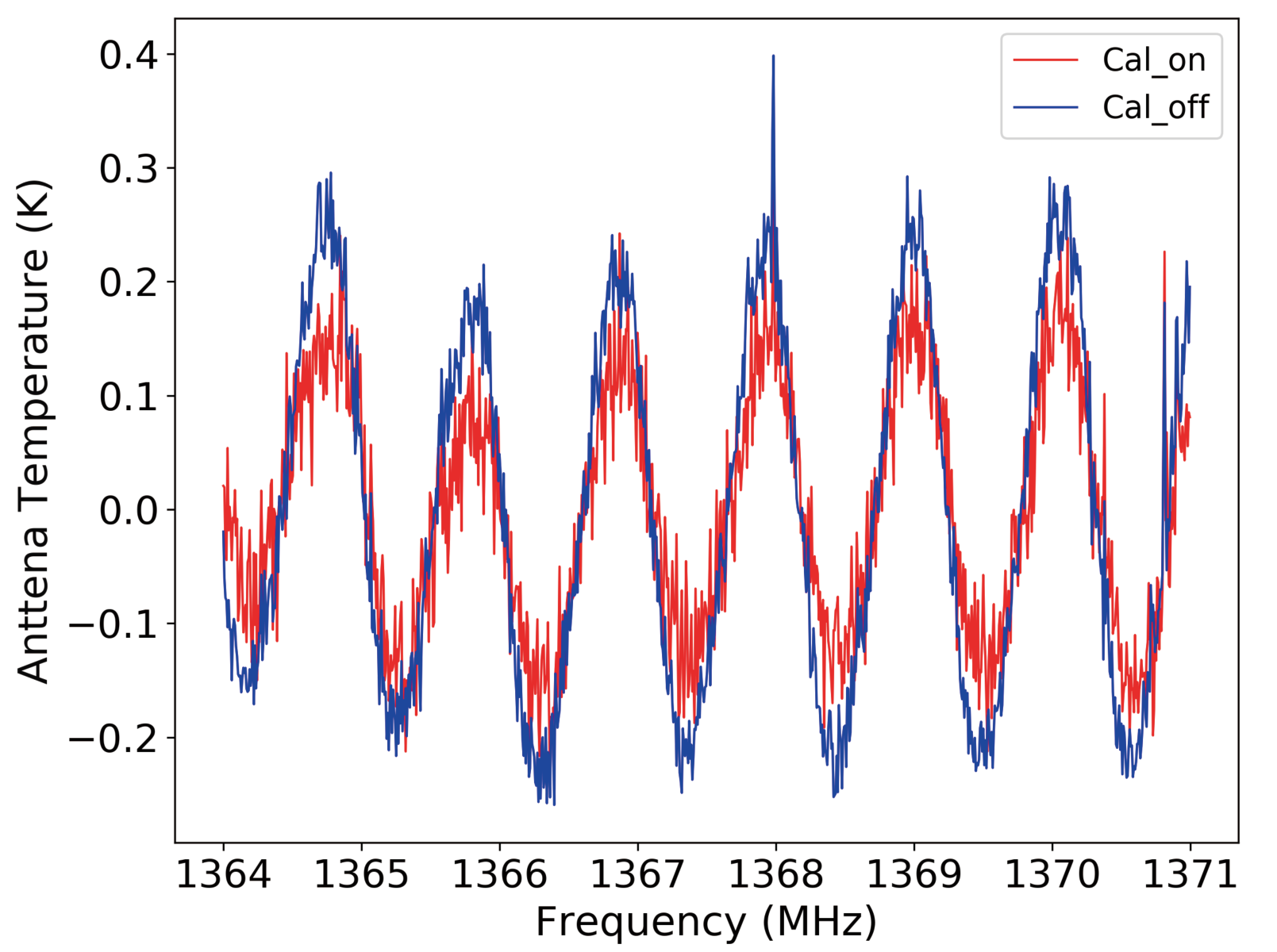}
\includegraphics[width=0.49\textwidth, angle=0]{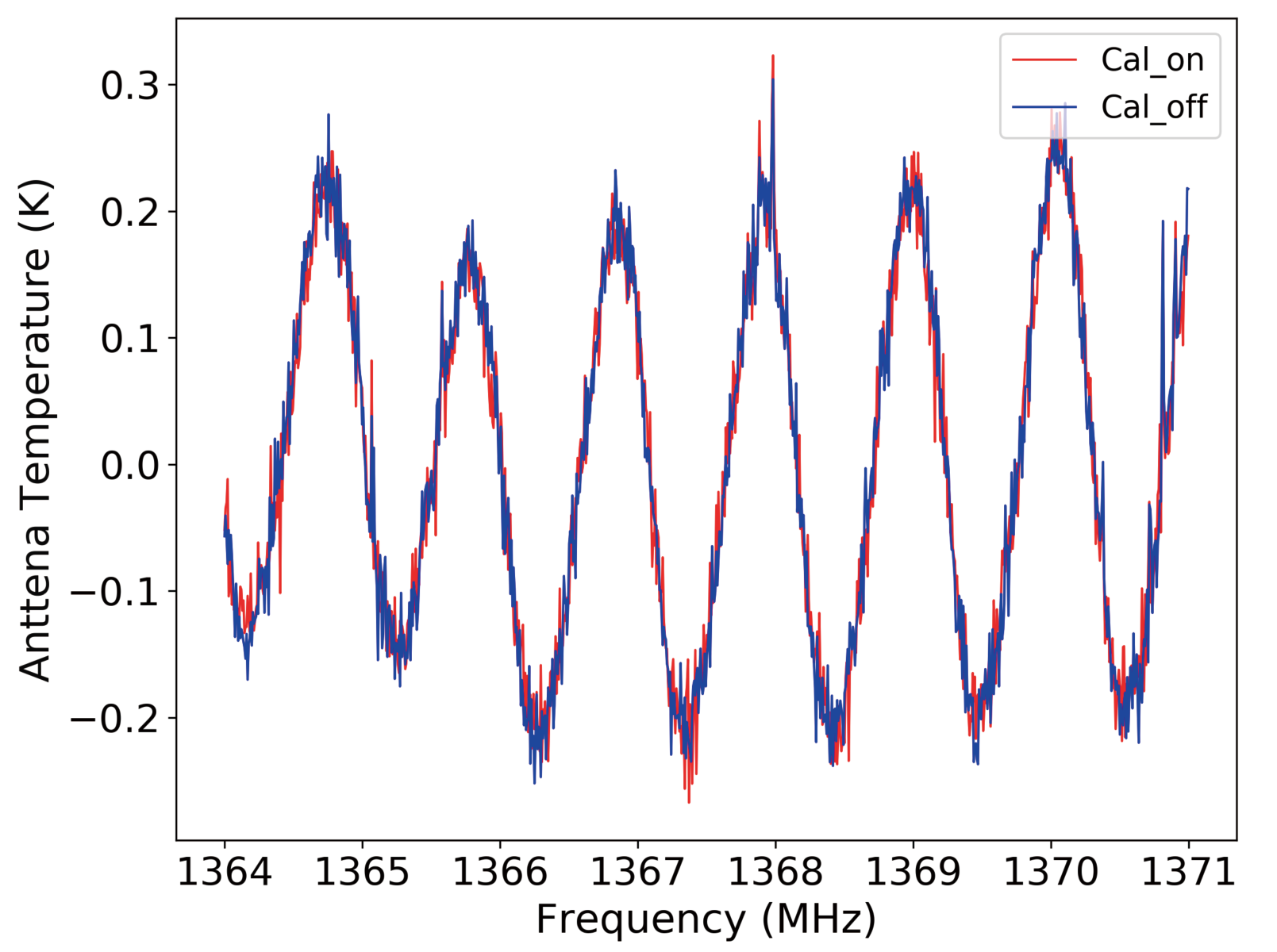}
\includegraphics[width=0.49\textwidth, angle=0]{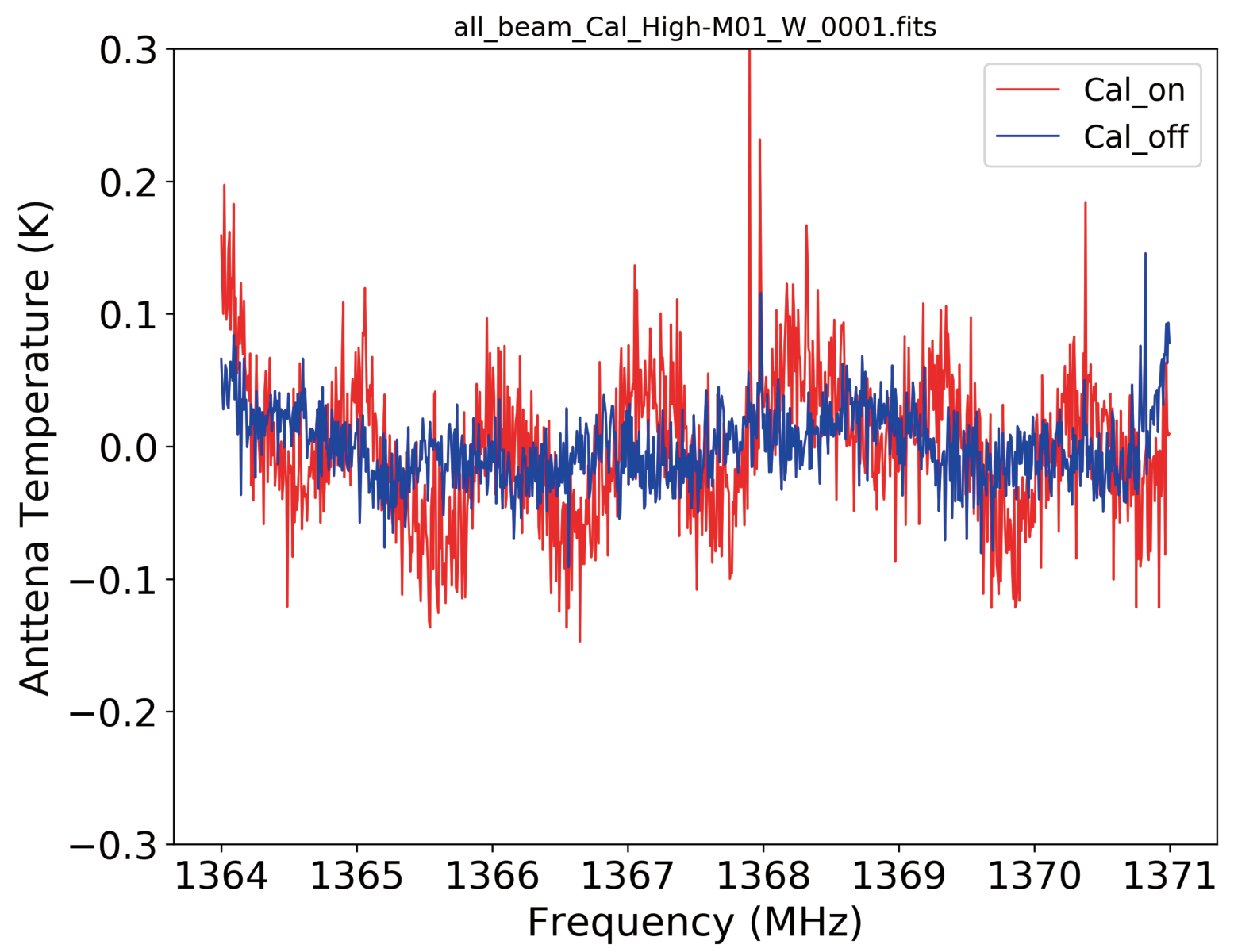}
\includegraphics[width=0.49\textwidth, angle=0]{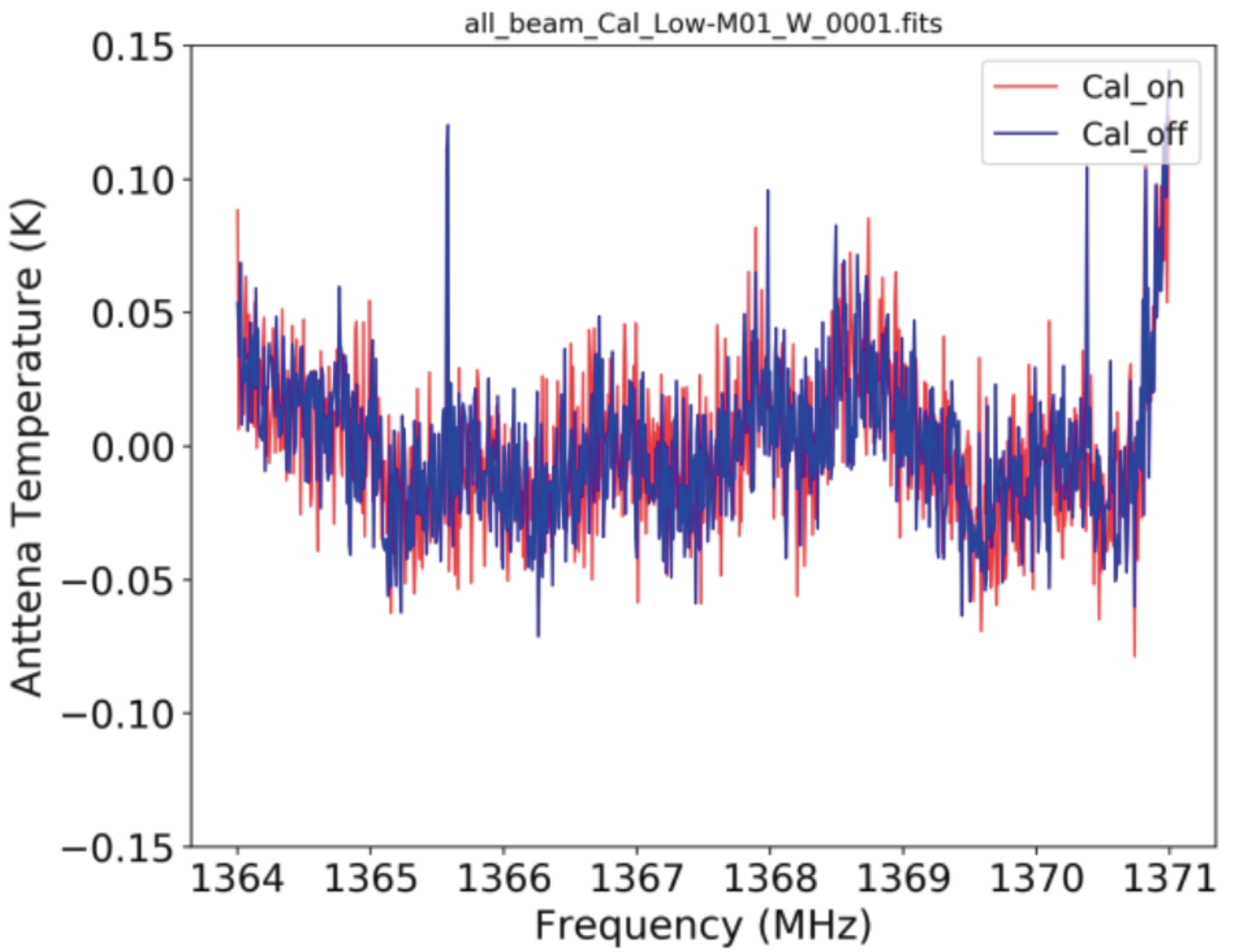}
\caption{The comparison of the band-pass ripples before (upper panel plots) and after (lower panel plots) the cabin dock platform is covered with metal mesh. The left side plots are with high power noise diode injection and the right side plots are with the low noise power.}
\label{fig:sdw}
\end{figure}

Efforts have been put onto analyzing and minimizing the standing wave effects of FAST. To identify the major contribution to the FAST standing wave, we checked the band-pass during the receiver cabin rising. The width of the frequency ripple varies with the cabin height following the relation given by Equation \ref{equ:sdw}. This experiment was done with the cabin dock uncovered. Thus the ground radiation passing through the central hole of the dish serves as the EM source for the standing wave. In Figure \ref{fig:sdw}, the figures from the upper panel shows the band-pass before the cabin dock is covered with metal mesh and those from the lower panel are after. The fluctuation declines significantly when the ground radiation is blocked out. 

The feed leakage emission can be another EM source for the standing wave. Since the horns emit EM wave when the calibration noise diode is fired up. Figure \ref{fig:sdw} shows the band-passes with high power noise diode (left column plots) and with low power noise diode (right column plots). It can be seen that the signal from the high power noise diode ($\sim 10$\,K) affects the band-pass fluctuation, whereas the low power signal ($\sim 1$\,K) does not.  

\begin{figure}
\centering
\includegraphics[width=0.6\textwidth, angle=0]{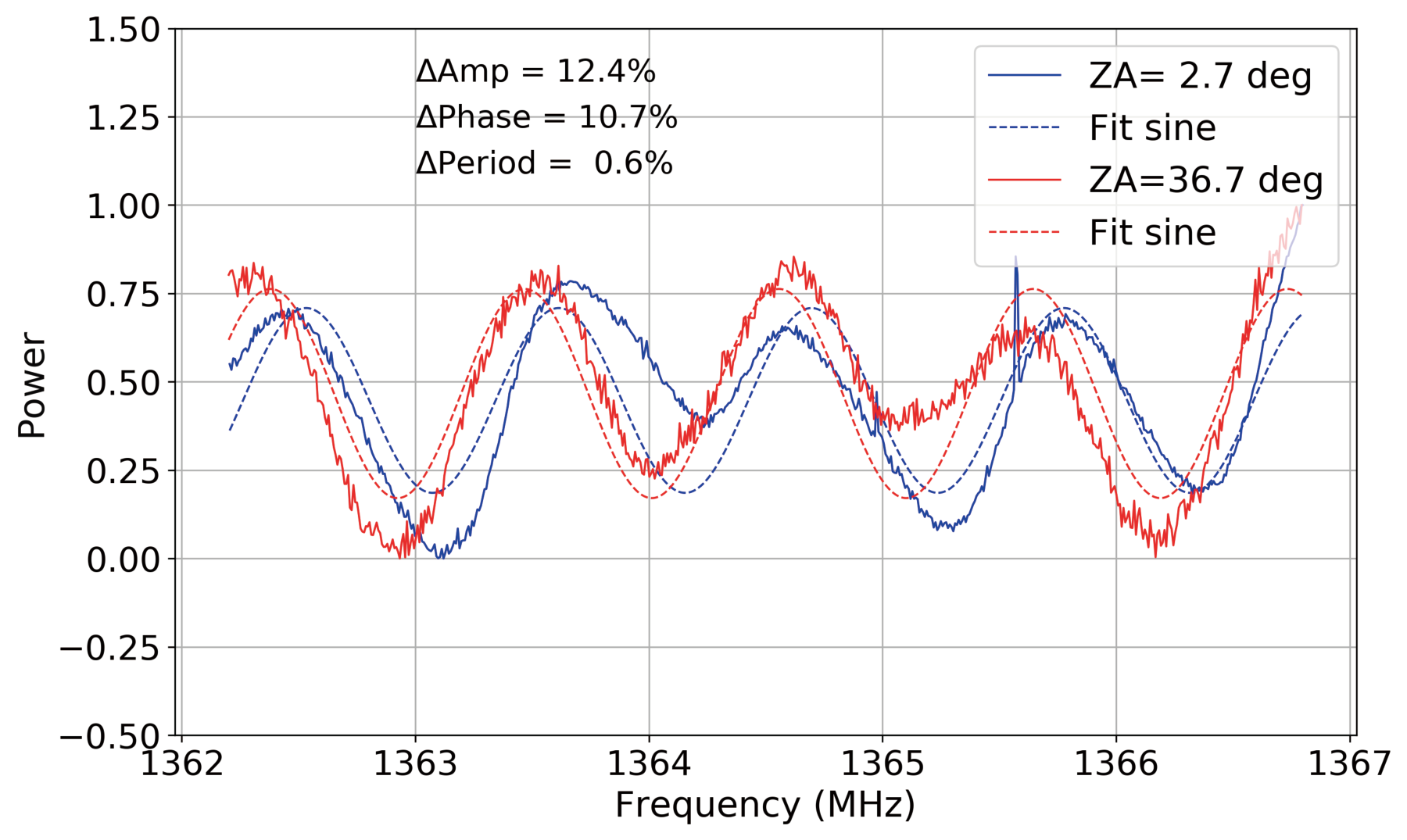}
\caption{Band-pass ripples at different zenith angles. The blue line is the spectrum obtained with drift-scan observation at ZA = 2.7\deg and the red line is at ZA = 36.7\deg. The dashed lines are the sine function fitting results. A linear baseline removal was processed before hand. The intensity of the spectra are normalized and the unit is relative power.}
\label{fig:sdw-za}
\end{figure}

The standing wave at different zenith angles (ZA) has also been checked and yet no significant difference was found. Figure \ref{fig:sdw-za} shows the comparison of the spectra band-pass from ZA = 2.7 \deg and 36.7\deg. The data are fitted to sine function. The major fitting parameters are amplitude, phase, and period. The relative differences of those parameters are 12.4\%, 10.7\%, and 0.6\%. The stable ripple patterns are expected since the distance between the receiver cabin and the apex of the parabolic dish should remain unchanged during observing. 

\begin{figure}
\centering
\includegraphics[width=0.6\textwidth, angle=0]{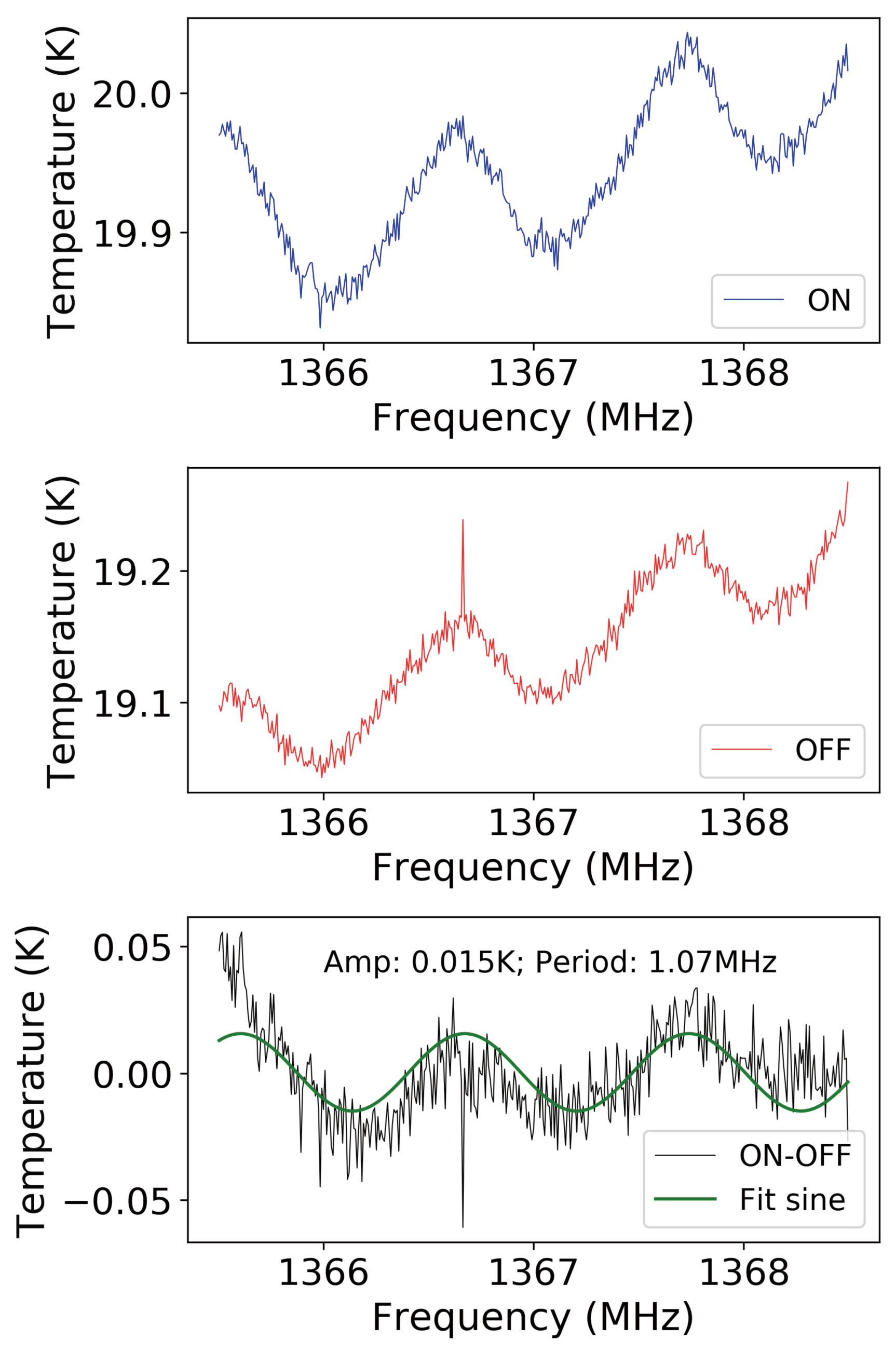}
\caption{Standing wave test from the ON/OFF observation towards J073631.82+383058.3. The top and middle panels are the calibrated band-passes from ON- and OFF- position. The bottom panel is the ON-OFF calibrated band-pass with a linear baseline removal. The sine function is fit to the spectrum, which gives the amplitude of the standing wave $\sim 15$ mK.}
\label{fig:sdw-onoff}
\end{figure}

The band-pass difference is smaller when the ZA difference is smaller. So ON/OFF position switch technique can be used to calibrate the baseline and to decrease the standing wave ripples. Figure \ref{fig:sdw-onoff} shows the standing wave test from the ON/OFF observation towards J073631.82+383058.3. The top and middle panels are the calibrated band-passes from ON- and OFF- position. The bottom panel is the ON-OFF calibrated band-pass and a sine function fitting result, which shows the ripple amplitude of $\sim 15$ mK. 

\begin{figure}
\centering
\includegraphics[width=0.48\textwidth]{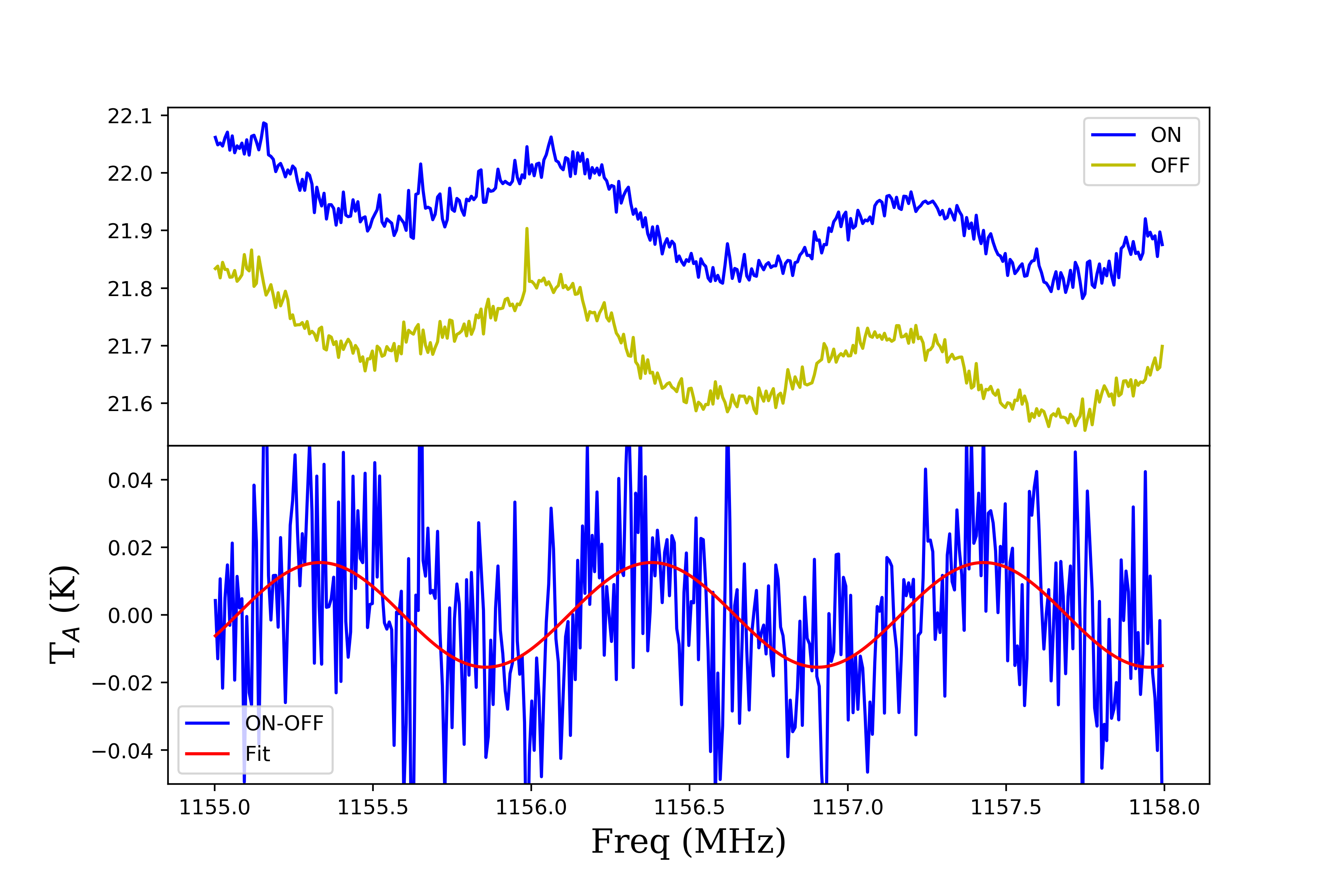}
\includegraphics[width=0.48\textwidth]{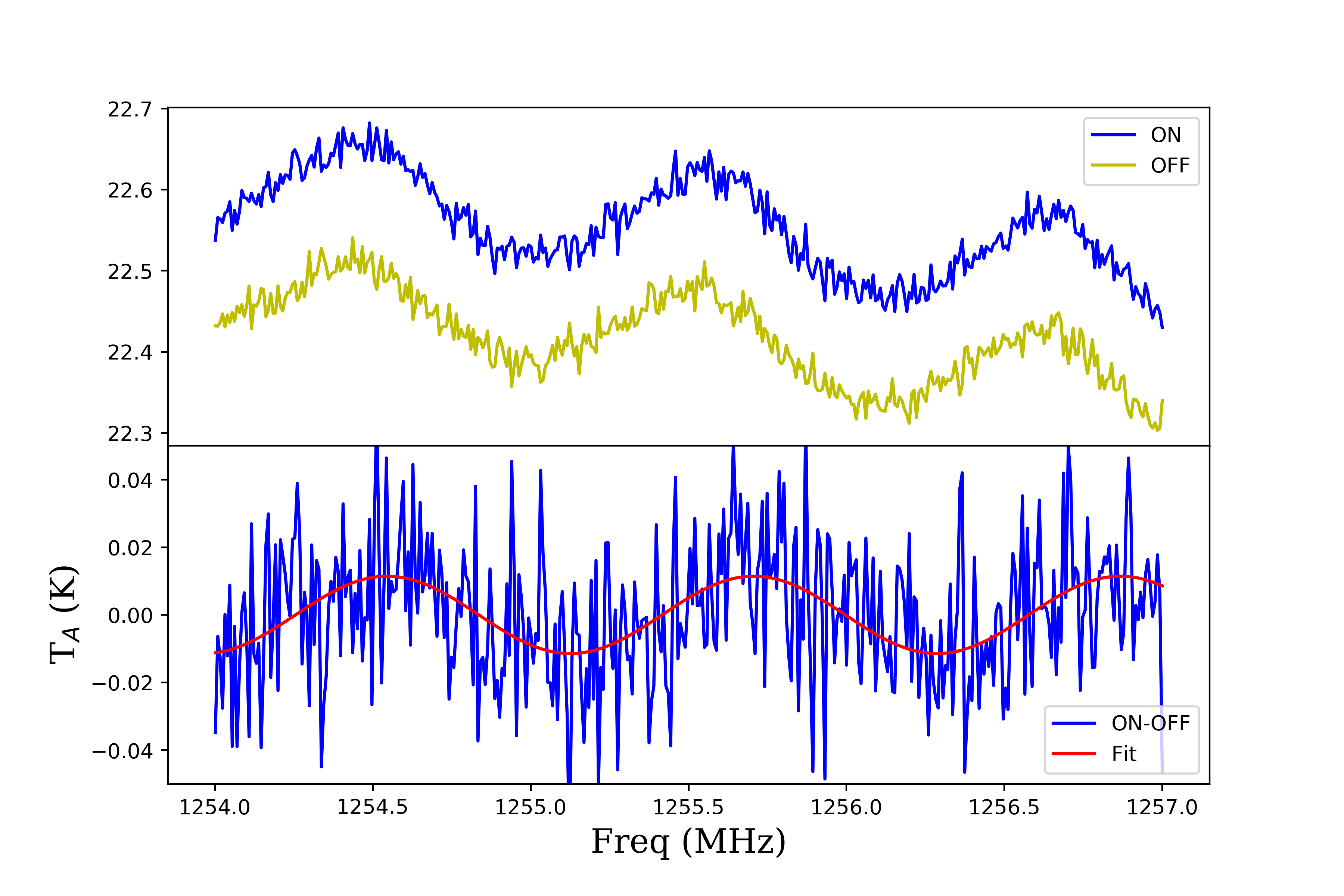}
\includegraphics[width=0.48\textwidth]{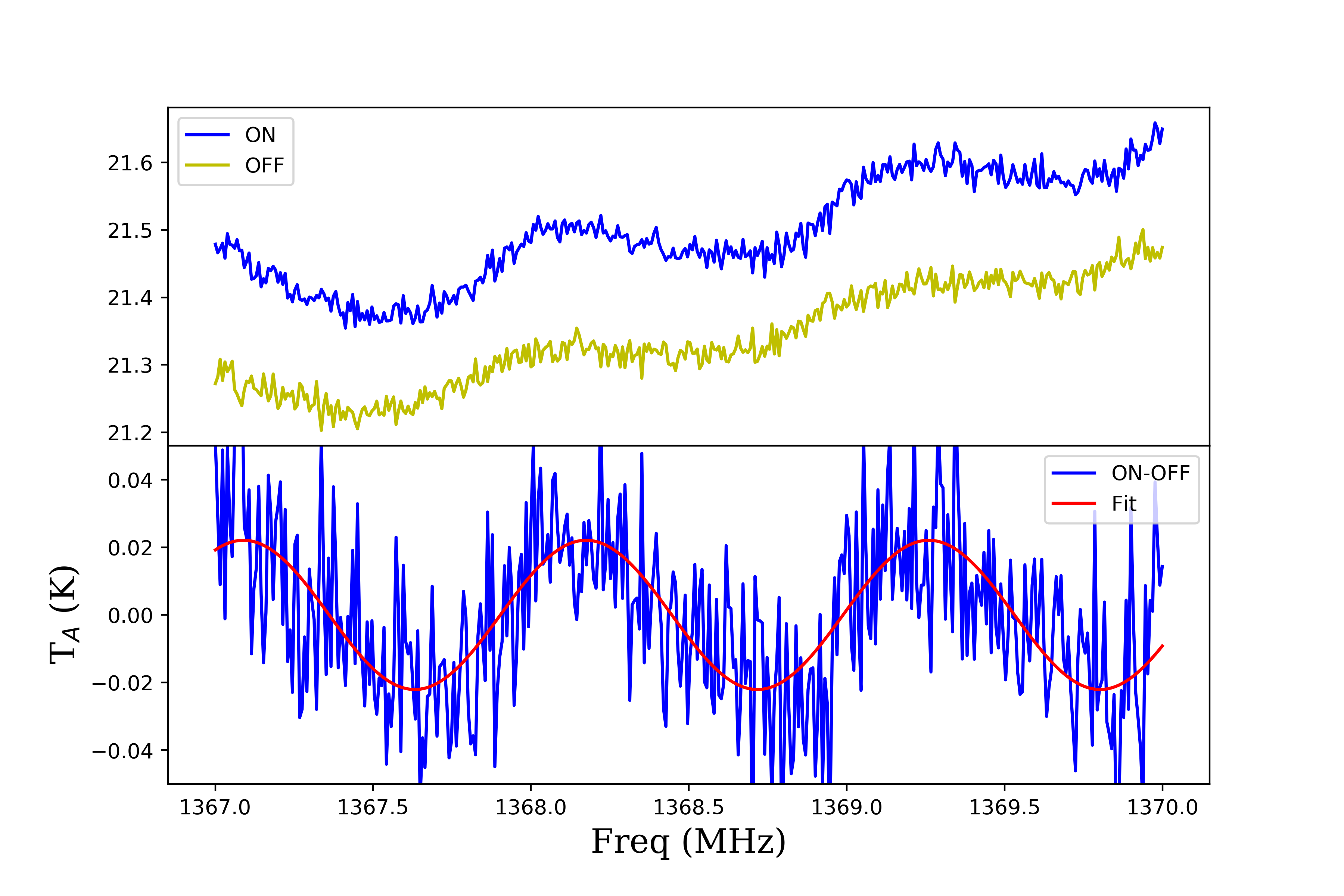}
\includegraphics[width=0.48\textwidth]{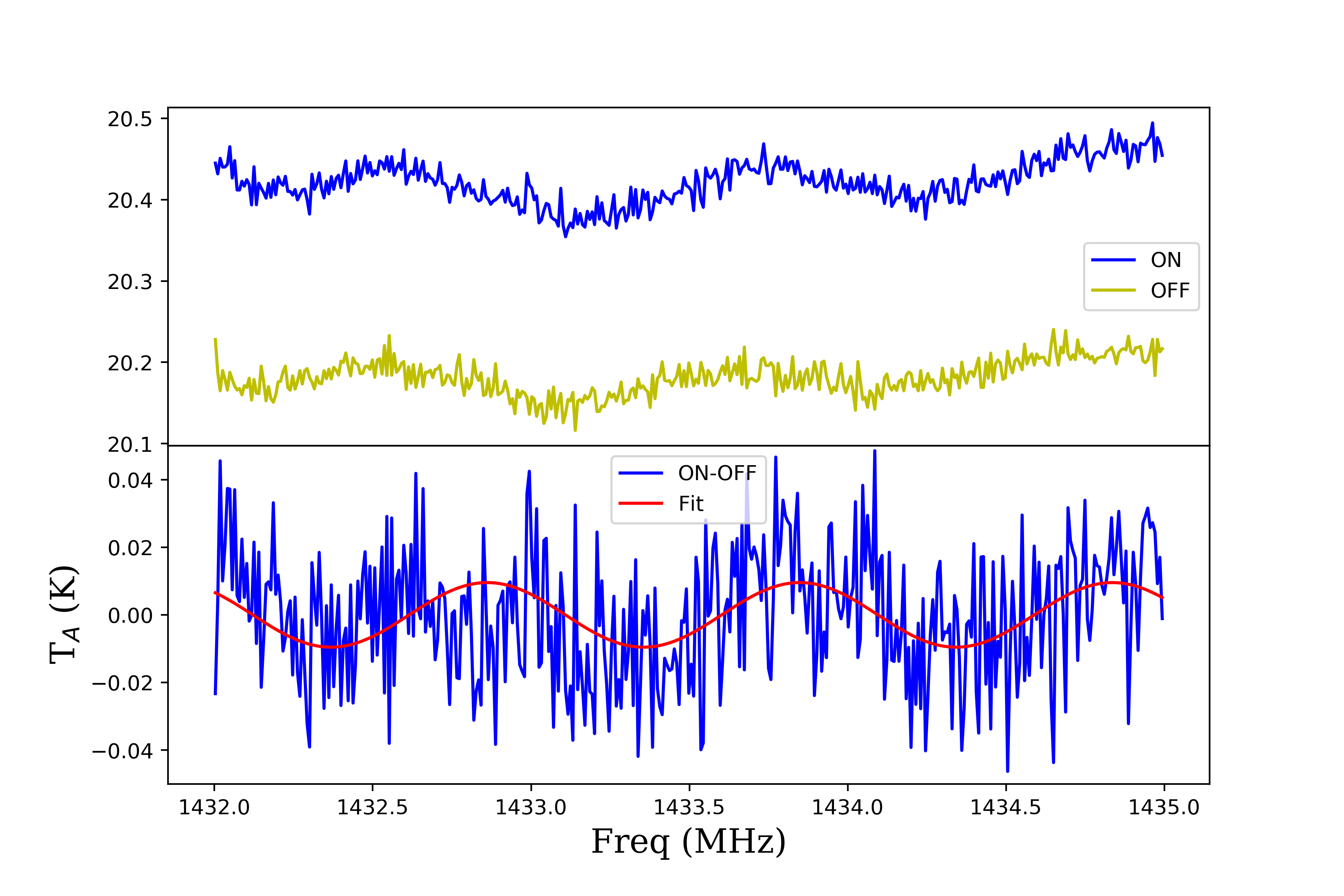}
\caption{Bandpass spectrum at different frequency ranges during  ON-OFF observations towards NGC 2718. The amplitudes of standing wave in frequency range of [1150,1158],[1254,1257], [1367,1370] and [1432,1435] MHz are  15, 11, 22, and 9.5 mK, respectively. }
\label{fig:NGC2718-onoff}
\end{figure}

Another testing observations were obtained toward NGC2718 under ON-OFF mode. As shown in Figure \ref{fig:NGC2718-onoff}, the amplitude of standing wave varies as frequency.

\subsection{Polarization}
\label{subsec:polariztion}

\begin{figure}[htp]
\centering
  \includegraphics[width=0.9\textwidth]{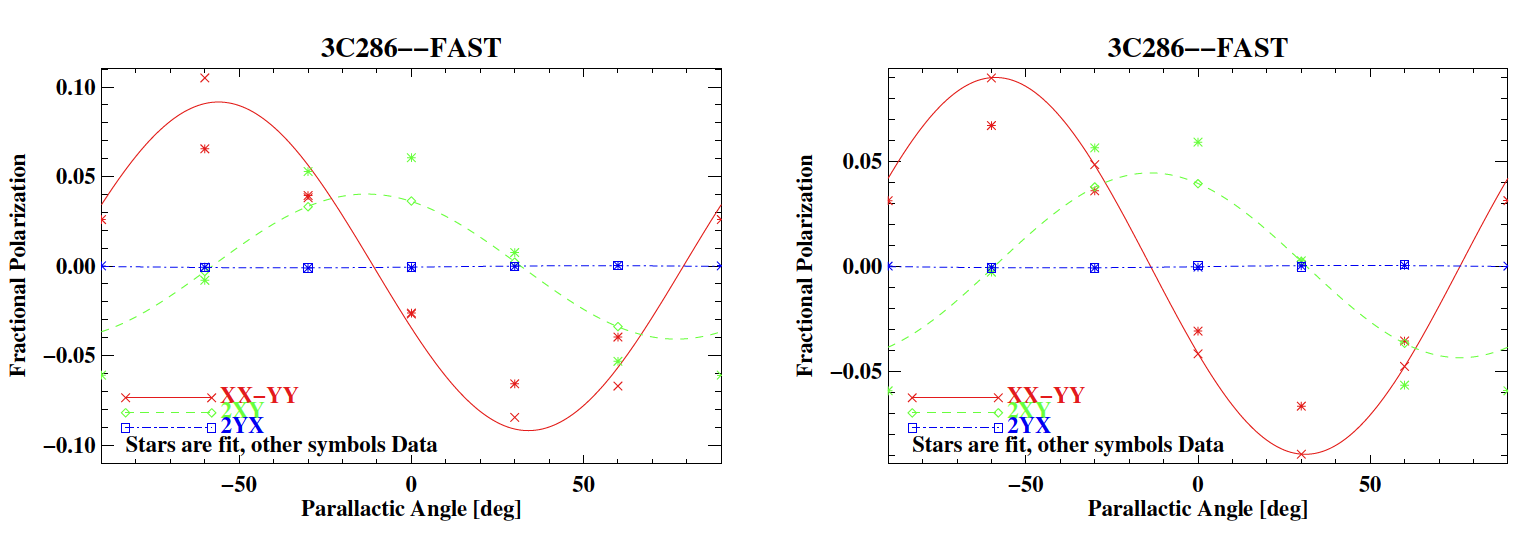}
  \caption{The measurements and the fits of the Stokes Q, U, and V parameters of 3C286 at 5 parallactic angles of the 500 MHz band (left) and the 32 MHz band (right). The red, green, and blue data sets represent the Stokes Q (XX-YY), U (2XY), and V (2YX) signals normalized with respect to the Stokes I signals.}
\label{fig:pol}
\end{figure}

	We carried out FAST polarization observations of 3C286 to measure the instrumental polarization of the telescope in Oct. 2018. 3C286 is a standard polarization calibrator with stable polarization degrees and polarization angles from 1 to 50 GHz (Perley \& Butler 2013). 3C286 was drifted at parallactic angles of -60, -30, 0, 30, and 60 degrees through the central beam of the 19-beam receiver. The strength of the noise diode was set to 1K with on-off period of 0.2 sec. The four correlations of the XX, YY, XY, and YX signals were simultaneous recorded with the ROACH backends in both the spectral line modes of 500 MHz bandwidth and 32 MHz bandwidth. The data reduction including the gain and phase calibration of the system, the calibration of the four correlated spectra, and the derivation of the instrumental polarization using the data at 5 parallactic angles was carried out with the RHSTK package (Heiles et al. 2012). 
	Fig. \ref{fig:pol} shows the measurements of the Stokes Q, U, and V parameters of 3C286 at 5 parallactic angles. By fitting the sinusoidal behaviors of the Stokes parameters as functions of parallactic angles, the polarization degree and polarization angle of 3C286 are 6.6 +/- 1.5 \% and 33.4 +/- 6.4 degrees, respectively. Our calibrated polarization results of 3C286 are close to the values of 9.47 +/- 0.02 \% and 33 +/- 1 degrees in Perley \& Butler (2013). The instrumental polarization parameters of FAST obtained from the data were close to being unitary, indicating a good isolation between the two signal paths of linear polarization feeds and a good performance of the polarization facility of FAST. We expect that with better characterization of the polarization properties including the pointing accuracy, beam width, beam squint, and beam squash of FAST, the polarization data will be calibrated to an accuracy of 0.1-0.01 \%, in order to carry out scientific spectral line polarization observations in the near future.

\section{Observation }
\label{sec:observation}

\subsection{Observation Modes}
\label{subsec:obs_mode}

Now there are 8 observation modes in the FAST.  The details are shown in the follows.

\begin{figure}
\centering
  \includegraphics[width=0.8\textwidth]{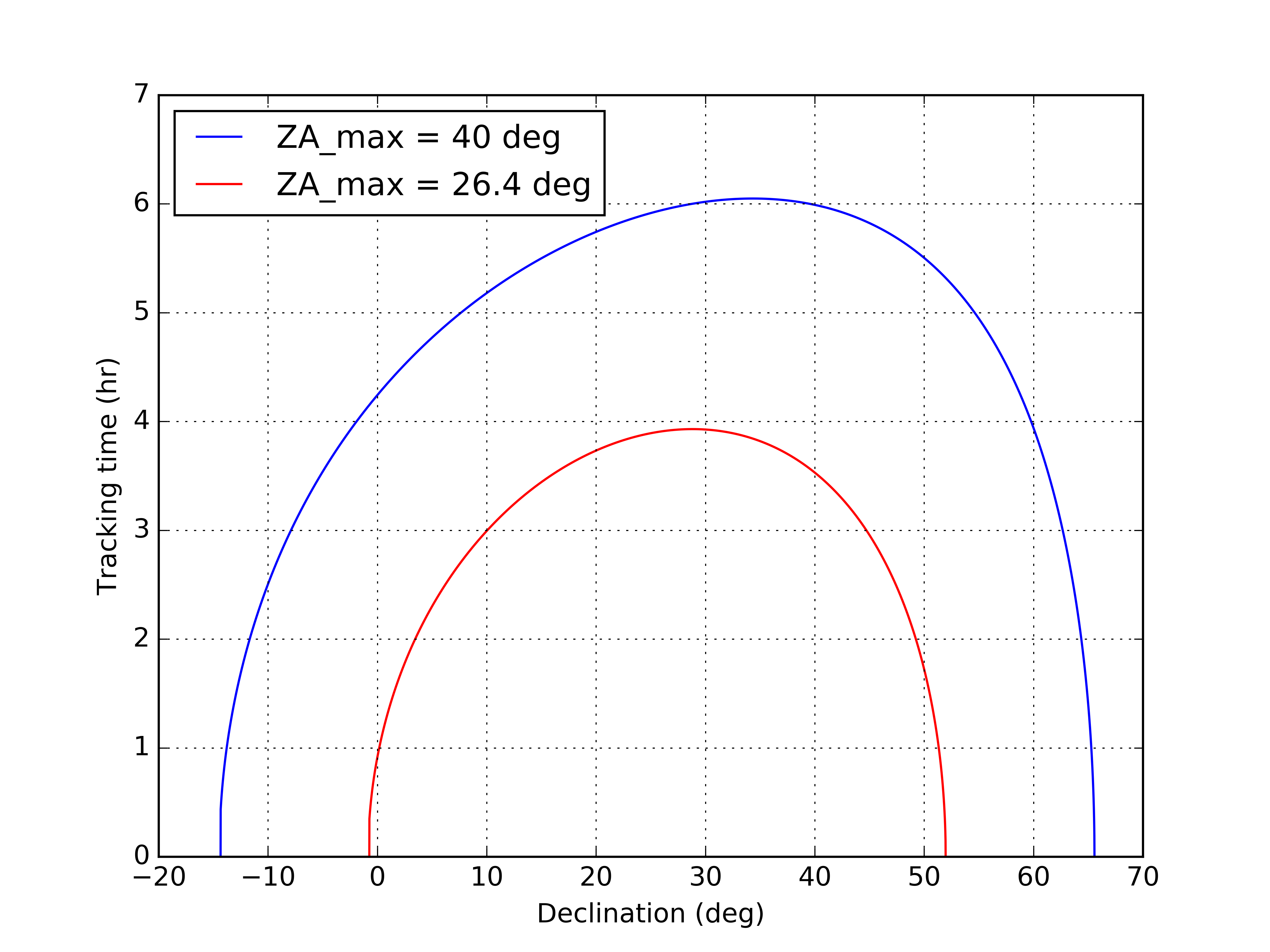}
  \caption{Maximum tracking time for source at specific declination. Results for zenith angle of 40$^{\circ}$ and 26.4$^{\circ}$ are represented with blue and red color, respectively.}
\label{fig:trackingtime}
\end{figure}

\begin{figure}
\centering
  \includegraphics[width=0.8\textwidth]{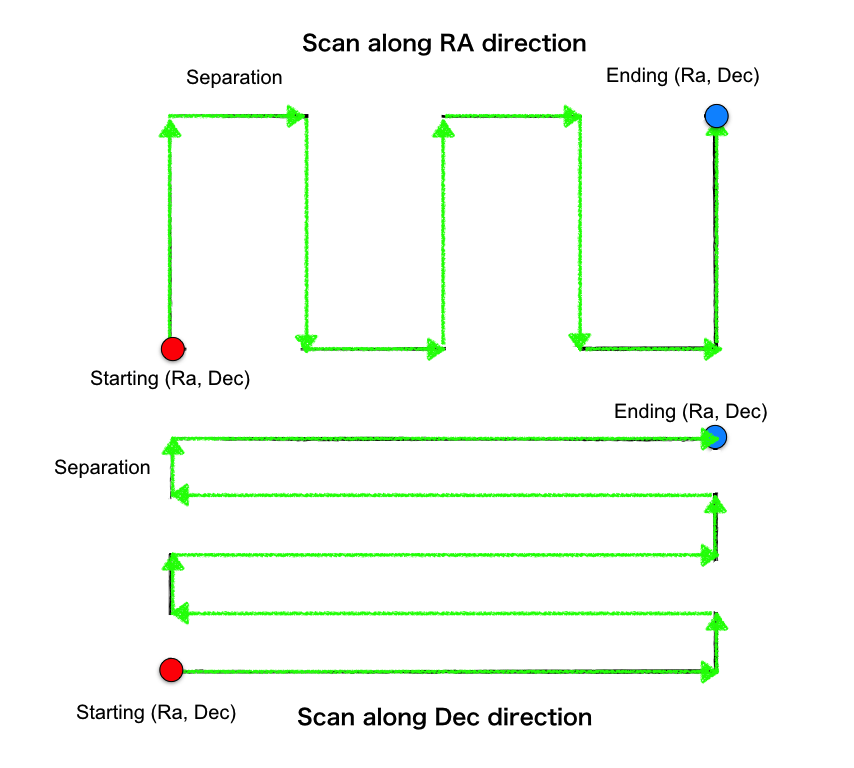}
  \caption{Schematic diagram of OTF mode. The blue line represents scanning trajectory.  The top and bottom figure shows scan information along RA and Dec direction, respectively. }
\label{fig:otf}
\end{figure}

\begin{figure}
\centering
  \includegraphics[width=0.8\textwidth]{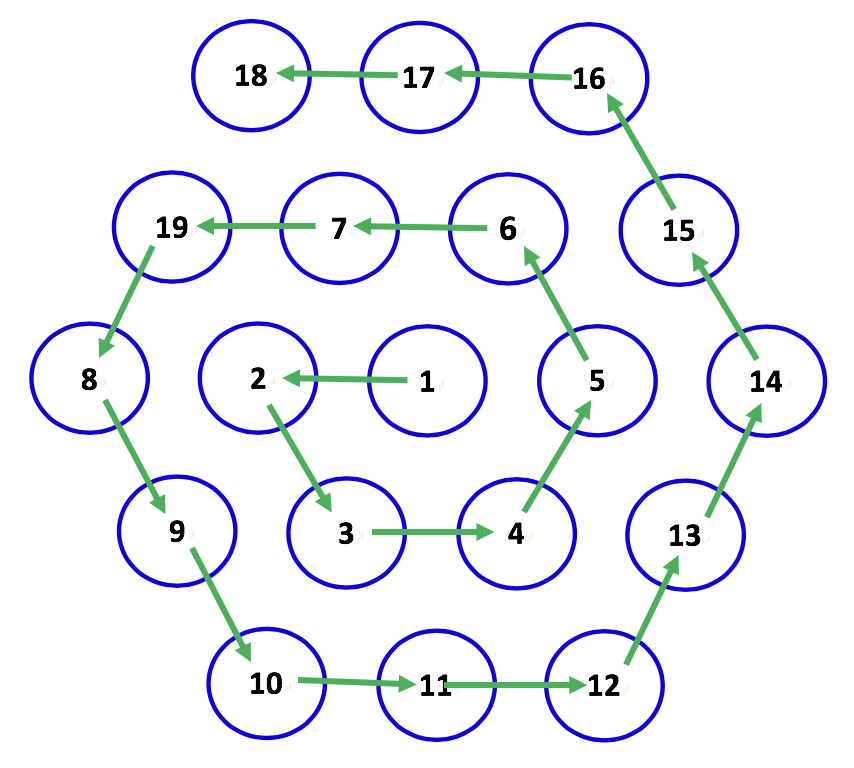}
  \caption{ Schematic diagram of MutiBeamCalibration mode. Green arrow line represents the beam sequence for observing the calibrator.  In this plot, the  }
\label{fig:multibeamcal}
\end{figure}

\begin{figure}
\centering
  \includegraphics[width=0.8\textwidth]{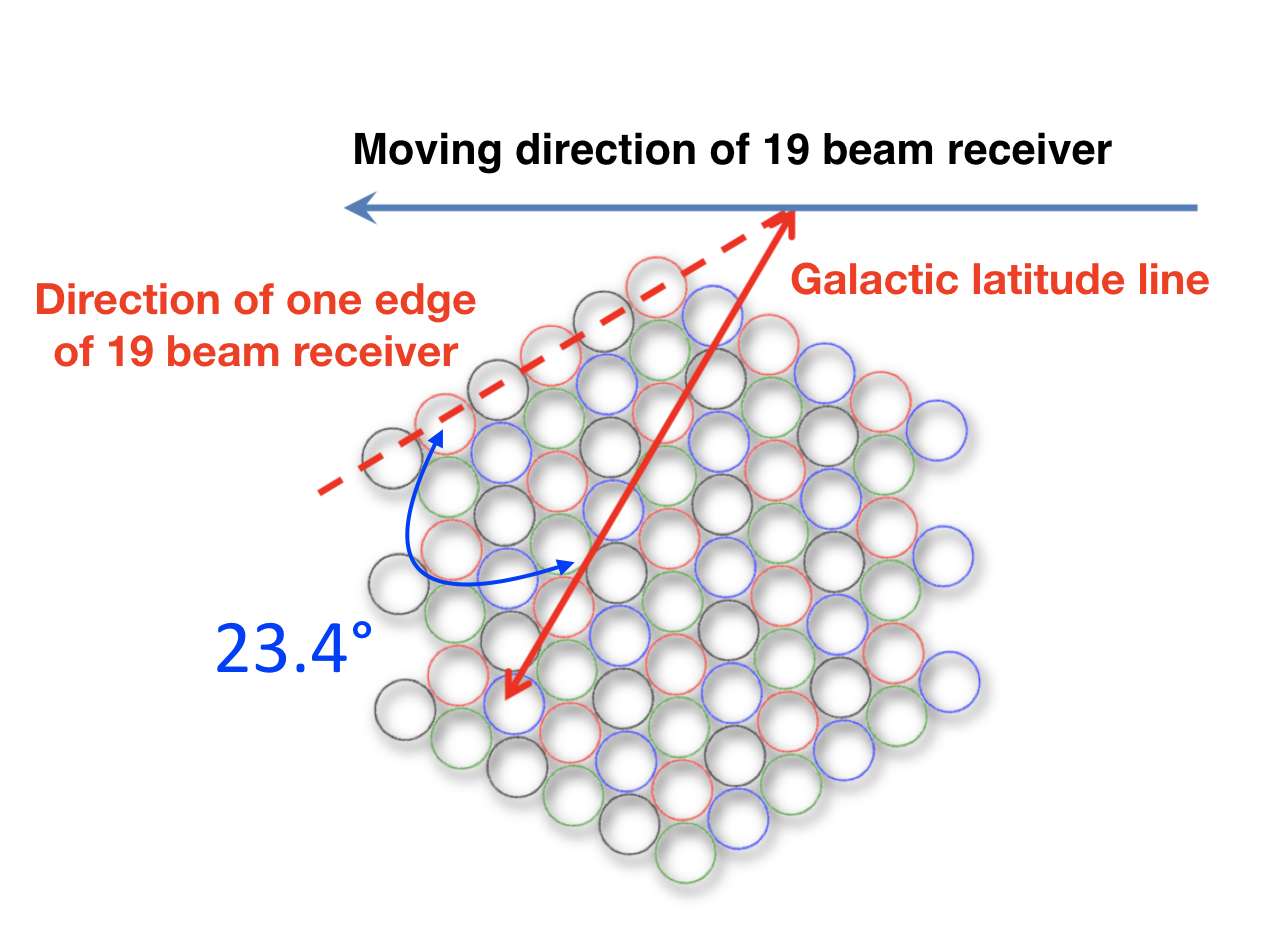}
  \caption{Schematic diagram of snapshot mode. Blue line represents moving direction of 19 beam receiver. Dashed red line represents the direction of one edge of 19 beam receiver. Galactic latitude line is represented with solid red line.  The angle between dashed and solid red line is 23.4$^{\circ}$. }
\label{fig:snapshot}
\end{figure}

\begin{enumerate}
\item  Drifting scan. In this mode, the position of feed cabin is fixed. To avoid pointing offset by the cooling of oil in the actuator, the actuators  used for shaping the paraboloid surface  are adjusted continually to keep same paraboloid in this mode. The telescope points across the sky as the earth rotates. There are four parameters in this mode, source name, source coordinate in J2000 system, observational time range, and rotating angle that represents cross angle between the line along beam 8, 2, 1, 5, and 14 and line of declination. The rotating angle has a limit of [-80,80] degree. 

\item  Total power.  In this mode, the source can be tracked continuously.   Tracking time for source at different latitude is shown in Fig. \ref{fig:trackingtime}. Three parameters including source name, coordinate and integration time are needed.  Derived rms of averaged spectrum of both polarization is estimated with, 

\begin{equation}
\sigma  = \frac{T_{sys}}{\sqrt{2\beta \tau}} \rm\  K,
\end{equation}

where  $\beta$=1.2B$\rm_{chan}$. B$\rm_{chan}$ is channel width of spectrometer in unit of Hz. 

\item  Position switch.  The design of position mode is to achieve quick switch between source ON and source OFF in order to reduce baseline variation.  There are five parameters in this mode. 

\begin{itemize}
\item  Coordinate of ON position. 
\item  Coordinate of OFF position.   The position of OFF source is designed to be within 1 degree from that of ON source.   
\item  Integration time of  ON source.
\item  Integration time of OFF source. 
\item  Times of ON-OFF cycle.
\end{itemize}

Overhead time between ON and OFF position depends on separation of ON and OFF position, $\Delta\theta$. It is 30 s for $\Delta\theta$ < 20' and is 60 s for $20'$ $\leq \Delta\theta$ < $60'$.  

Derived rms of each spectrum with both polarizations is estimated with, 
\begin{equation}
\sigma  = \frac{T_{sys}}{\sqrt{\beta \tau}} \rm \  K.
\end{equation}

where $\beta$=1.2B$\rm_{chan}$. B$\rm_{chan}$ is channel width of spectrometer in unit of Hz.

\item  On-the-fly (OTF) mapping.  This mode is designed for mapping a sky area with Beam 1 only.  Six parameters are necessary for observation: source name, source position, observational time range, sky coverage (e.g., 7$'\times$7$'$ of the mapping region), scanning separation (e.g., 1$'$) between two parallel scanning lines, and scanning direction (along RA or Dec). Scanning speed is 15 arcsec/s in default. The schematic diagram of this mode is shown in Fig. \ref{fig:otf}. 

\item MultiBeamOTF mapping.  This mode is proposed to map the sky with 19 beams simultaneously.  Compared to the OTF mapping, the MultiBeamOTF mapping mode has similar scanning trajectory but a larger separation (e.g., 20 arcmin) between parallel scans.  Besides, the parameter of rotation angle is available  in this mode. 

\item MutiBeamCalibration.  In this mode, 19 beams will be switched in sequence  to track the calibrator, allowing for quick calibration of  the gain of 19 beams in 30 minutes. Switching time between two beams is 40 s. The integration time of each beam is a parameter for setting.  The schematic diagram of this mode is shown in Fig. \ref{fig:multibeamcal}. 

\item BasketWeaving.   This mode is to scan the sky along a meridian line.  Scanning speed ranges from 5 to 30 arcsec/s. Setting parameters include starting time, starting declination, ending declination,  duration time. 

\item  Snapshot. This mode is used to fully map the sky in grid. This type of mapping is not a Nyquist sampling, but could map a region with relatively deep  integration time.  This is beneficial especially for pulsar searching. As shown in Fig. \ref{fig:snapshot},  the movement of 19 beam receiver would ensure fully cover the sky along same Galactic latitude. Necessary parameters include source name, beginning and ending coordinate (RA and Dec), observational time range, and scanning speed (less than 30$'$/s). 

\end{enumerate}

The above modes are available for observation at FAST now.  

\subsection{Effect of Radio Frequency Inference}
\label{subsec:RFI}

The current observation hints that RFI in FAST data is mainly divided into three types: narrow-band RFI, 1 MHz wide RFI and some wider fixed frequency RFI. Every type is discussed in more detail below.

\begin{enumerate}
	\item The narrow-band RFI has ever been ubiquitous through the FAST data. We speculate that there are many origins, like the interference from instruments or the local influence of the telescope. For the FAST spectral data with  a frequency resolution of 0.48kHz (divided into one million channels in 500\,MHz), they are extremely narrow and mostly only appear on one or several channels. Some of the narrow-band RFI could have high strength in a short  integration time, but some of the others are just like a faint bulge without the Gaussian profile. 
	In some cases, the narrow-band RFI tend to exhibit the periodic variations in the time domain, which might be found in the pulsar data if its time resolution is less than one second. The narrow-band RFI used to bring a lot of trouble for the FAST data processing, but it has been solved now.

	\item While the 1 MHz wide RFI is caused by the standing wave, which looks like regular sinusoidal wave or the single bump in FAST spectra. The big bump originates from the superpose of some standing waves with different amplitudes and periods, sometimes approaching to Gaussian profile. The typical width of this type of RFI is 1 or a few MHz. The existing L-band observation shows that their distribution in time domain and frequency domain is not regular.
	
	\item The fixed frequency RFI is due to satellite or civil aviation from the sky. It usually have a fixed frequency in a wider distribution, and have the strongest intensity in the whole band-pass, even improve the baseline there. Its spectral profile has a complex multi-peak structure, and the width of one peak might be around 20 MHz. The profile and intensity of it varies slightly in different observation time. In addition, the existence of such a wide enough RFI may also be due to the higher sensitivity of FATS. It receives some very strong signal, which results in the frequency width of the response signal exceed the range of satellite signal on both side.
	
\end{enumerate}

\begin{figure}[!ht]
	\centering
	\includegraphics[width=0.9\textwidth, angle=0]{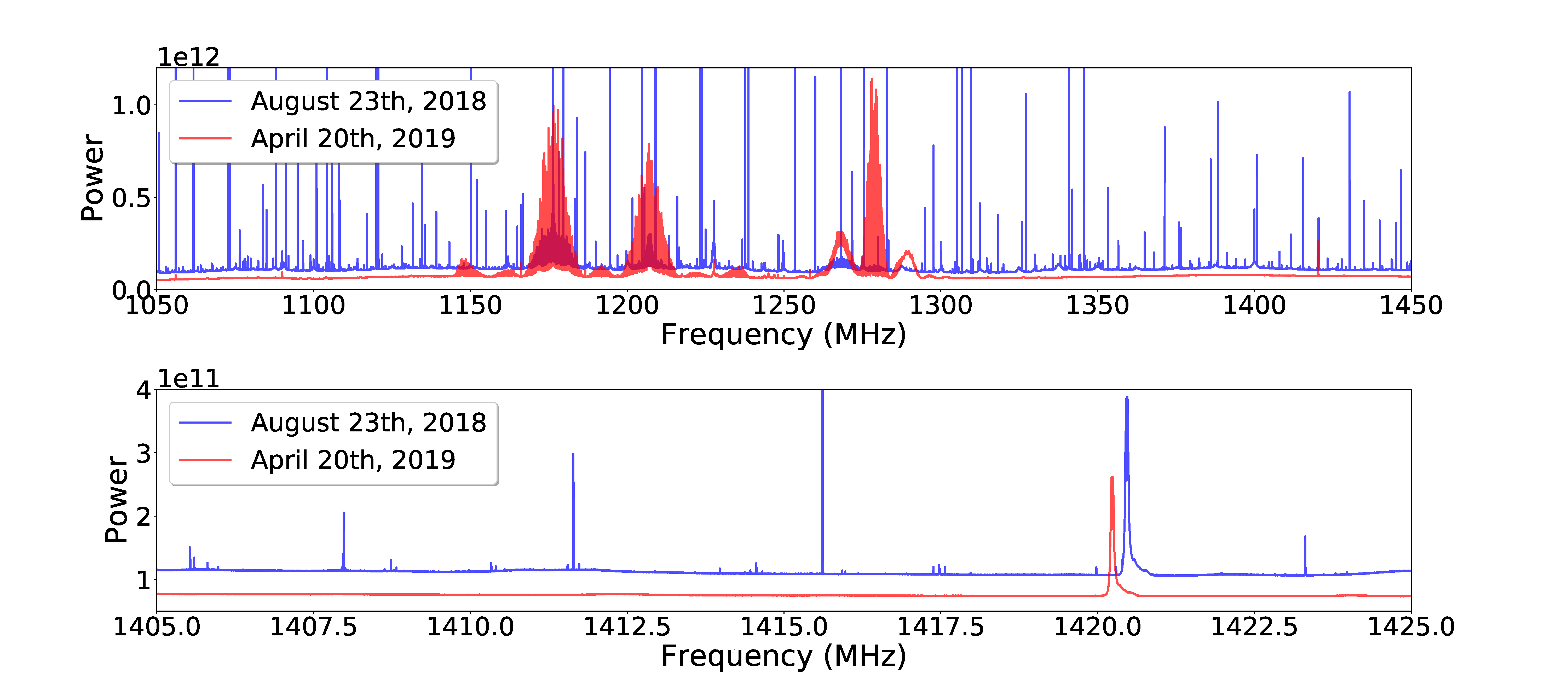}
	\caption{Comparison of the spectral band-pass in 1000-1500\,MHz (upper panel plot) and for \hi spectrum in the frequency of 1420\, MHz (lower panel plot) toward TMC-1 in different days, August 26th, 2018 (marked as blue line) and April 20th, 2019 (marked as red line).}
	\label{rfi1}
\end{figure}

Since the successful installation of the electromagnetic shielding in April 2019, narrow-band RFI has been much reduced. Fig.~\ref{rfi1} shows the comparison between the spectral results  before and after the installation of the electromagnetic shielding. They are observed on August 23, 2018, and April 20, 2019 in 10 minutes integration each day toward TMC-1, without periodic noise injection. These two days' results exhibit a significant decrease in RFI, suggesting the excellent result of the electromagnetic shielding. For some narrow line-width sources, such as Taurus, their molecular line width is within several or dozens of frequency channels and their RFI problem are very confusing previously. It's really satisfactory that the tremendous reduction provides great convenience for the spectral line observation and identification. However, the present data still exist some 1 MHz wide RFI and fixed frequency RFI.

\begin{figure}[!ht]
	\centering
	\includegraphics[width=0.9\textwidth, angle=0]{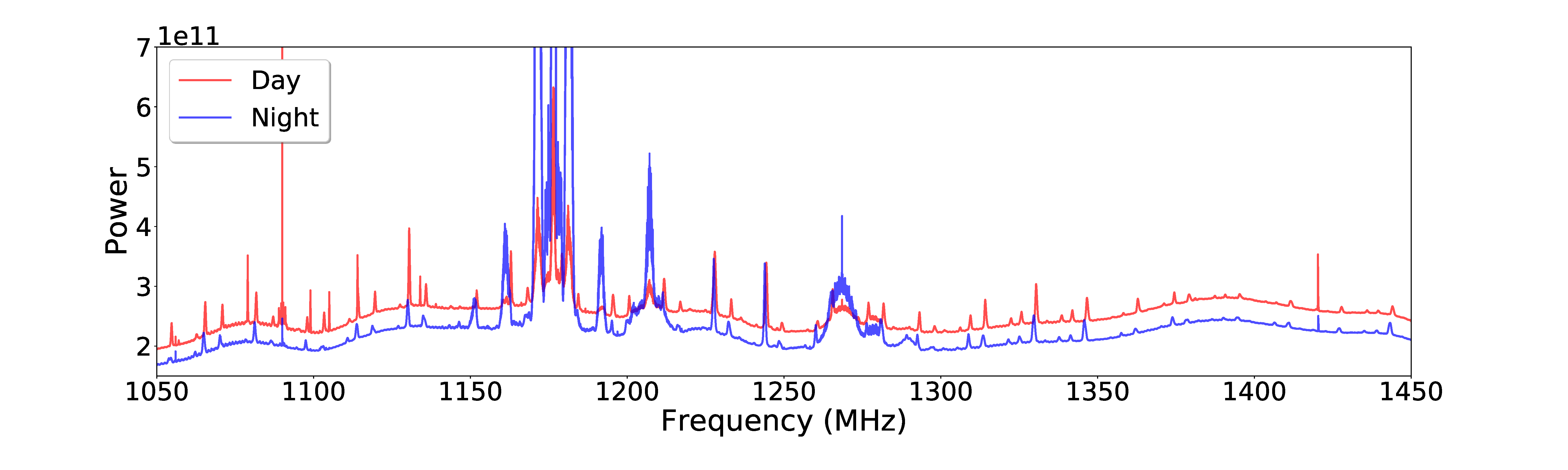}
	\caption{Comparison of the band-pass in the daytime and nighttime with each 10 minutes integration. The nighttime spectrum marked as blue line and the daytime marked as red line.}
    \label{rfi2}
\end{figure}

\begin{figure}[!ht]
	\centering
	\includegraphics[width=0.64\textwidth, angle=0]{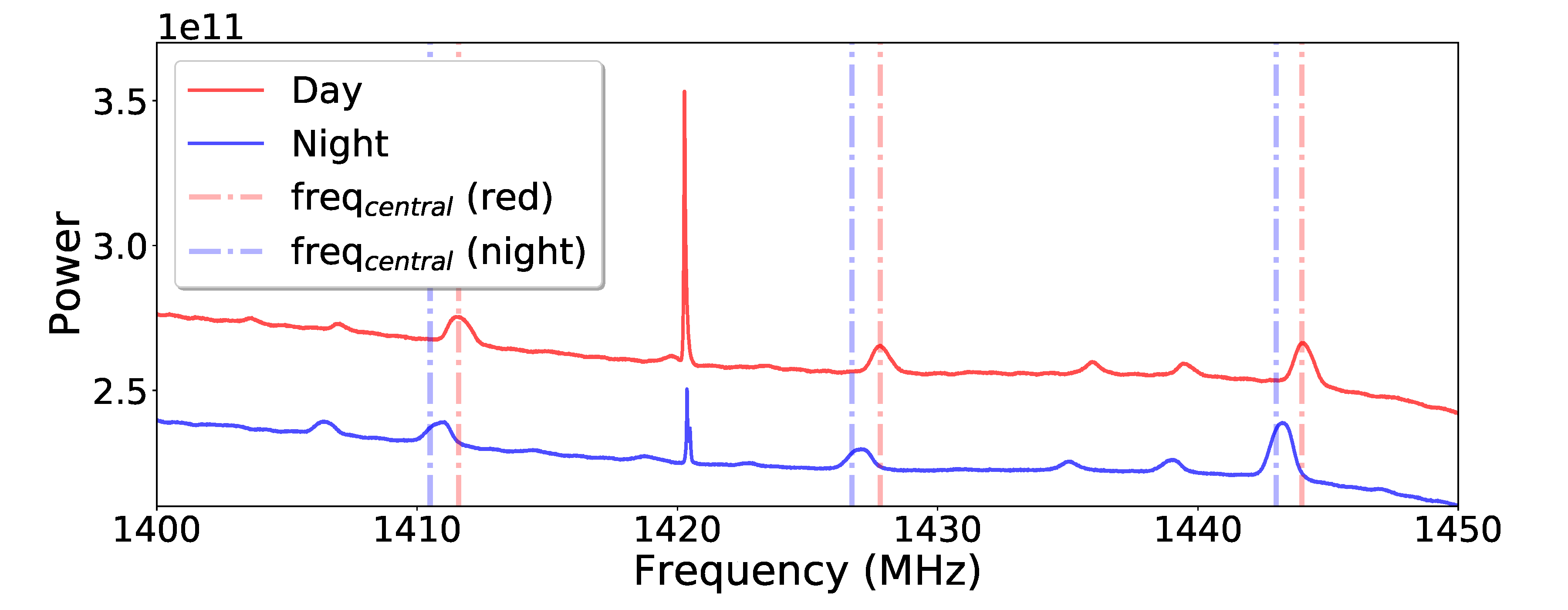}
	\caption{Zooming in the X and Y axes in the Fig.~\ref{rfi2}, the systemic shift of this wide RFI is plotted. Red lines represent daytime and blue lines represent nighttime. The vertical dash dot line shows the center frequency of a single RFI signal, making it easy to see the amplitude of the frequency shift from day to night.}
    \label{rfi3}
\end{figure}

In order to study the influence of human activities on FAST data and RFI signal, we observed 10 minutes at daytime and nighttime in one day, supposing that  human activity changes with day and night. The source of daytime observations is a quasar, observed at 15:00 on June 23, 2019. The night source is a star observed at 20:00 on the same day. Considering the beam dilution toward the point source and the extremely short integration time, we can assume that no signal from the day and night sources could be received except for \hi emission from the interstellar medium. Therefore, ignoring \hi, almost all of the emission in the obtained spectra should be RFI. Fig.~\ref{rfi2} shows the comparison of the daytime RFI with nighttime RFI in L band. We can hardly see the narrow-band RFI emission in this figure. While the 1 MHz wide RFI is always existed throughout the band-pass. The total number of its signal is almost invariant, no increase or reduce. But their central frequency slightly shifts from day to night. We magnified the axises of the Fig.~\ref{rfi2} to check the clearer movement, showing in the Fig.~\ref{rfi3}. At the same time, the wider RFI from satellites and civil aviation is always located at the same frequency, just with a little variation in intensity.

Hoping to study the small frequency shift of the 1 MHz wide RFI as a result of the standing wave, we have drawn this exact movement in the Fig.~\ref{rfi3}. The center frequency of RFI is marked as dash dot line. It reveals that the systematic ferquency shift of every RFI emission are all approximate 1 MHz in the figure, form daytime to nighttime. However, the 1 MHz shift doesn't apply to other FAST observation. We've checked the spectral data toward other sources. For the same source observed on different days, this type of RFI moves to different directions in frequency domain, but within a few MHz. While, this shift is uncorrelated with source selection, but only changes in time. The current results implies that the direction to higher or lower frequency and the amplitude  of such systematic shift in total band-pass is irregular. It's in line with our speculation: This type of RFI which looks like a single bump is caused by multiple standing waves superposition. And such a superposition would make the spectra ever-changing. We've used this sinusoidal superposition to fit the baseline and the RFI could be removed well. In the future, we will try to do more to solve the problem of standing wave and RFI.

\section{Summary}
\label{sec:summary}

The FAST has achieved its designed objectives. In this paper, we have presented current status of FAST performances. They are summarized as follows.

\begin{enumerate}
    \item The median power output of low and high noise diode are around 1.1 and 12.5 K. The measured temperature fluctuation of noise diode is $\sim$ 1\%, leading to $\sim$ 2\% accuracy in flux calibration when pointing accuracy and beam size are included. 
    
    \item Spatial distribution and power pattern of 19 beams within zenith angle of 26.4$^{\circ}$ were obtained by  mapping observations toward 3C454.3. Beam width of 19 beams is consistent with theoretical estimation. 

    \item The  pointing errors of 19-beam receiver in different sky position are less than 16$''$. The standard deviation of pointing errors is 7.9$''$.
     
    \item The aperture efficiency as a function of zenith angle could be fitted with two linear lines at specific frequency. It keeps almost constant with zenith angle of 26.4$^{\circ}$ and decreases by $\sim$ 1/3 at zenith angle of 40$^{\circ}$.
    
    \item The system temperature as a function of zenith angle could be fitted with a modified arctan function.  This fitting is valid for zenith angle within 40$^{\circ}$.
    
    \item The fluctuation of baseline is about $4$\% in 30 minutes. Rms of the baseline satisfies the expected sensitivity.
    
    \item Standing wave has an amplitude of $\sim$ 0.3\% compared to  continuum level. The amplitude, phase of standing wave would vary at different zenith angle during drifting. For position switch observations, standing wave of the residual ON-OFF spectrum would be suppressed with amplitude of $\sim$ 0.02 K, which varies in different frequencies. 
    
    \item  Derived polarization degree and polarization angle toward 3C286 with FAST are $6.6\pm 1.5$\% and 33.4$\pm$6.4 degrees, respectively. They are consistent with results from previous study.
    
    \item Eight observation modes including drifting scan, total power, position switch, On-the-fly, MultiBeamOTF, MultiBeamCalibrtation, BasketWeaving and snapshot are available for FAST observations now. 
    
    \item RFI environment has been greatly improved in the last 18 months. Narrow RFI with several spectral channels are reduced. More effort will be done for eliminating broad RFI with width of $\sim$ 1 MHz. 
    
\end{enumerate}

\vspace{12pt}

\normalem
\begin{acknowledgements}

We thank the beneficial discussion with Mao Yuan.  This work is supported by the National Key R \& D Program of China (NO. 2017YFA0402701), the National Natural Science Foundation (NNSF) of China  (No. 11803051,  No. 11833009). TNY was supported by the CAS "Light of West China" program.  LGH is additionally supported by the Youth Innovation Promotion Association CAS. 

\end{acknowledgements}

\appendix
\section{Electronics Gain fluctuations }

\begin{figure}[!ht]
\centering
\includegraphics[width=0.5\textwidth]{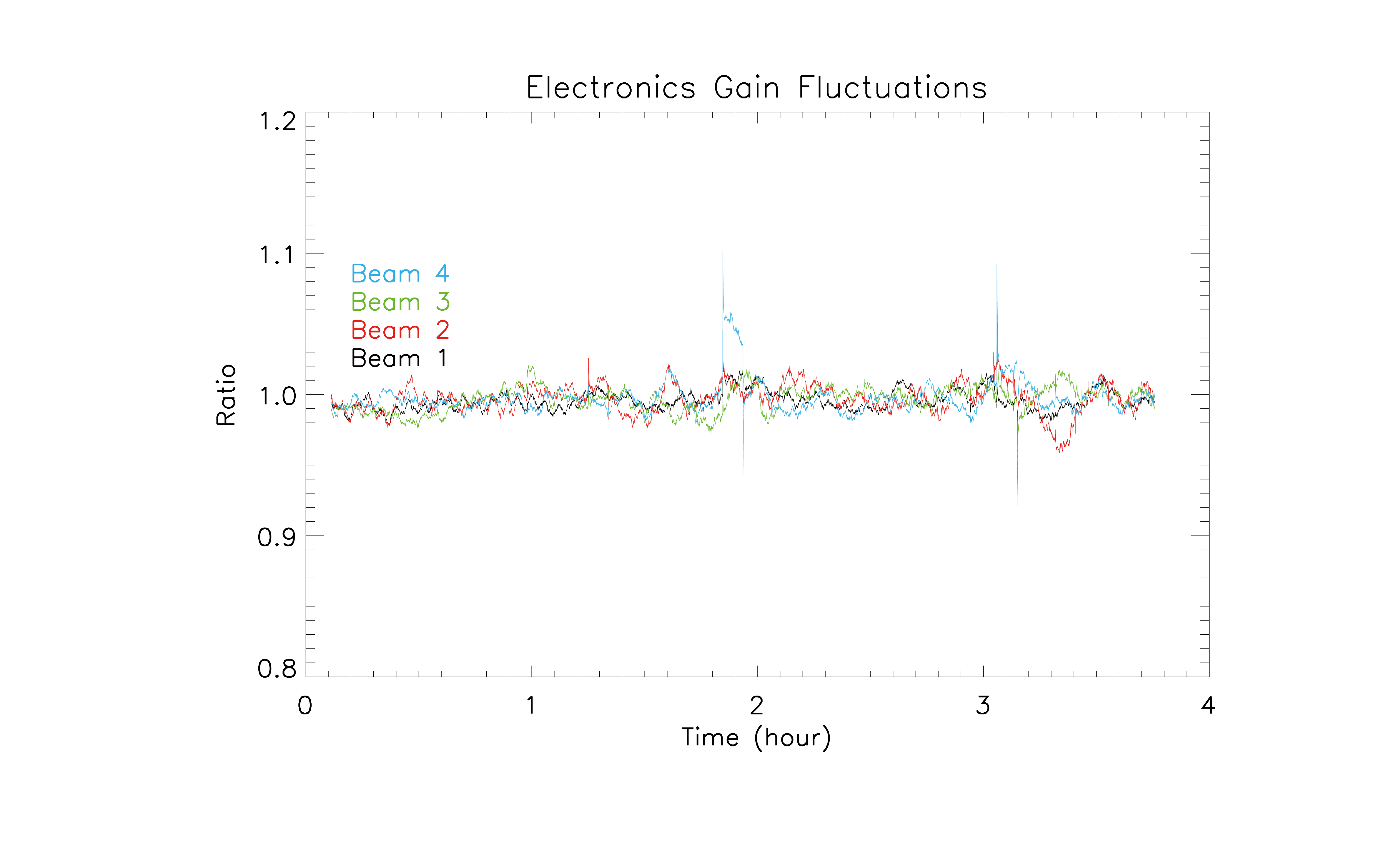}
\includegraphics[width=0.5\textwidth]{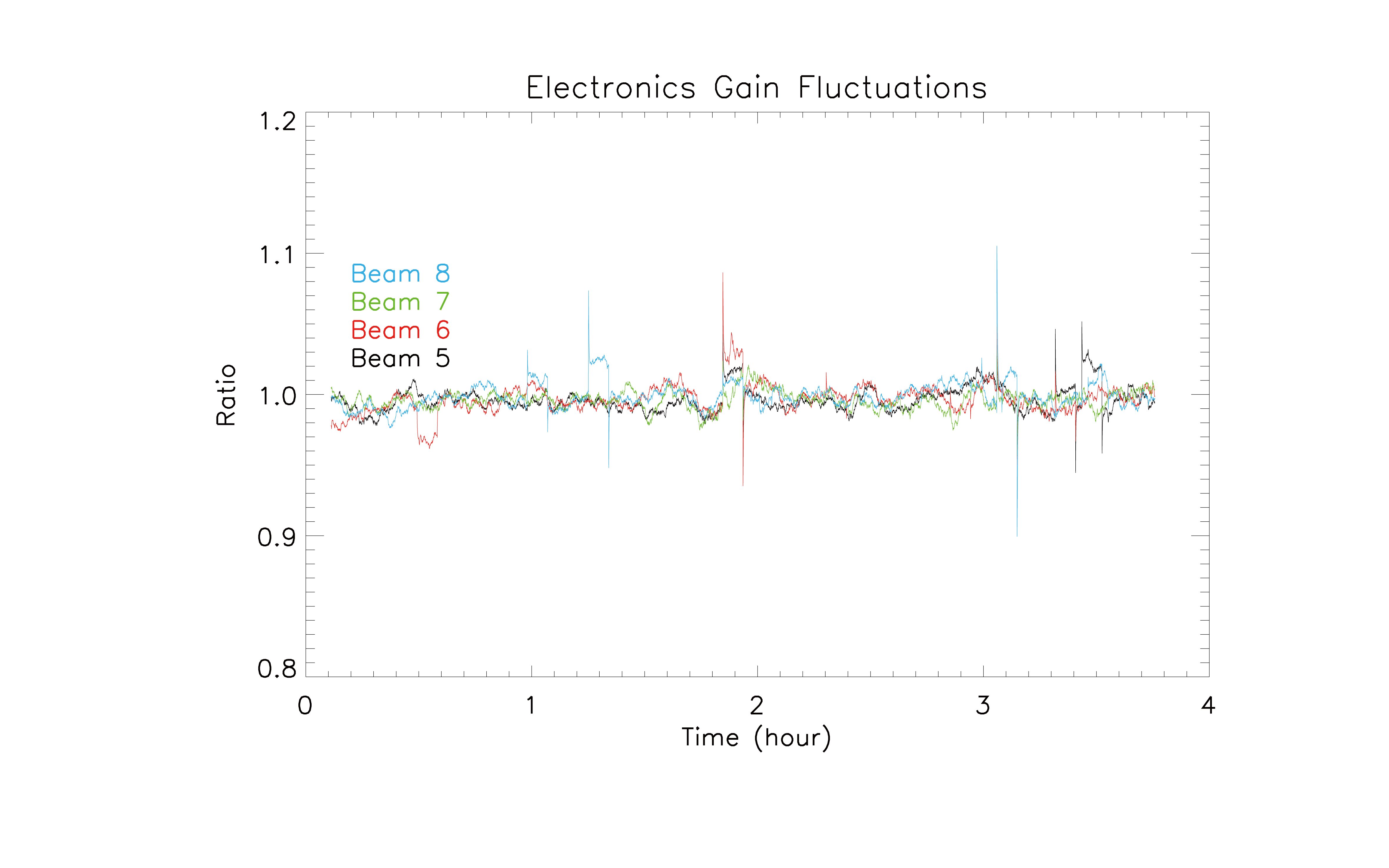}
\includegraphics[width=0.5\textwidth]{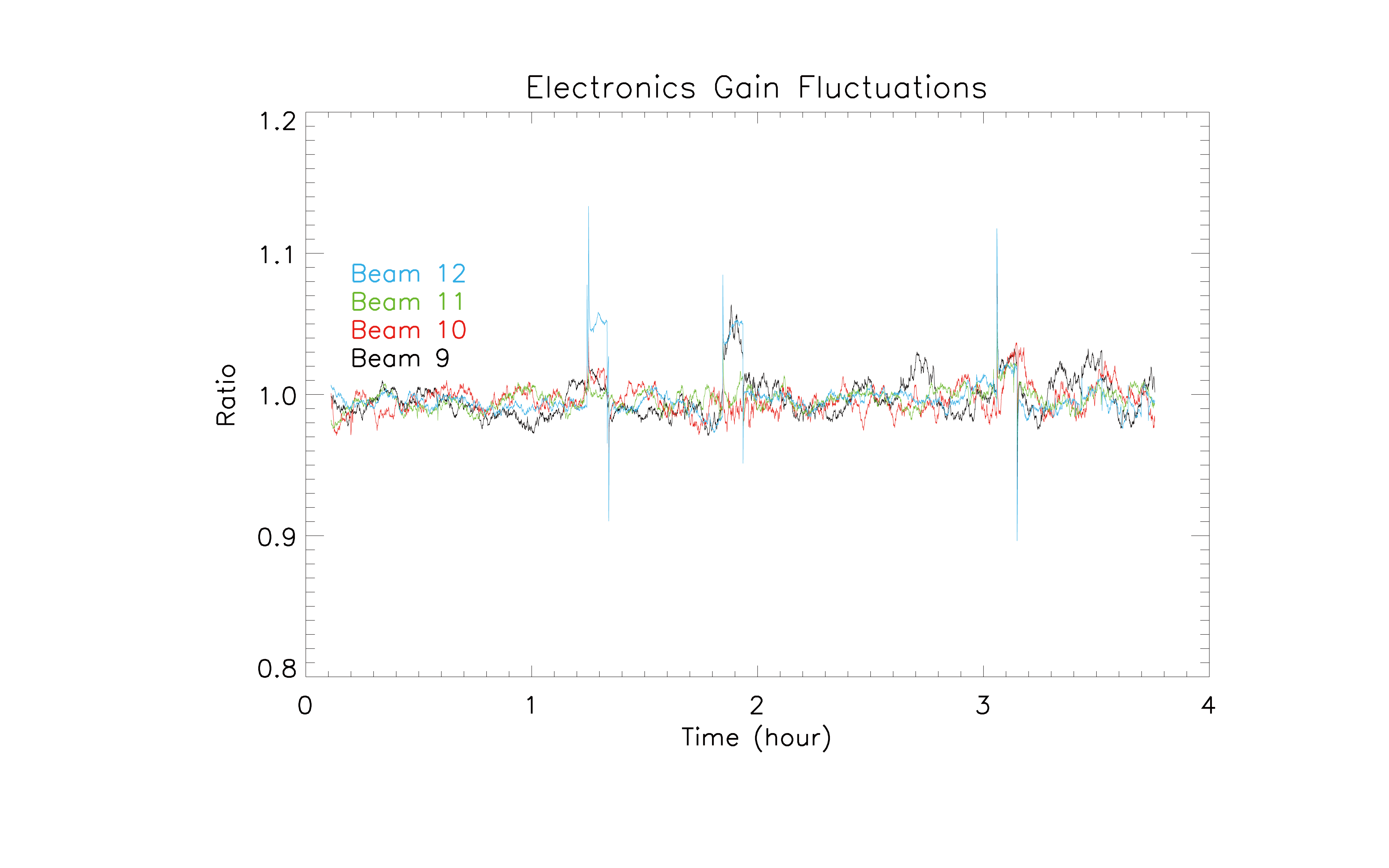}
\includegraphics[width=0.5\textwidth]{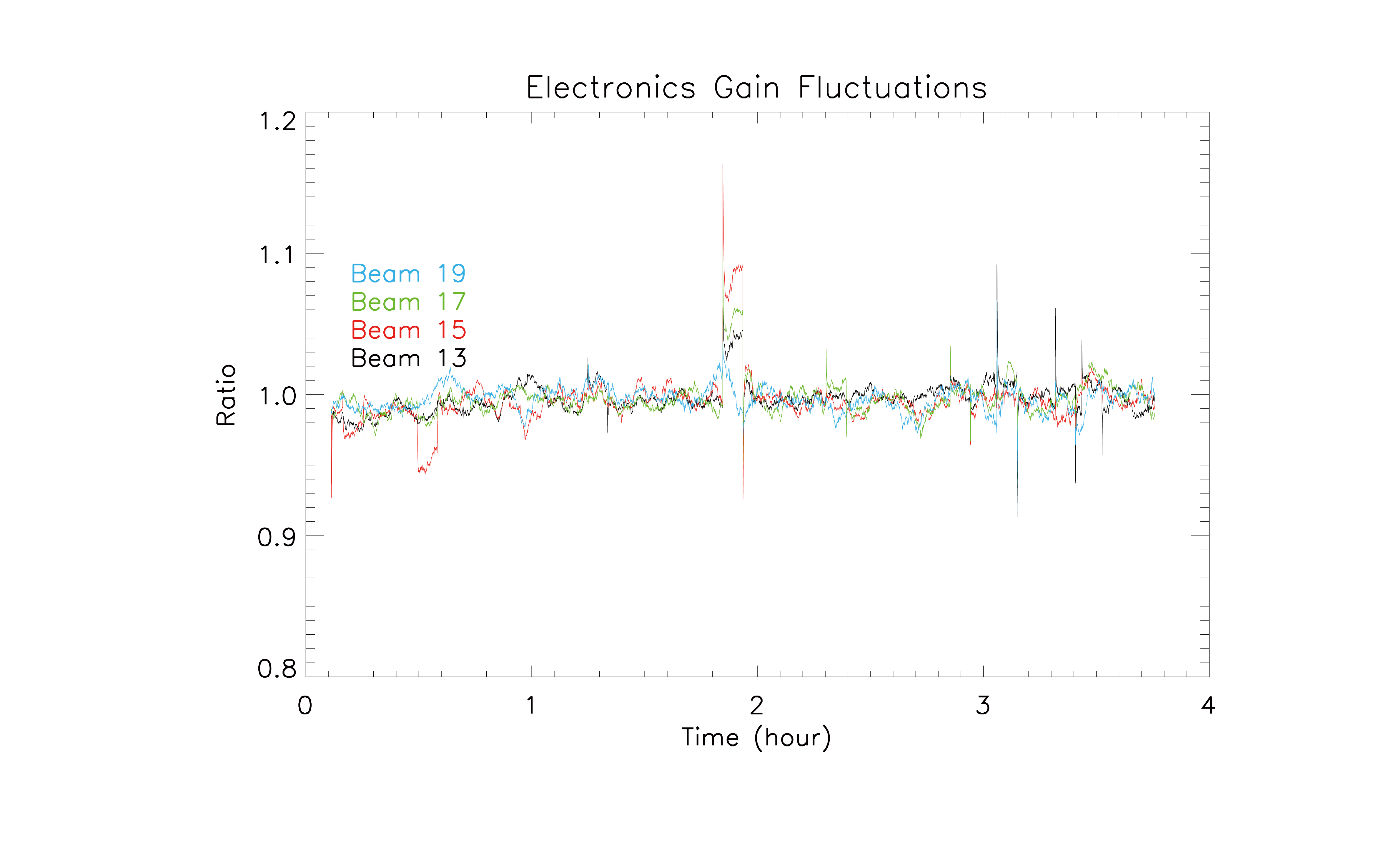}
\caption{The electronics gain fluctuations of the system over several hours. 
The vertical shows the ratio of fluctuations with respect to the mean temperature over time of Beam 1, 2, 3, 4, 5, 6, 7, 8, 9, 10, 11, 12, 13, 15, 17 and 19. 
}
\label{fig:electrongain_restbeams}
\end{figure}

\section{Aperture Efficiency}


\begin{figure}[!ht]
\centering
  \includegraphics[width=0.45\textwidth]{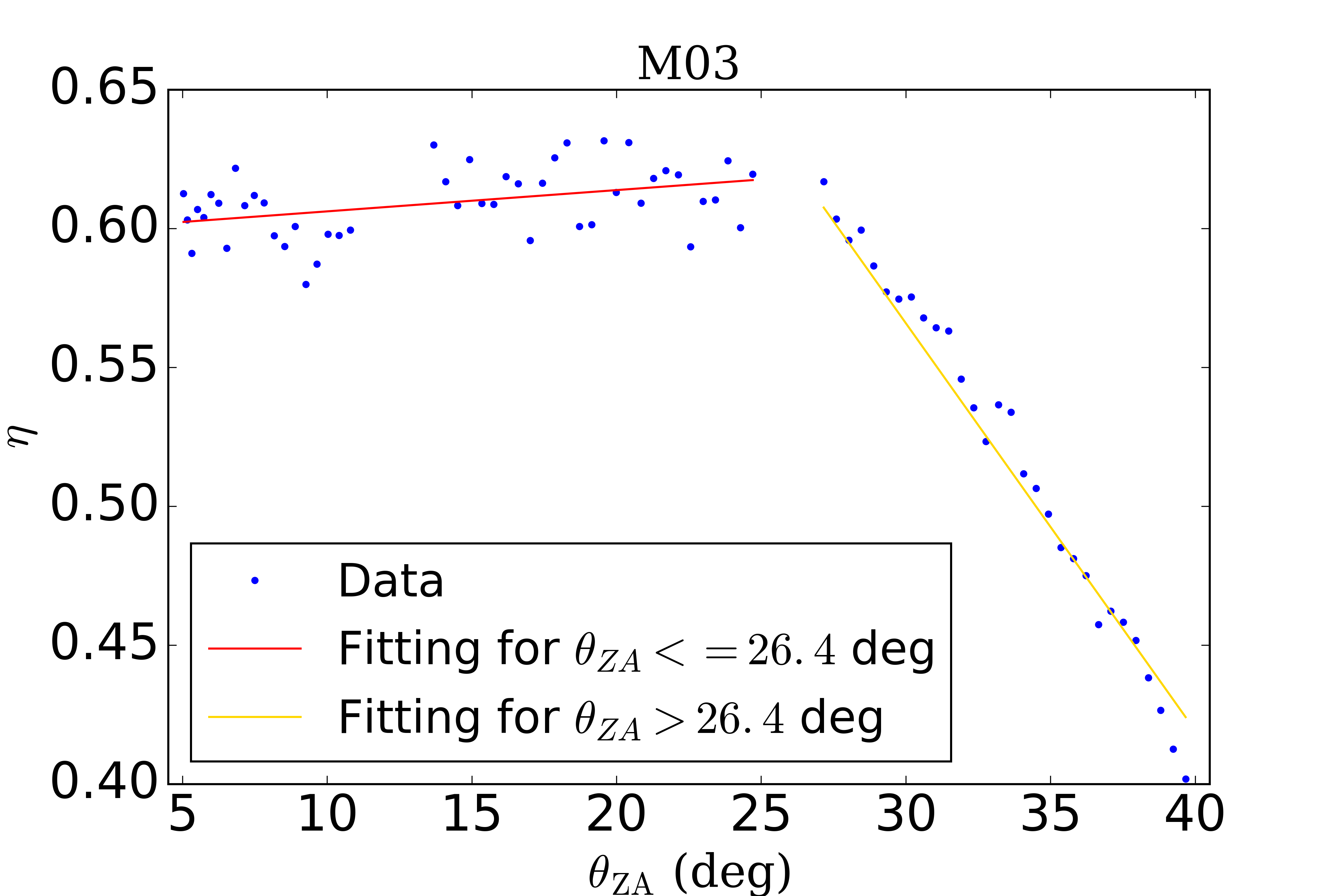}
  \includegraphics[width=0.45\textwidth]{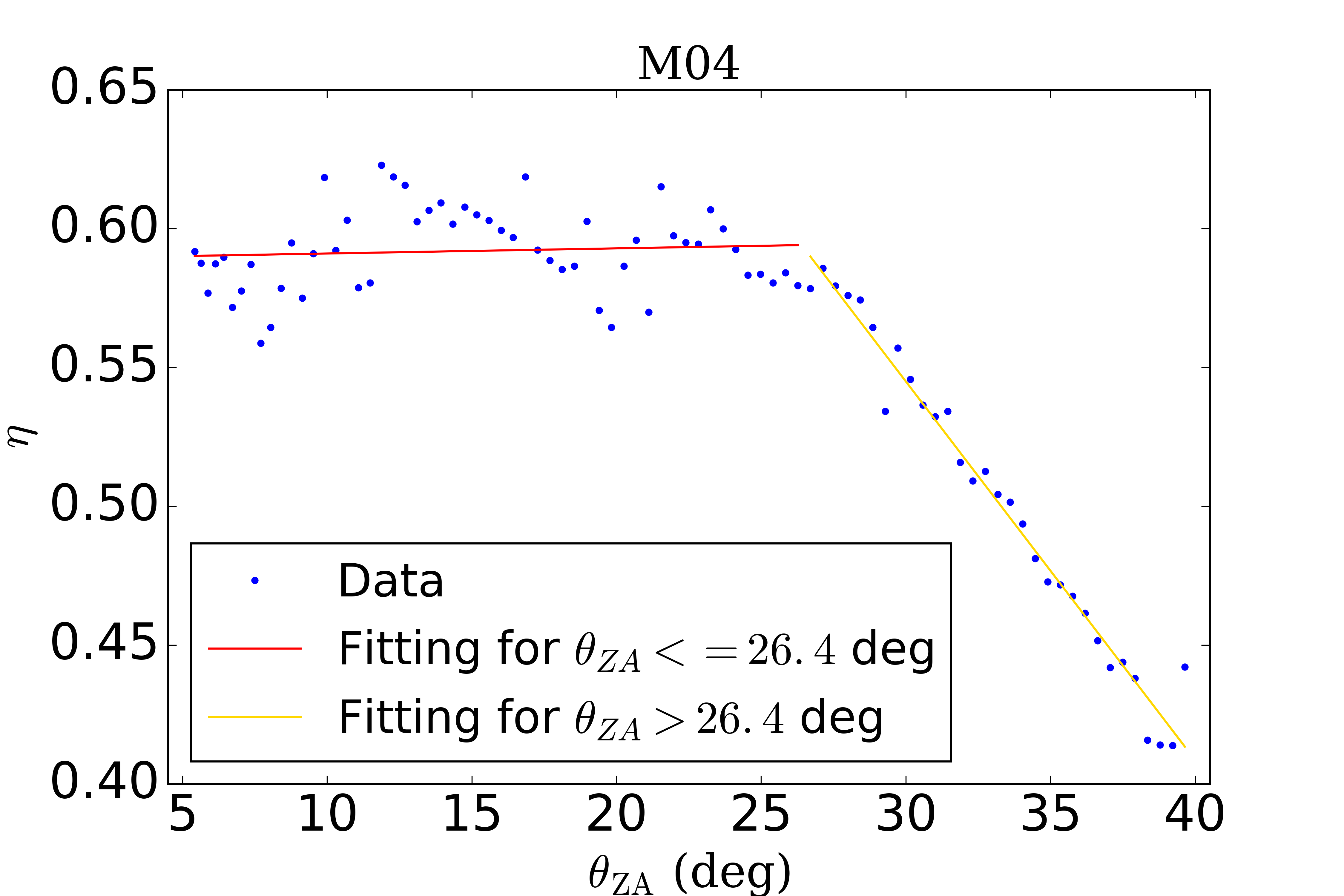}
  \includegraphics[width=0.45\textwidth]{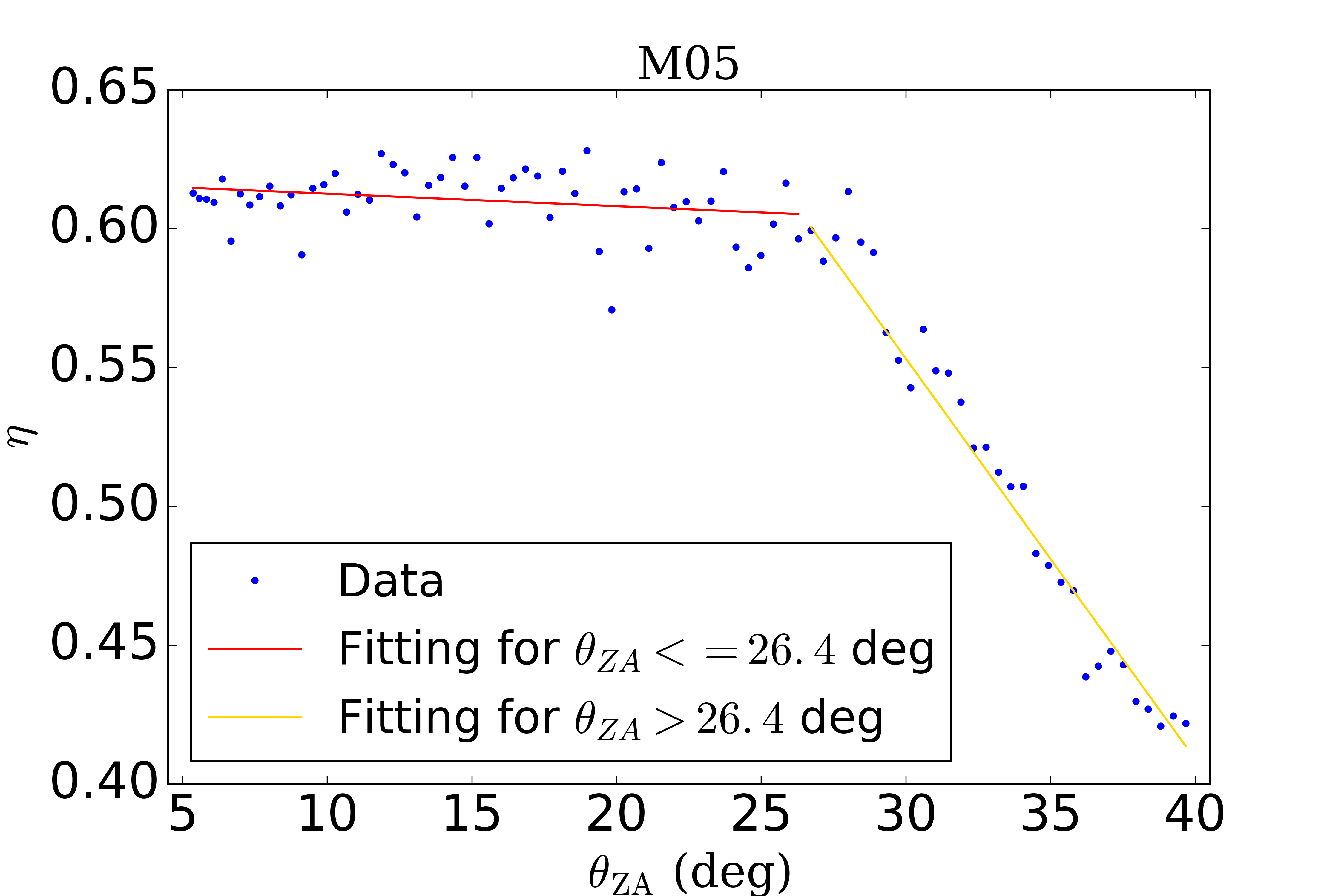}
  \includegraphics[width=0.45\textwidth]{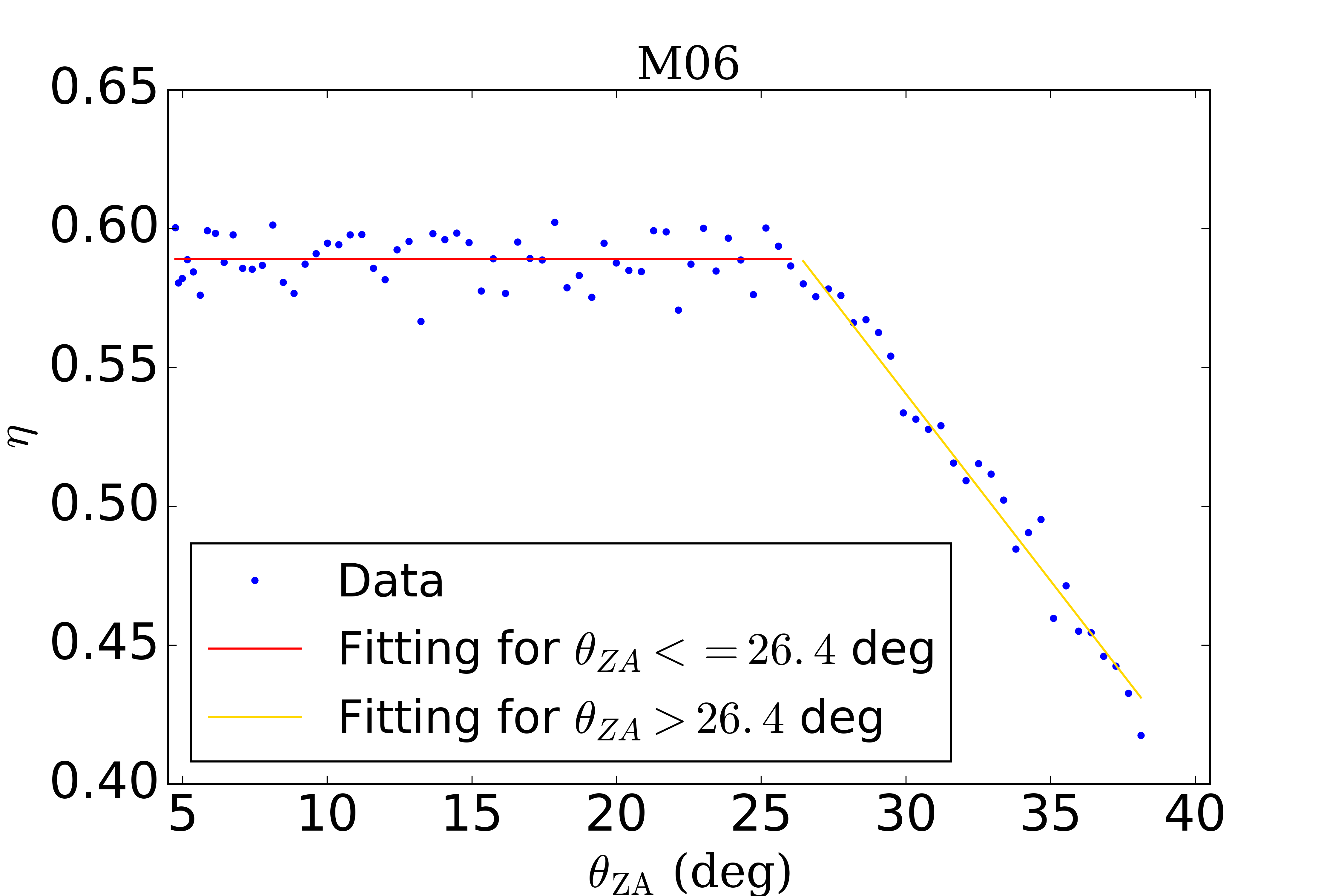}
   \includegraphics[width=0.45\textwidth]{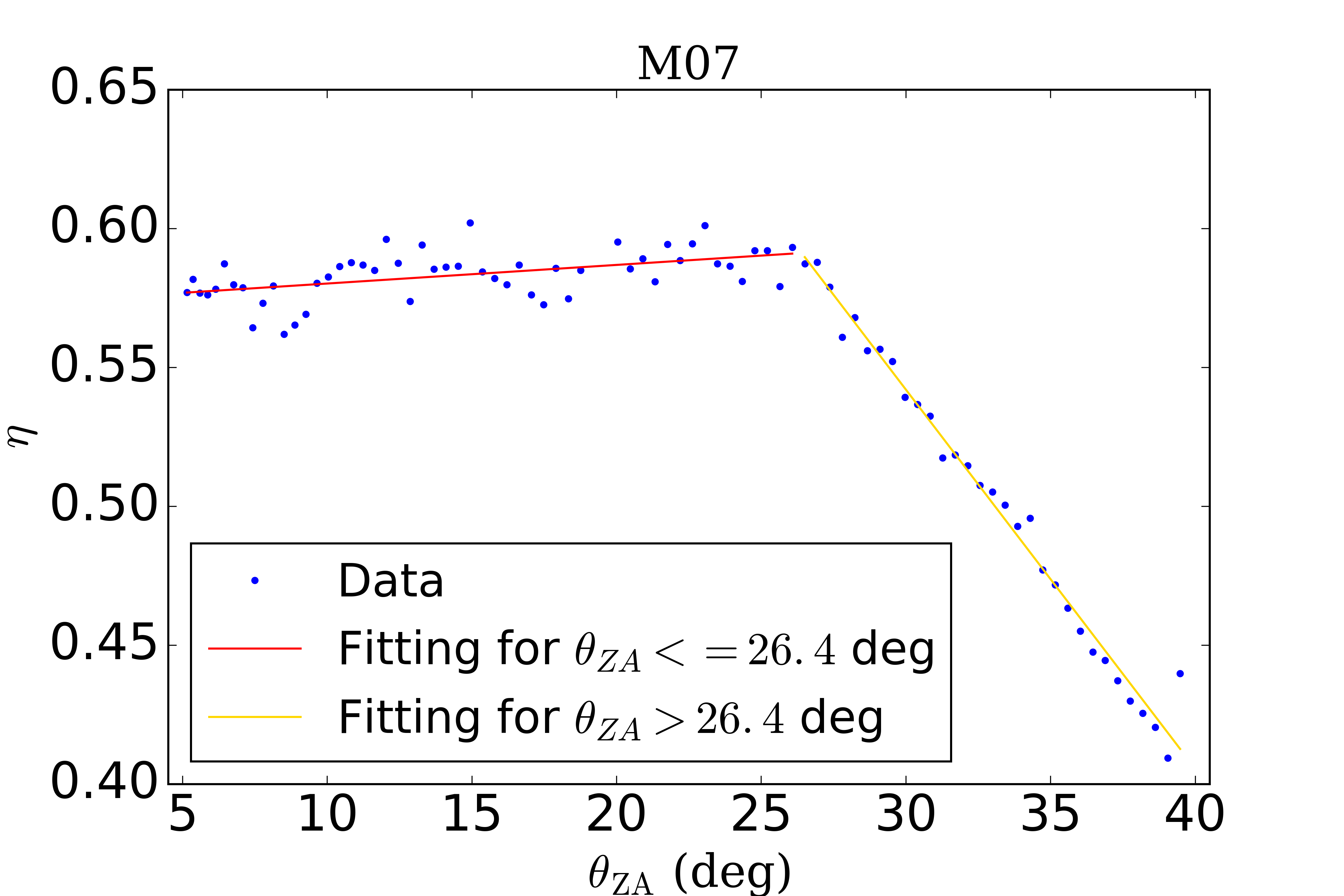}
  \includegraphics[width=0.45\textwidth]{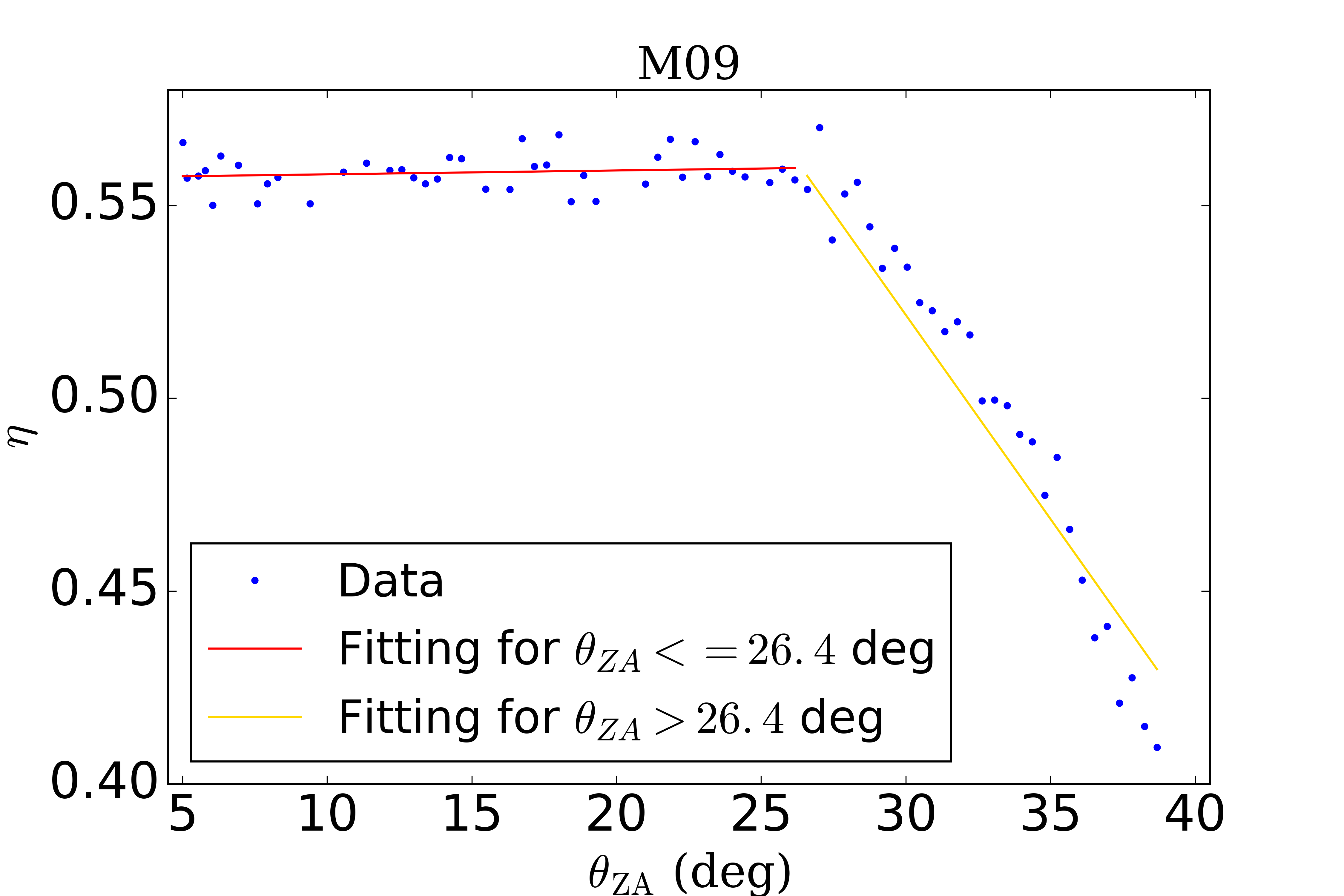}
  \includegraphics[width=0.45\textwidth]{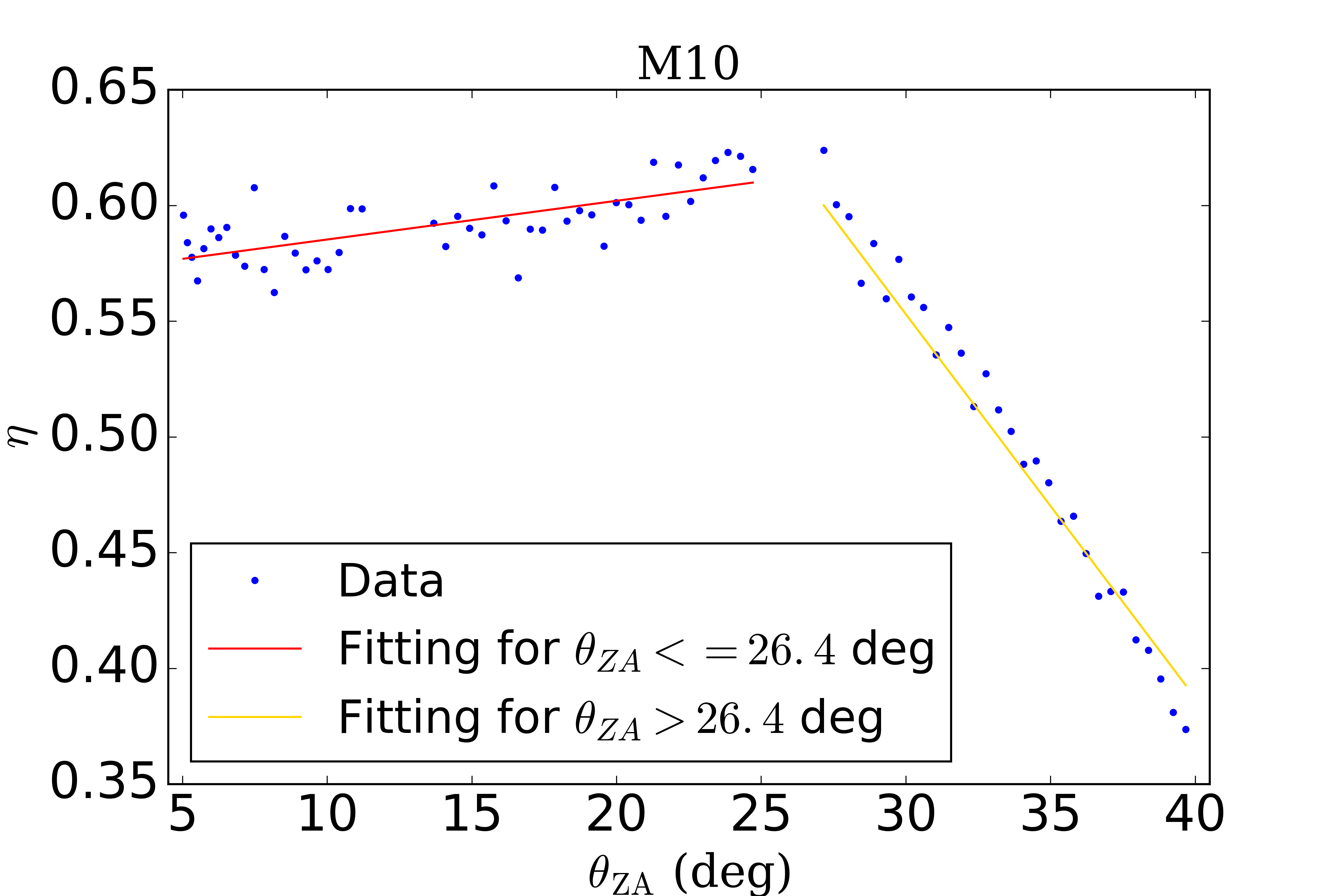}
  \includegraphics[width=0.45\textwidth]{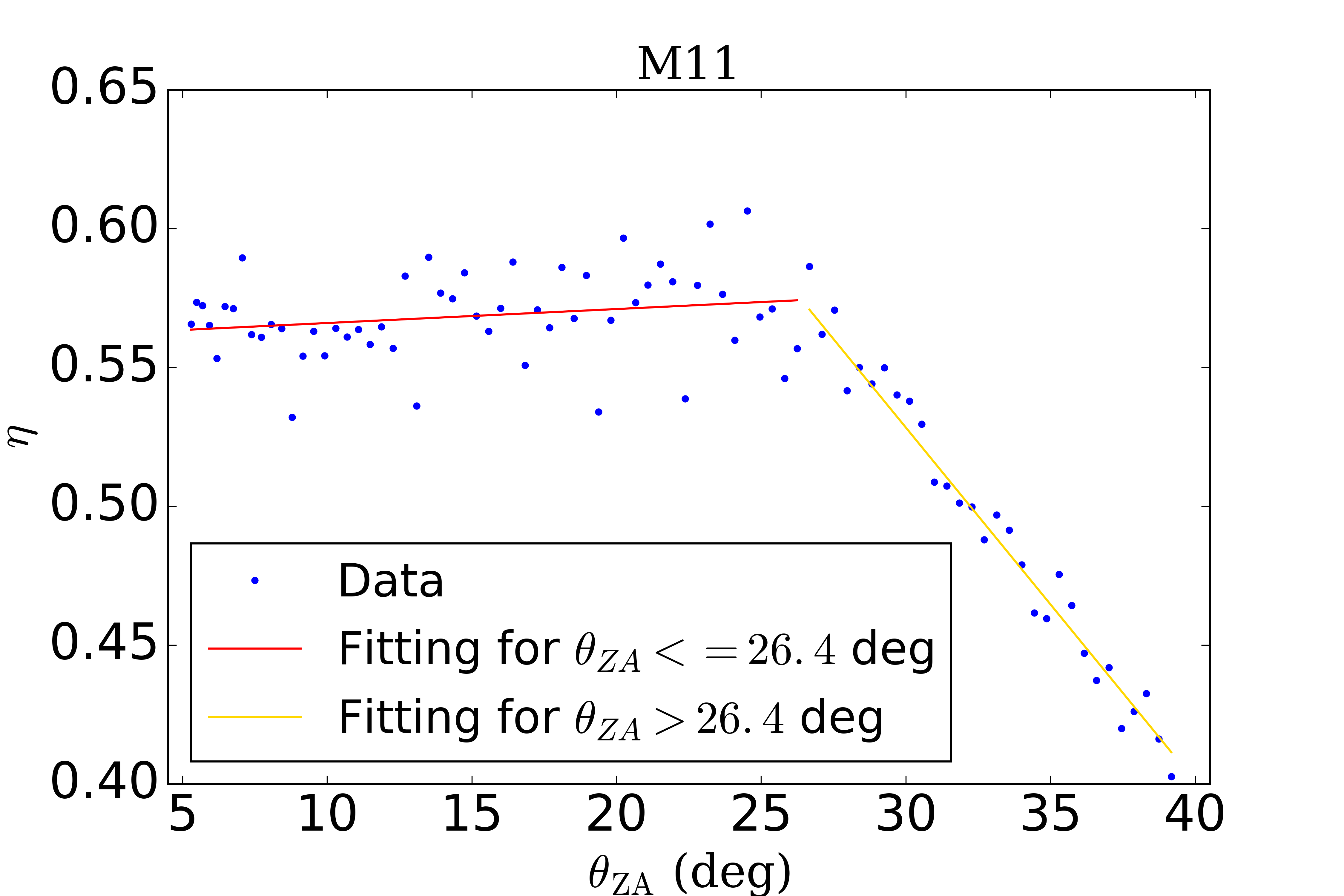}

\end{figure}
\begin{figure}[htb]
\centering
   \includegraphics[width=0.45\textwidth]{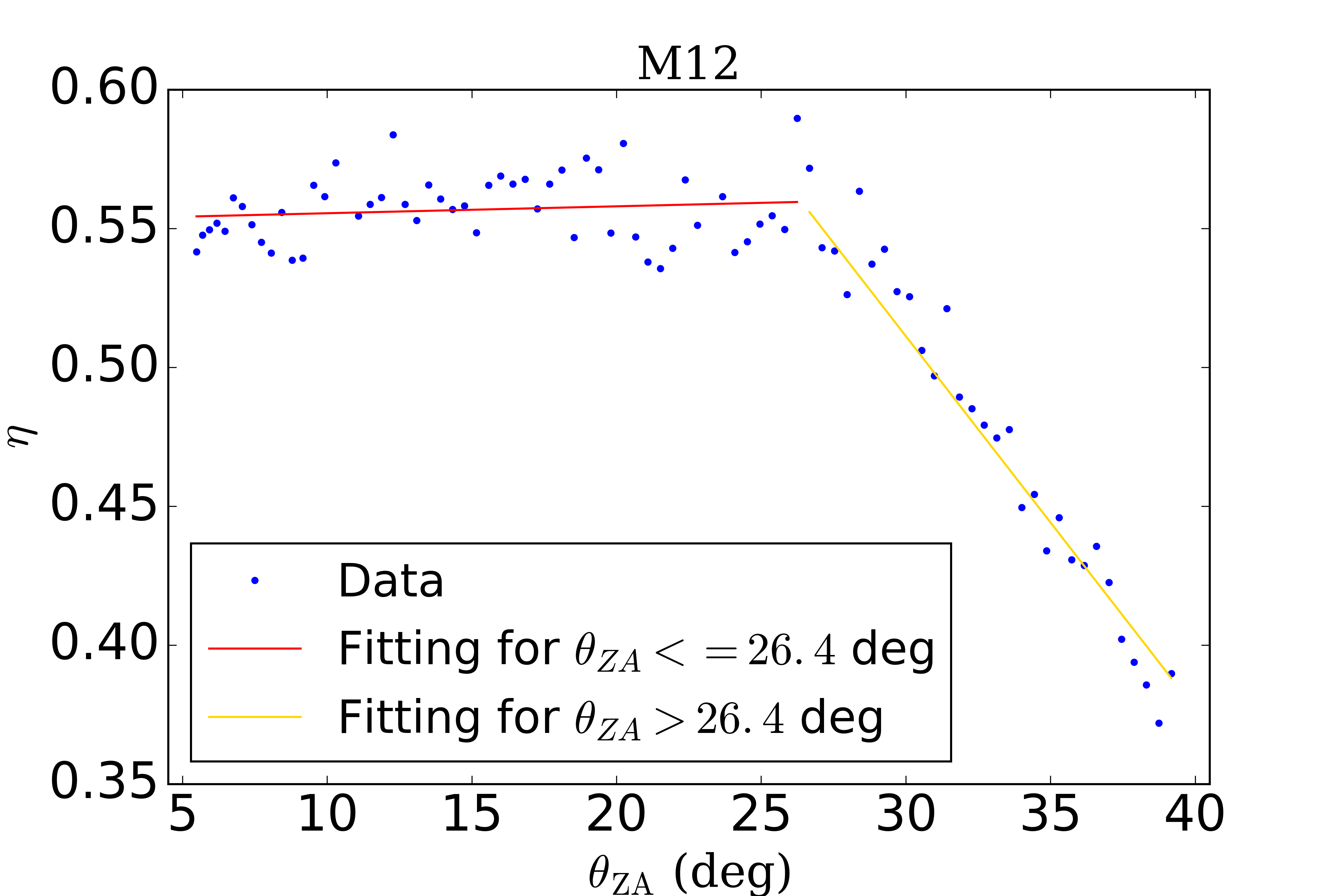}
  \includegraphics[width=0.45\textwidth]{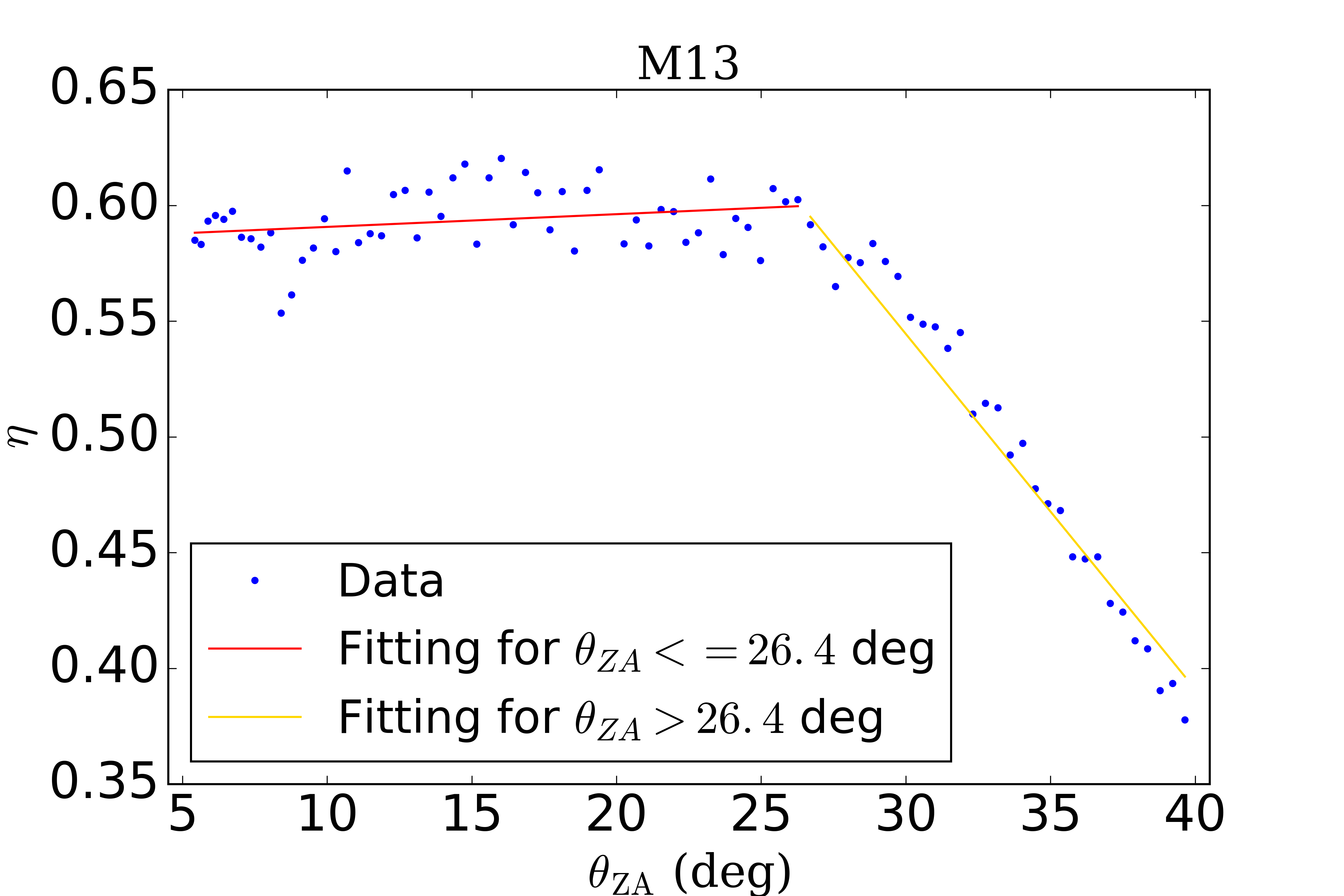}
  \includegraphics[width=0.45\textwidth]{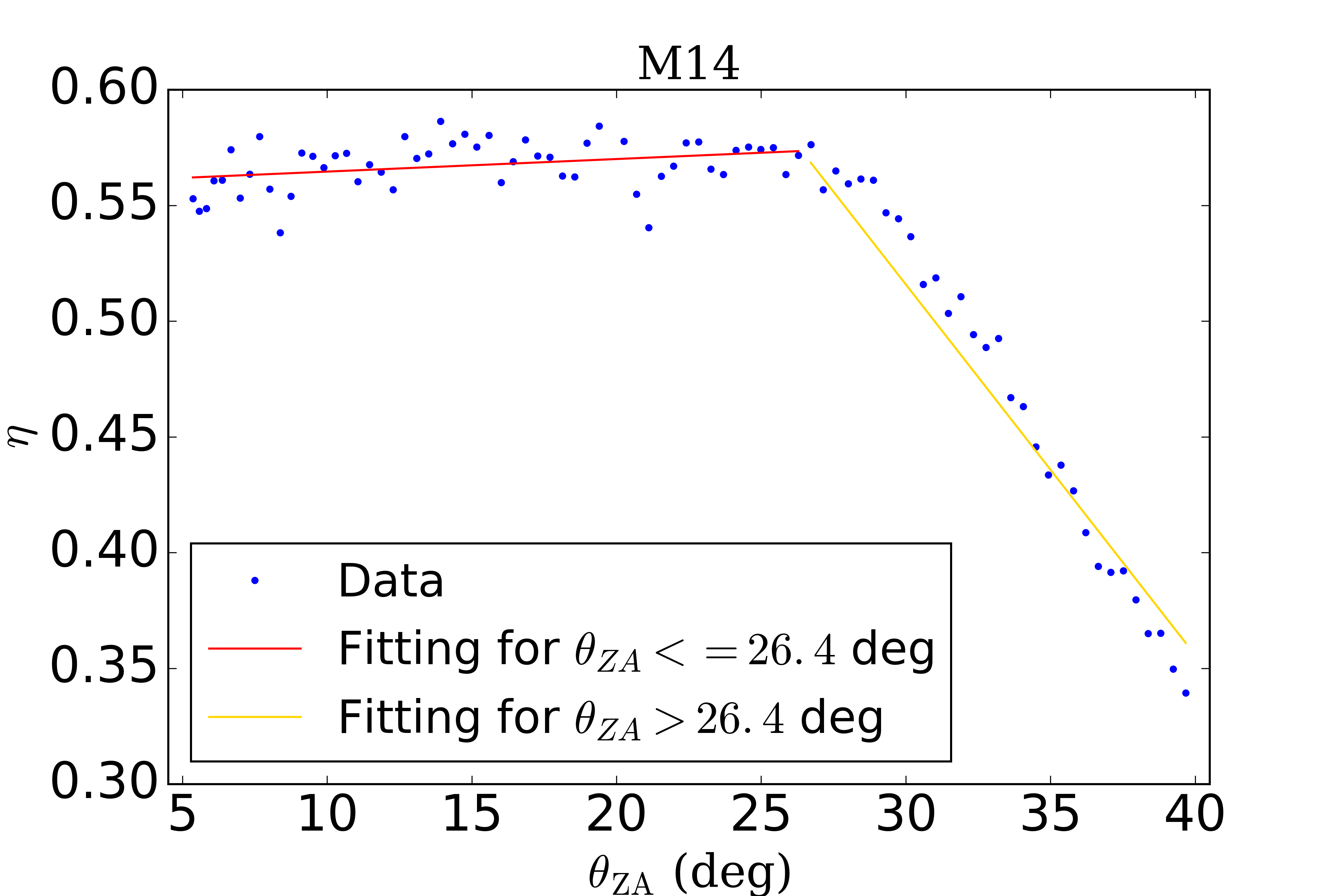}
  \includegraphics[width=0.45\textwidth]{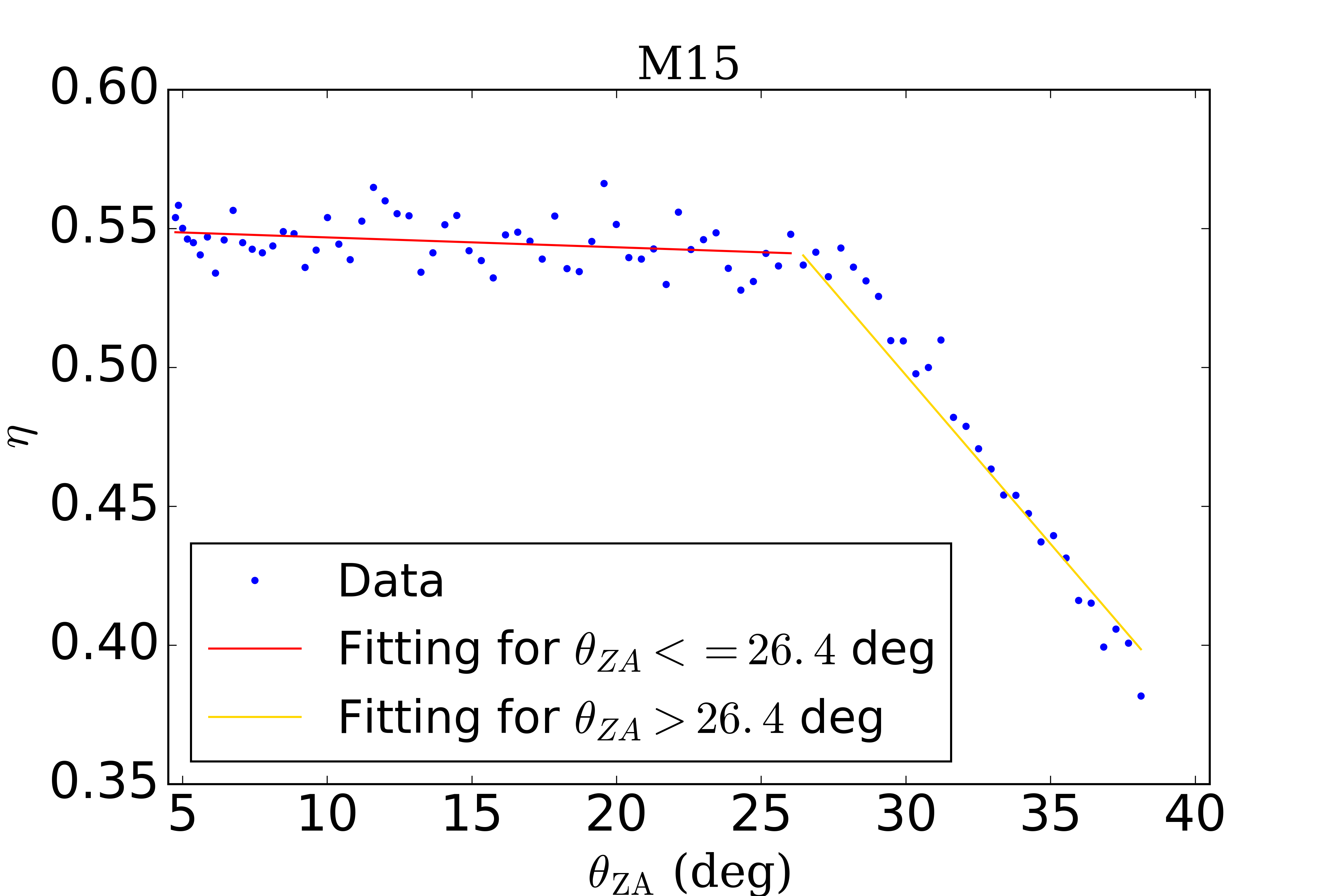}
  \includegraphics[width=0.45\textwidth]{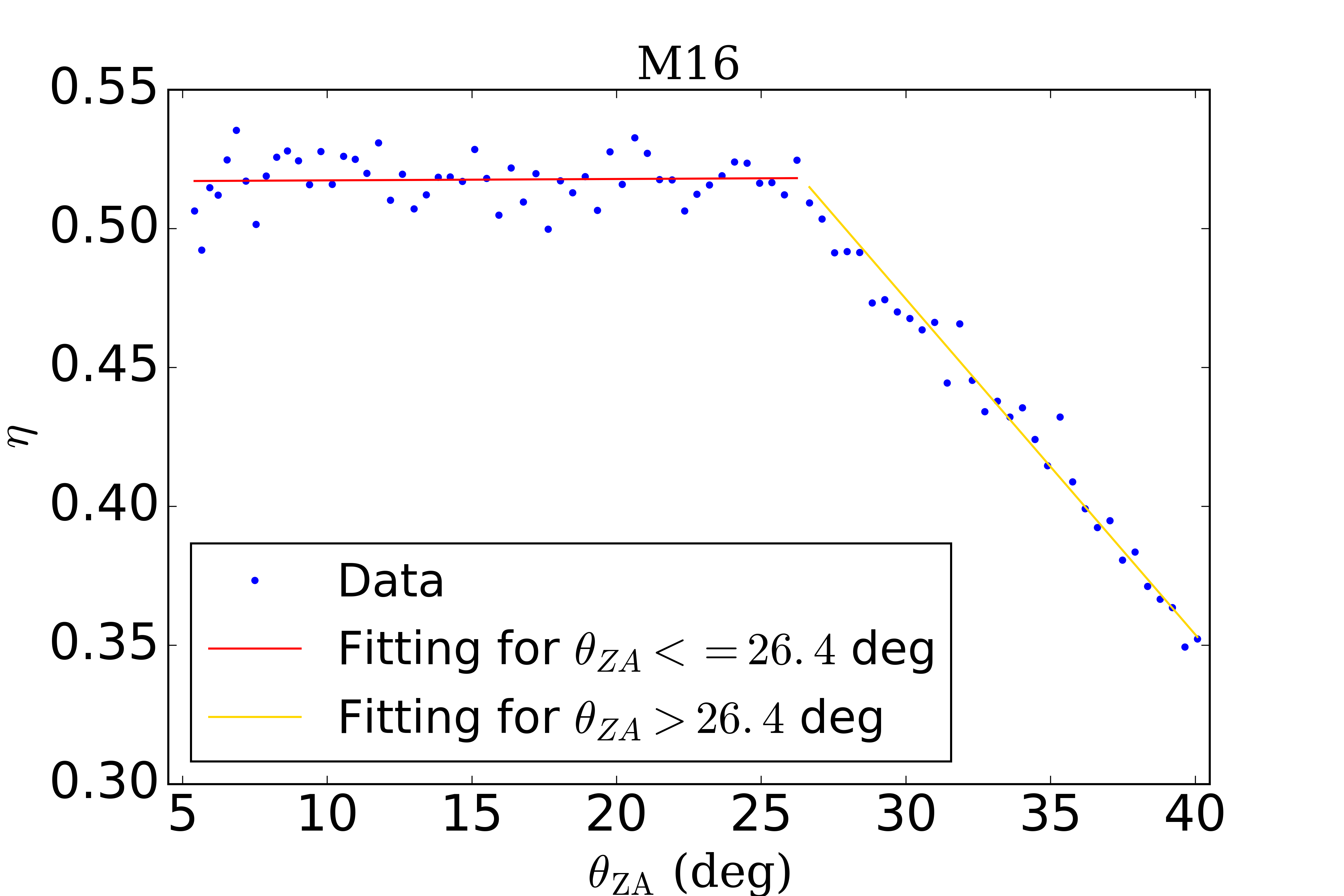}
  \includegraphics[width=0.45\textwidth]{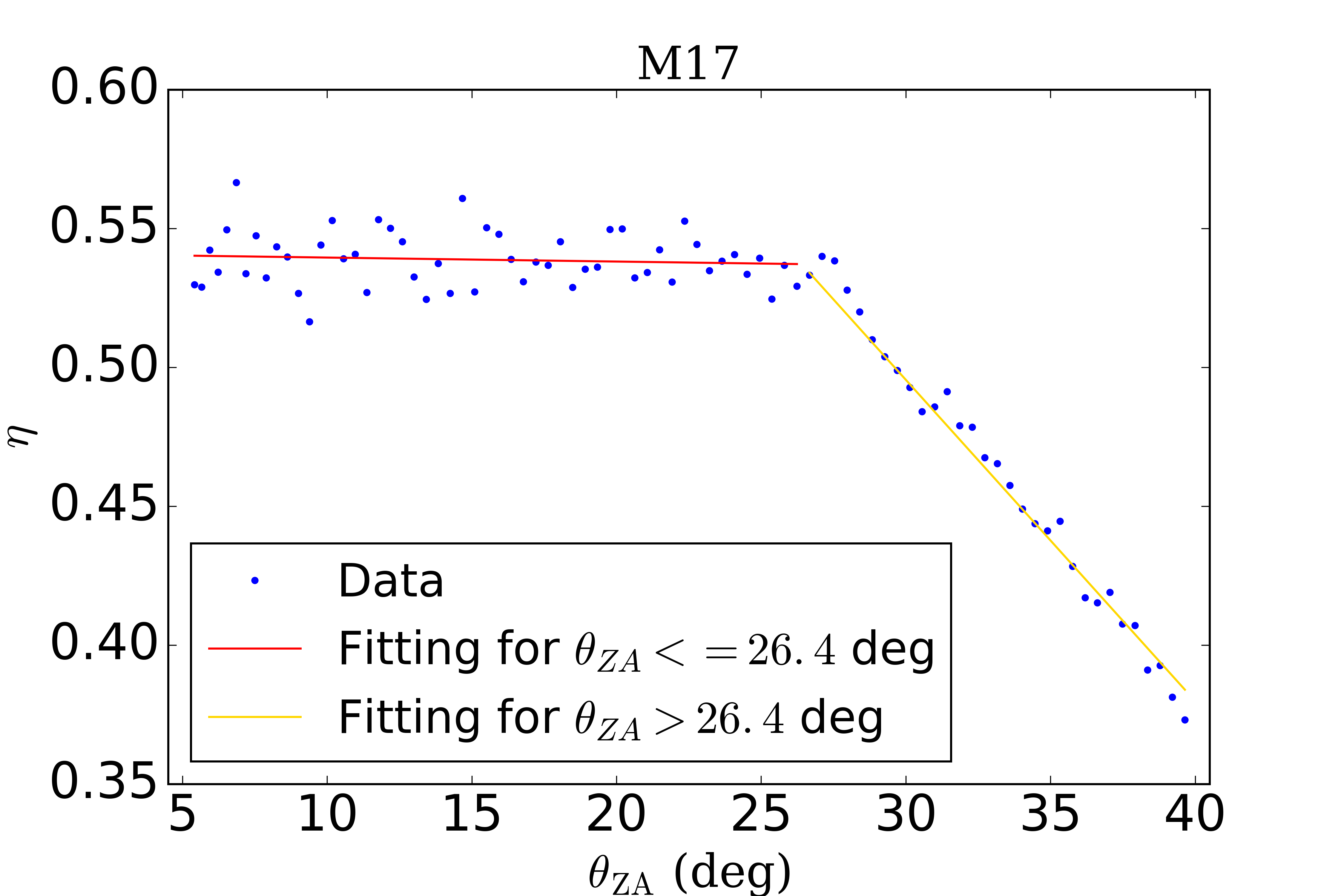}
  \includegraphics[width=0.45\textwidth]{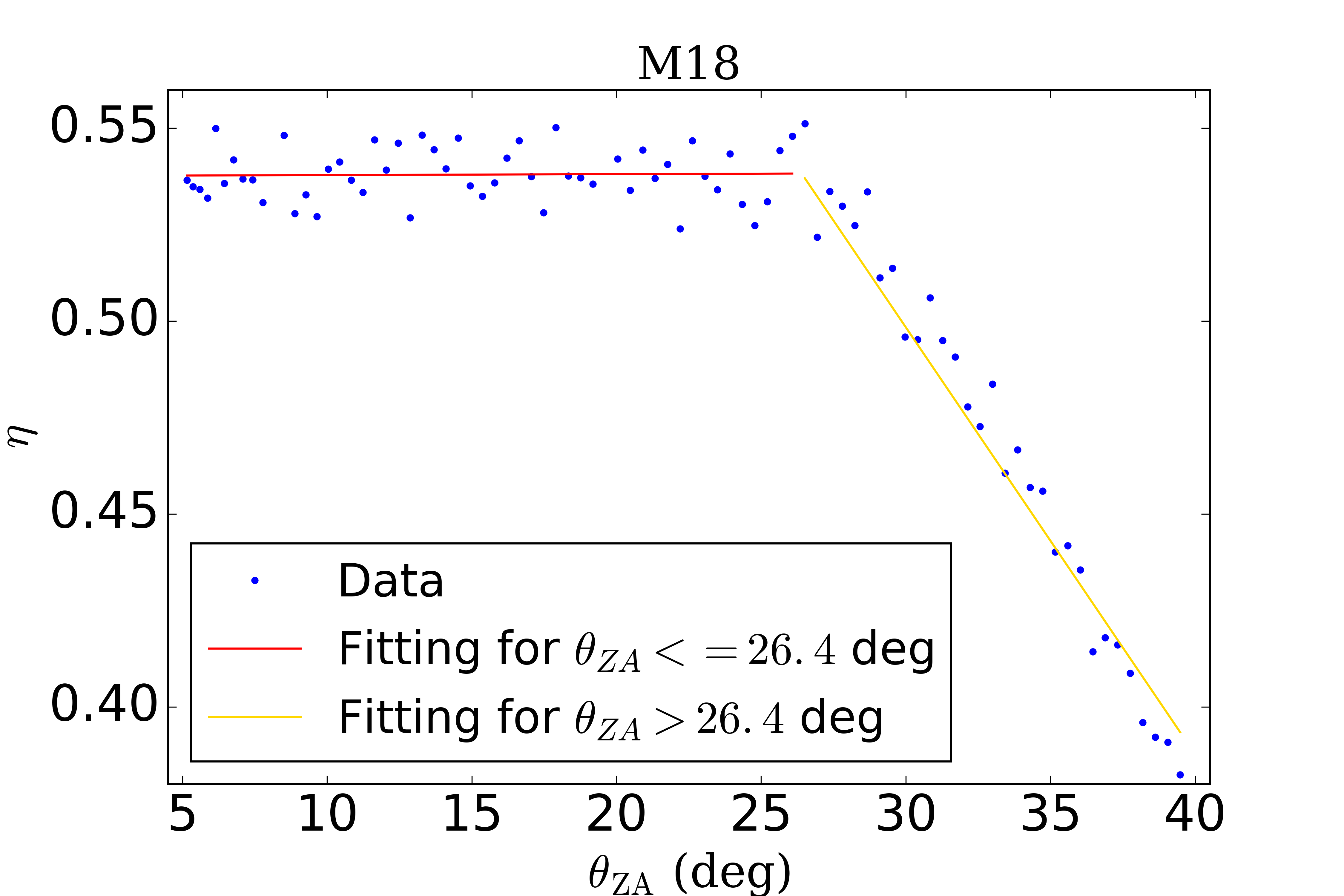}
 \caption{Measured $\eta$ curve as a function of zenith angle $\theta\rm_{ZA}$ at 1400 MHz for Beam 3, 4, 5, 6, 7, 9, 10, 11, 12, 13, 14, 15, 16, 17 and 18.   Fitting results when $\theta\rm_{ZA}$ $\leq$ 26.4$^\circ$ and $\theta\rm_{ZA}$ > 26.4$^\circ$ are represented with red and gold solid line, respectively.}
 \label{fig:etafit_restbeams}
\end{figure}

\section{System temperature}

\begin{figure}[!ht]
\centering
  \includegraphics[width=0.45\textwidth]{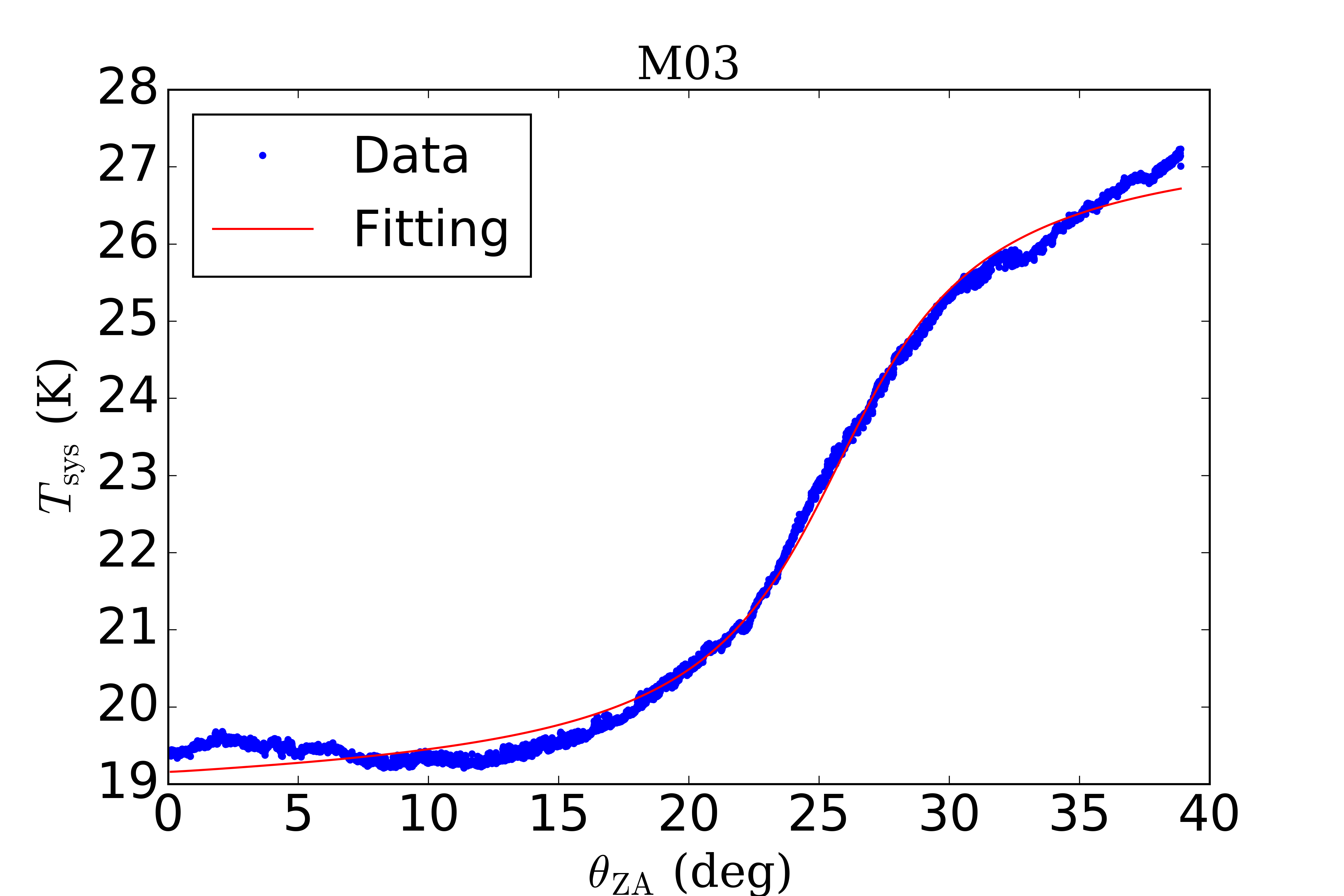}
  \includegraphics[width=0.45\textwidth]{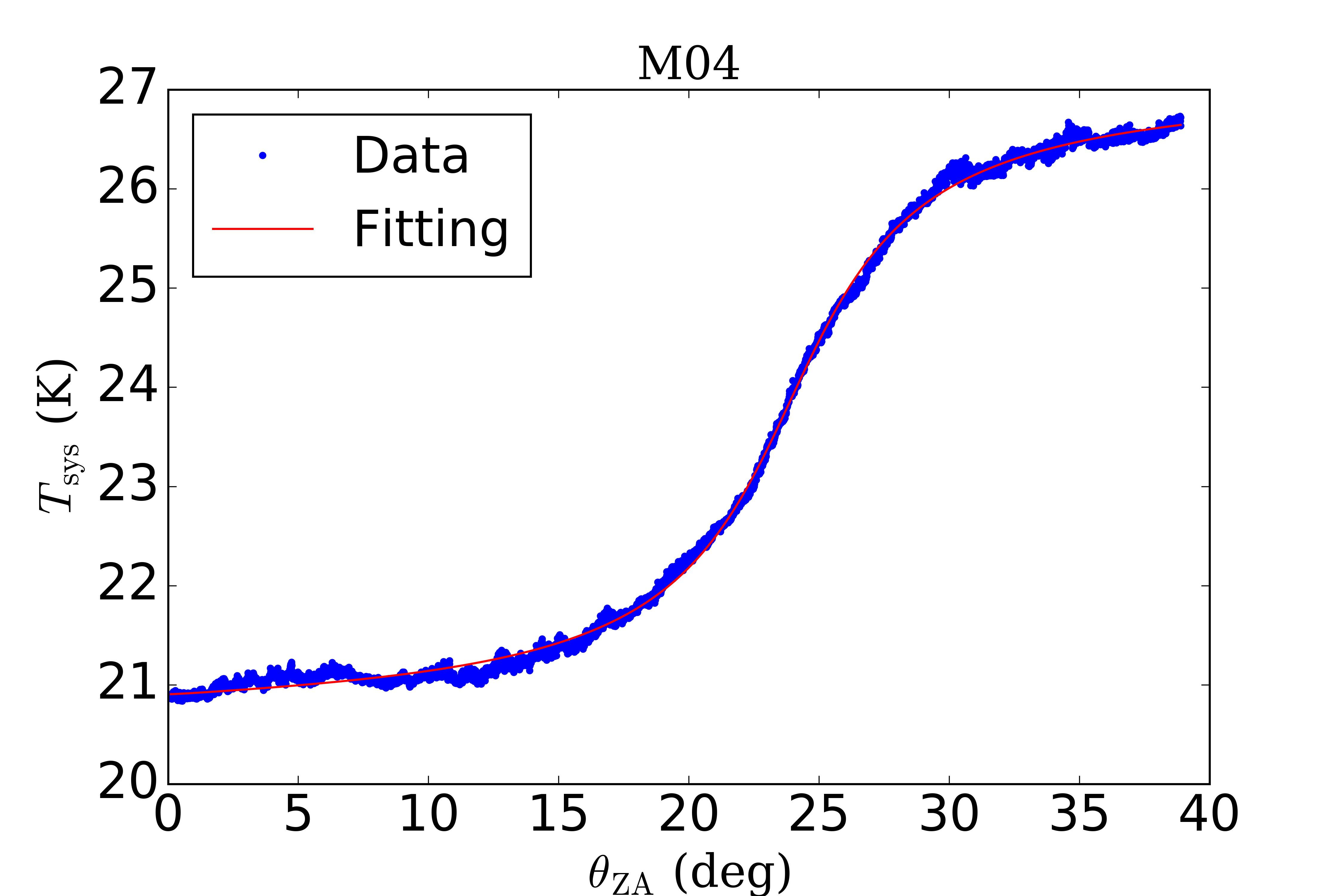}
  \includegraphics[width=0.45\textwidth]{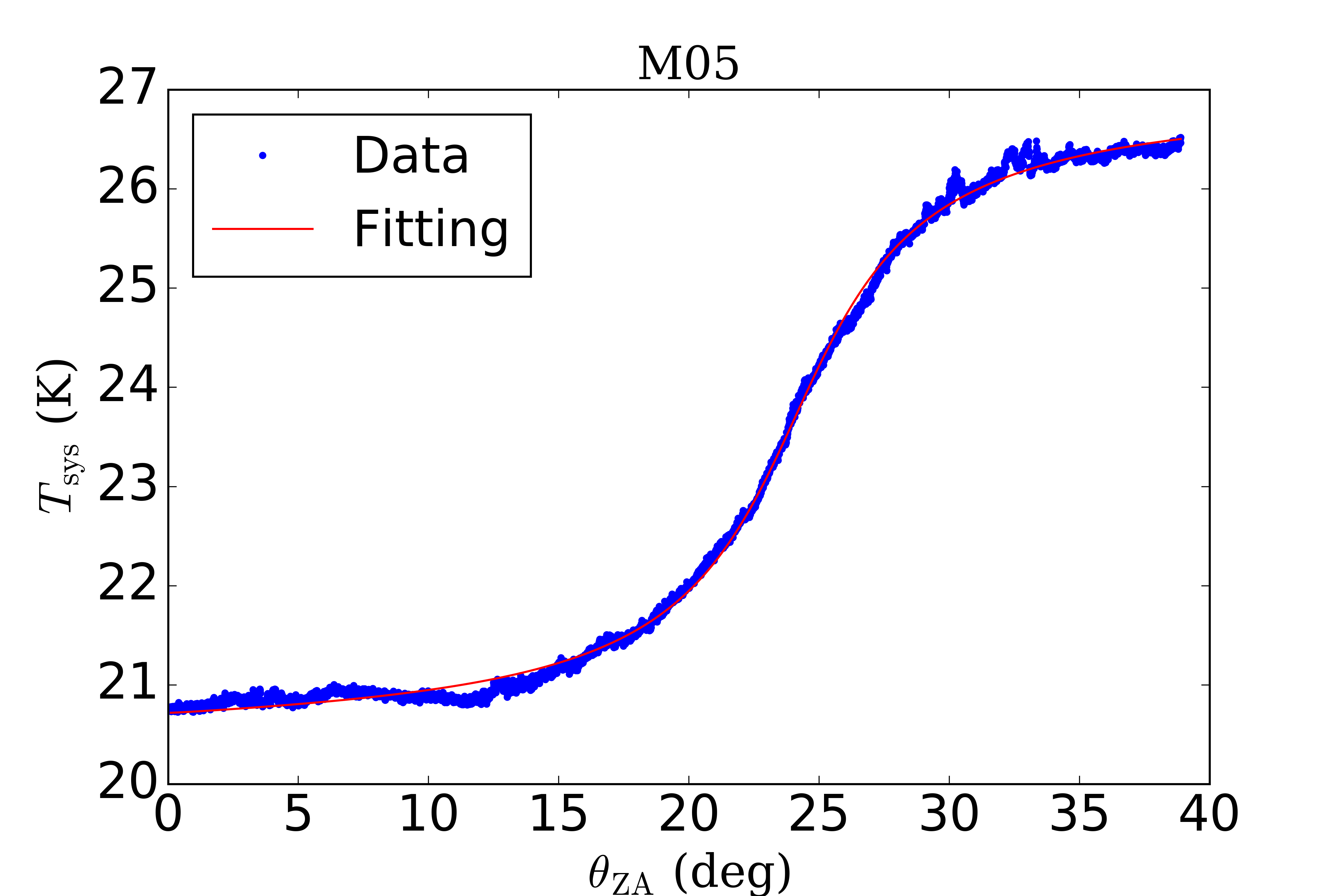}
  \includegraphics[width=0.45\textwidth]{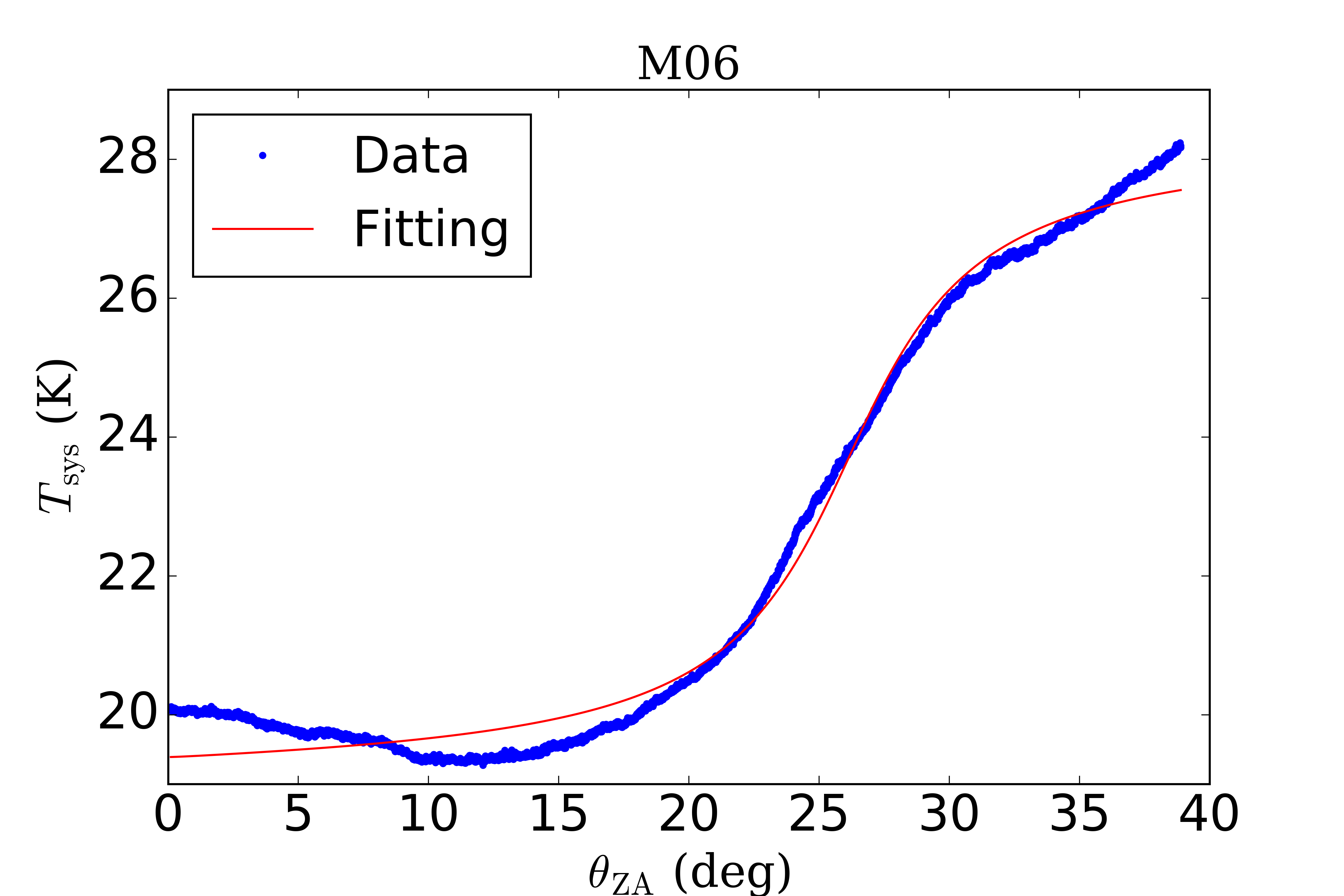}
   \includegraphics[width=0.45\textwidth]{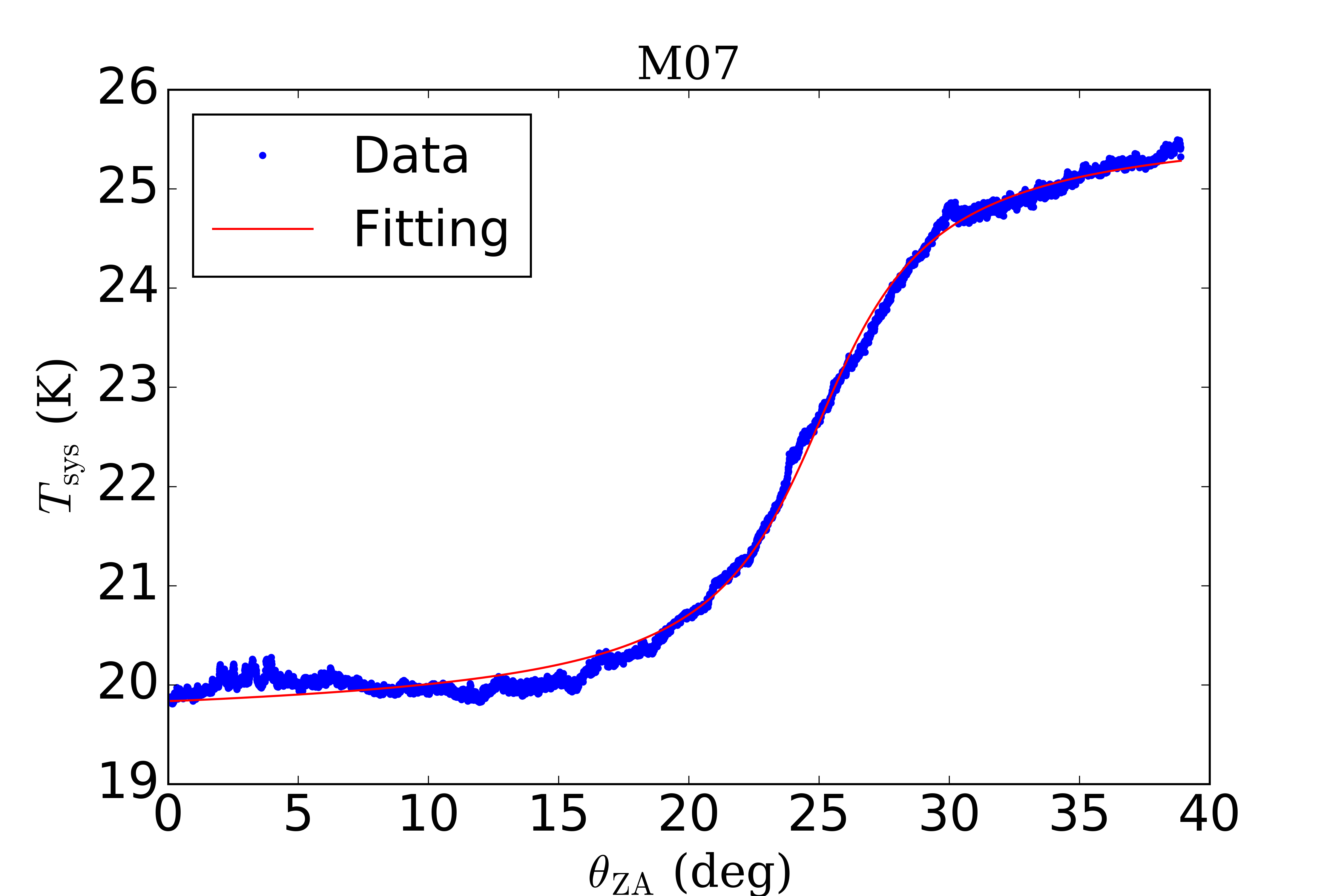}
  \includegraphics[width=0.45\textwidth]{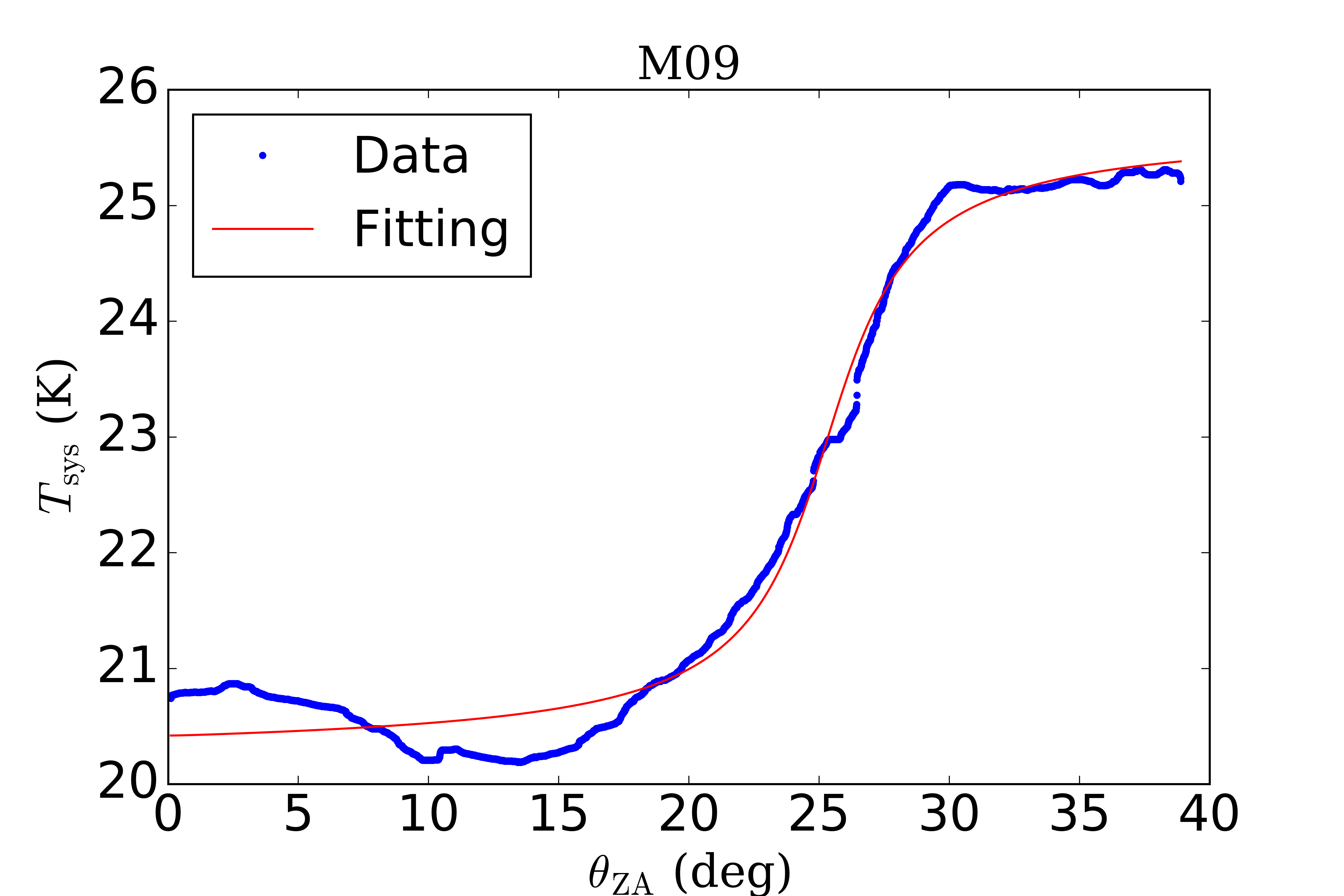}
  \includegraphics[width=0.45\textwidth]{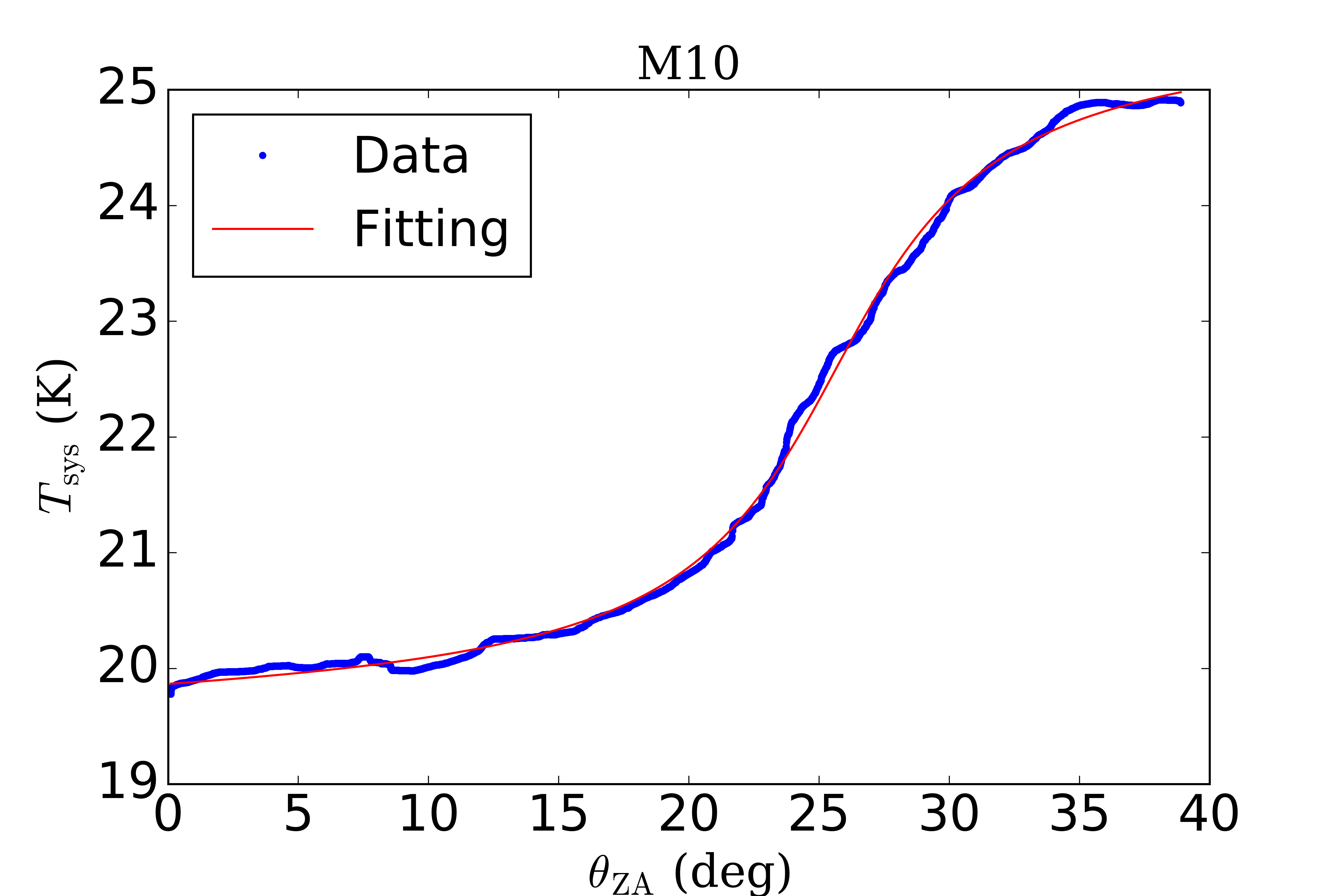}
  \includegraphics[width=0.45\textwidth]{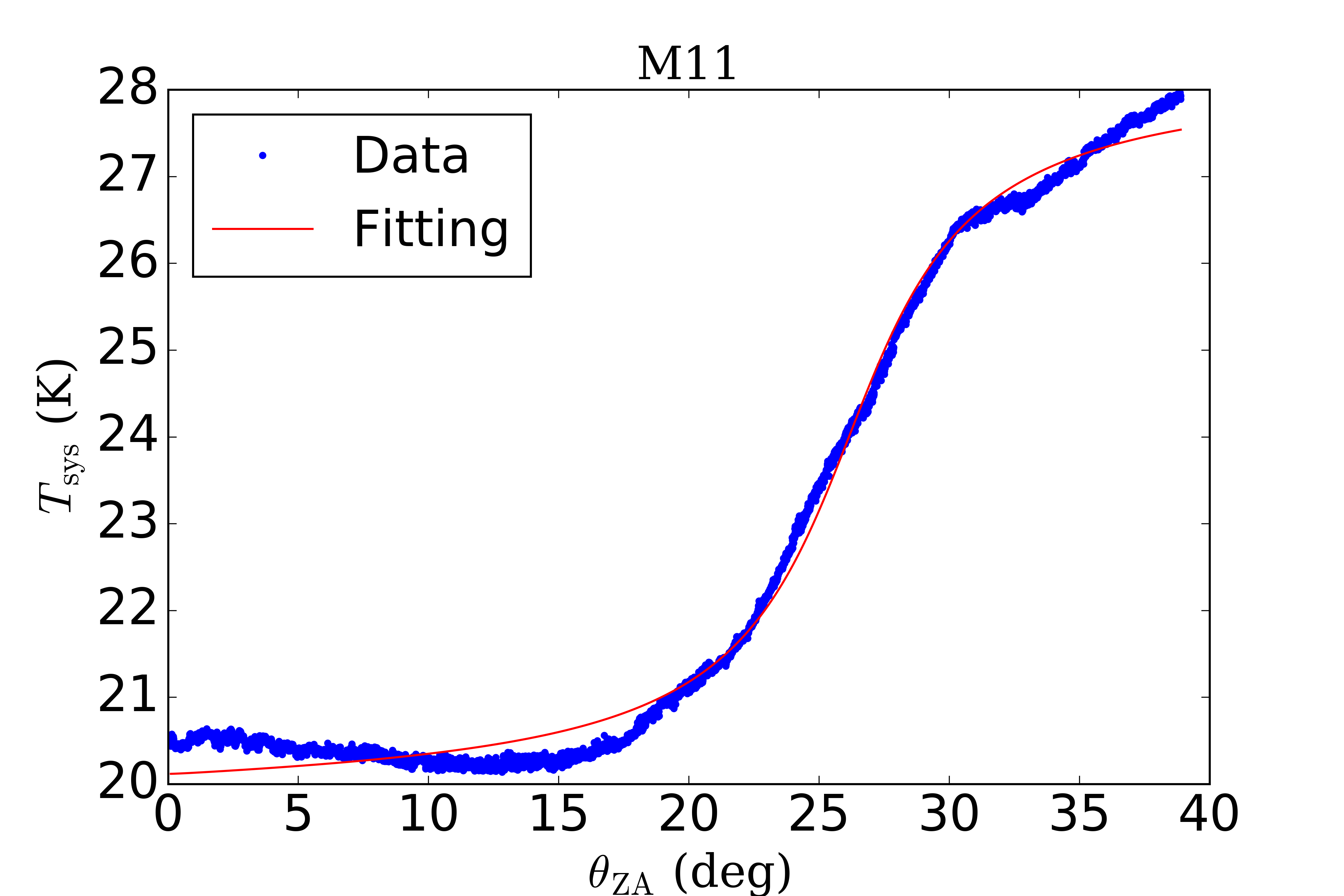}

\end{figure}
\begin{figure}[htb]
\centering
  \includegraphics[width=0.45\textwidth]{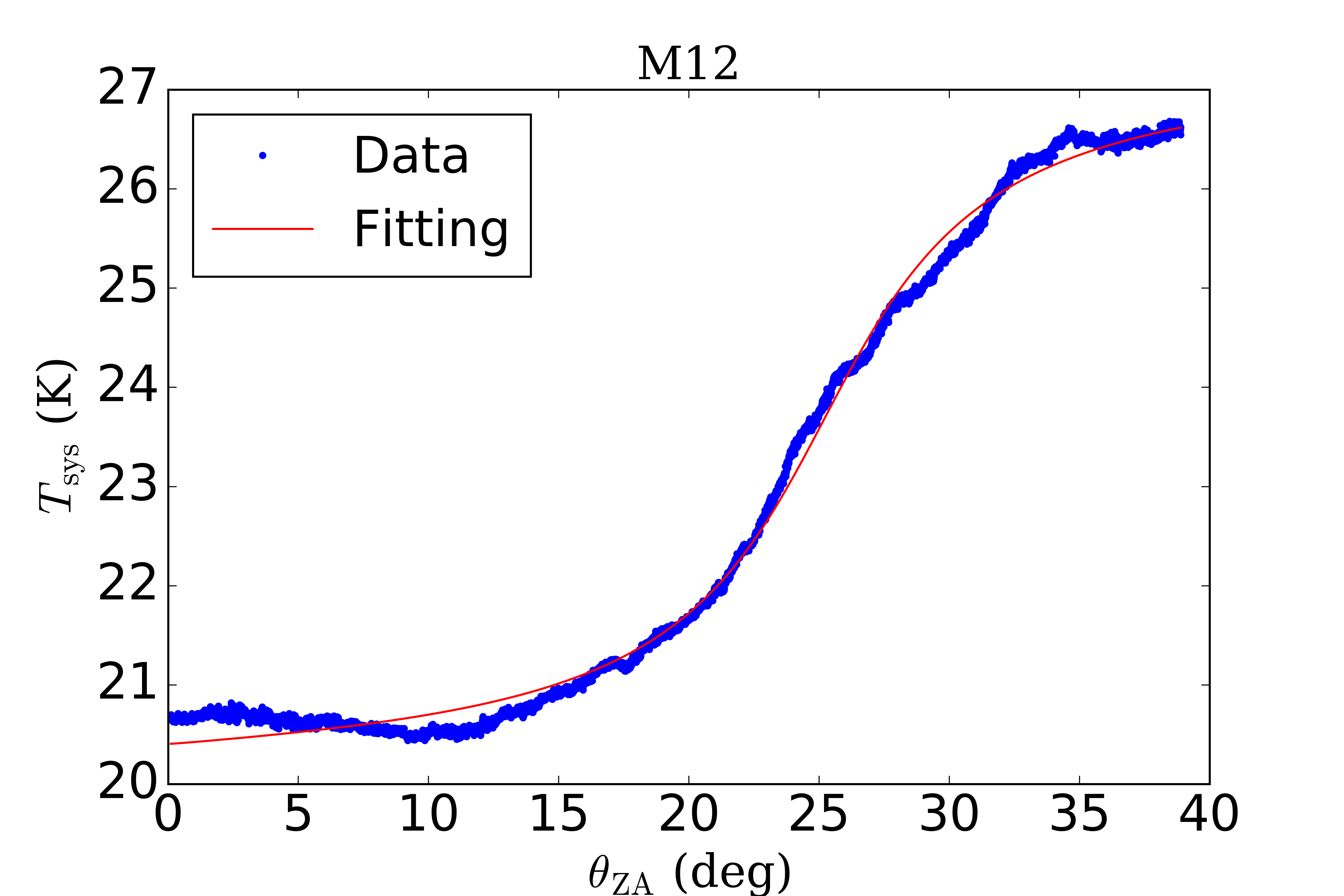}
  \includegraphics[width=0.45\textwidth]{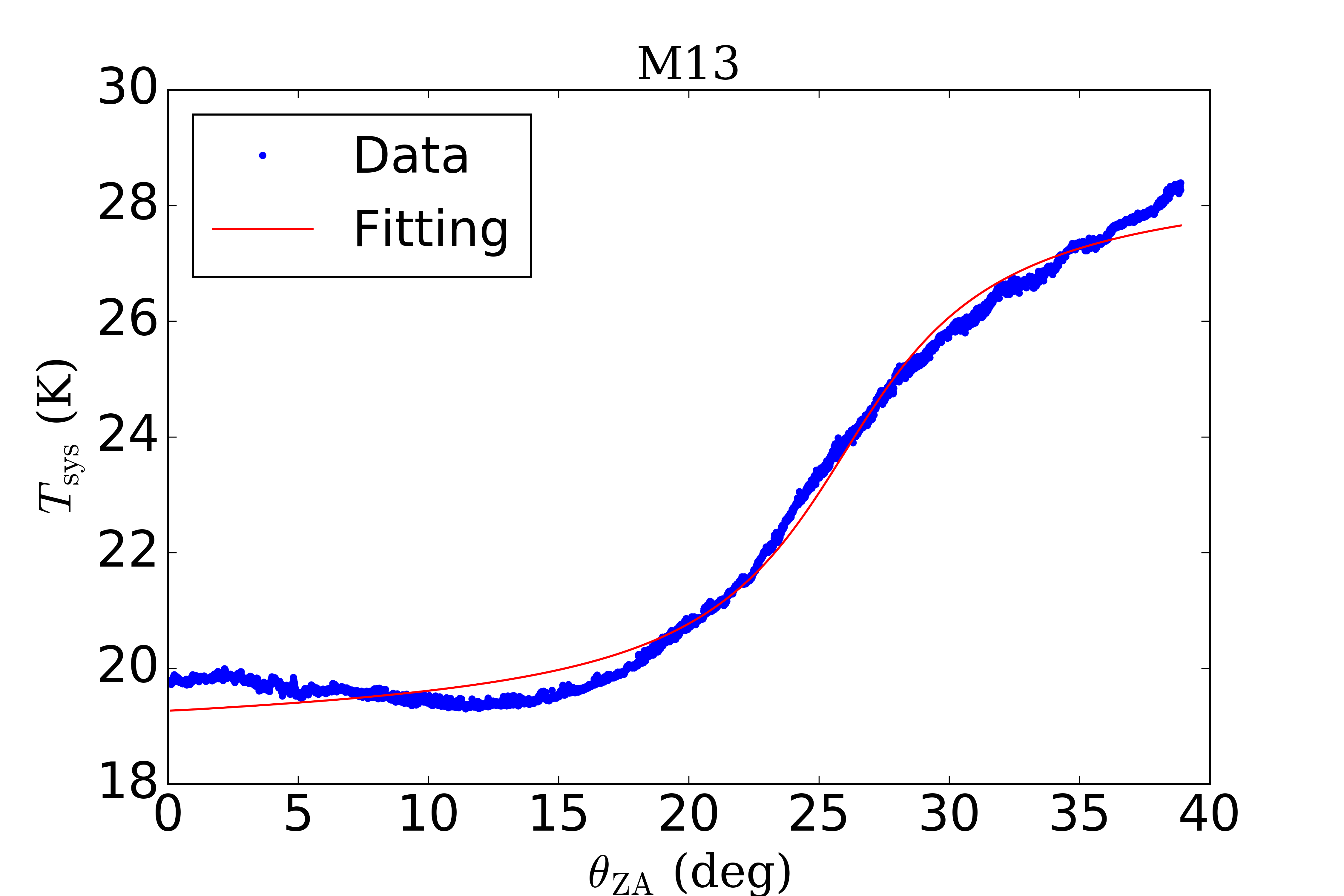}
  \includegraphics[width=0.45\textwidth]{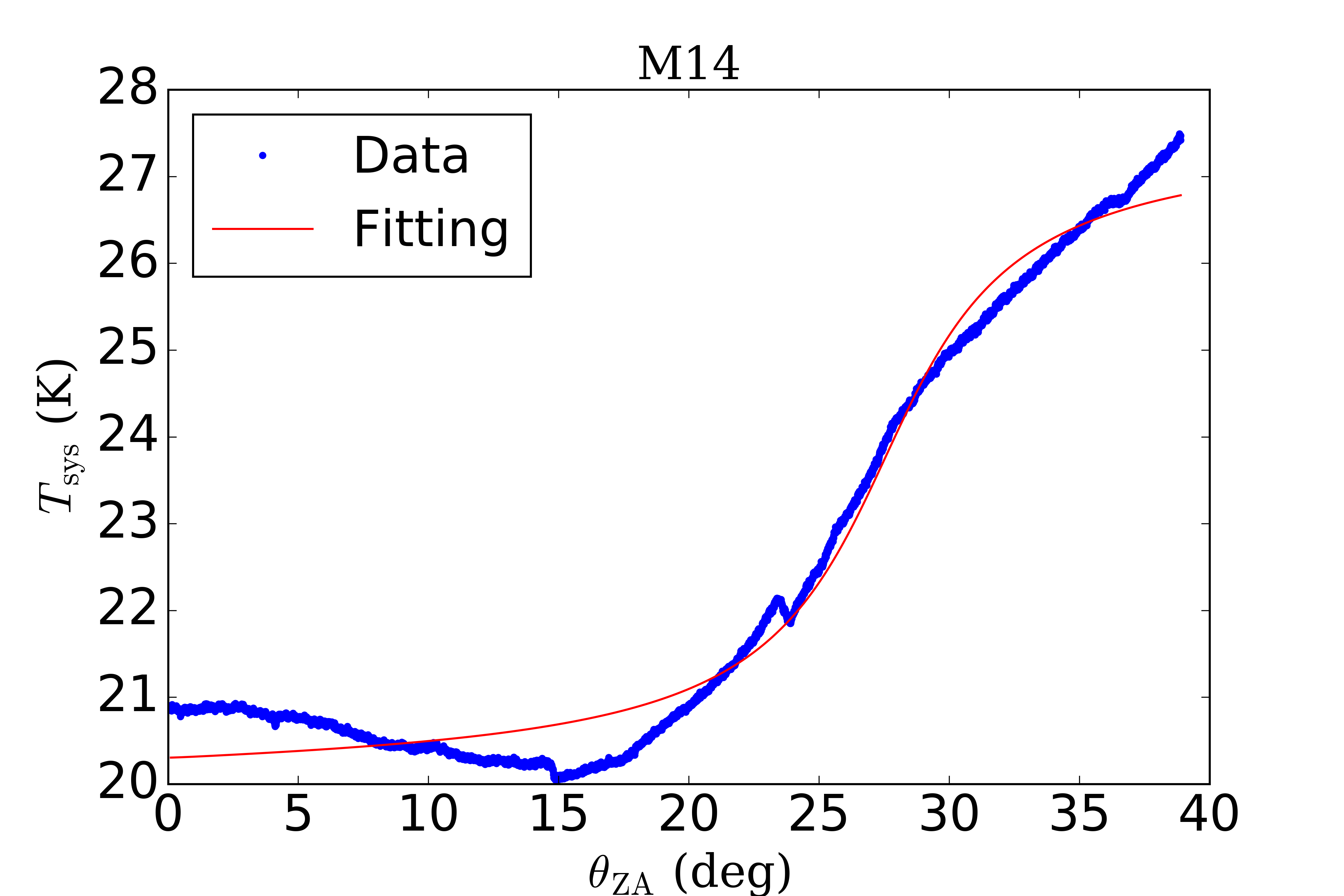}
  \includegraphics[width=0.45\textwidth]{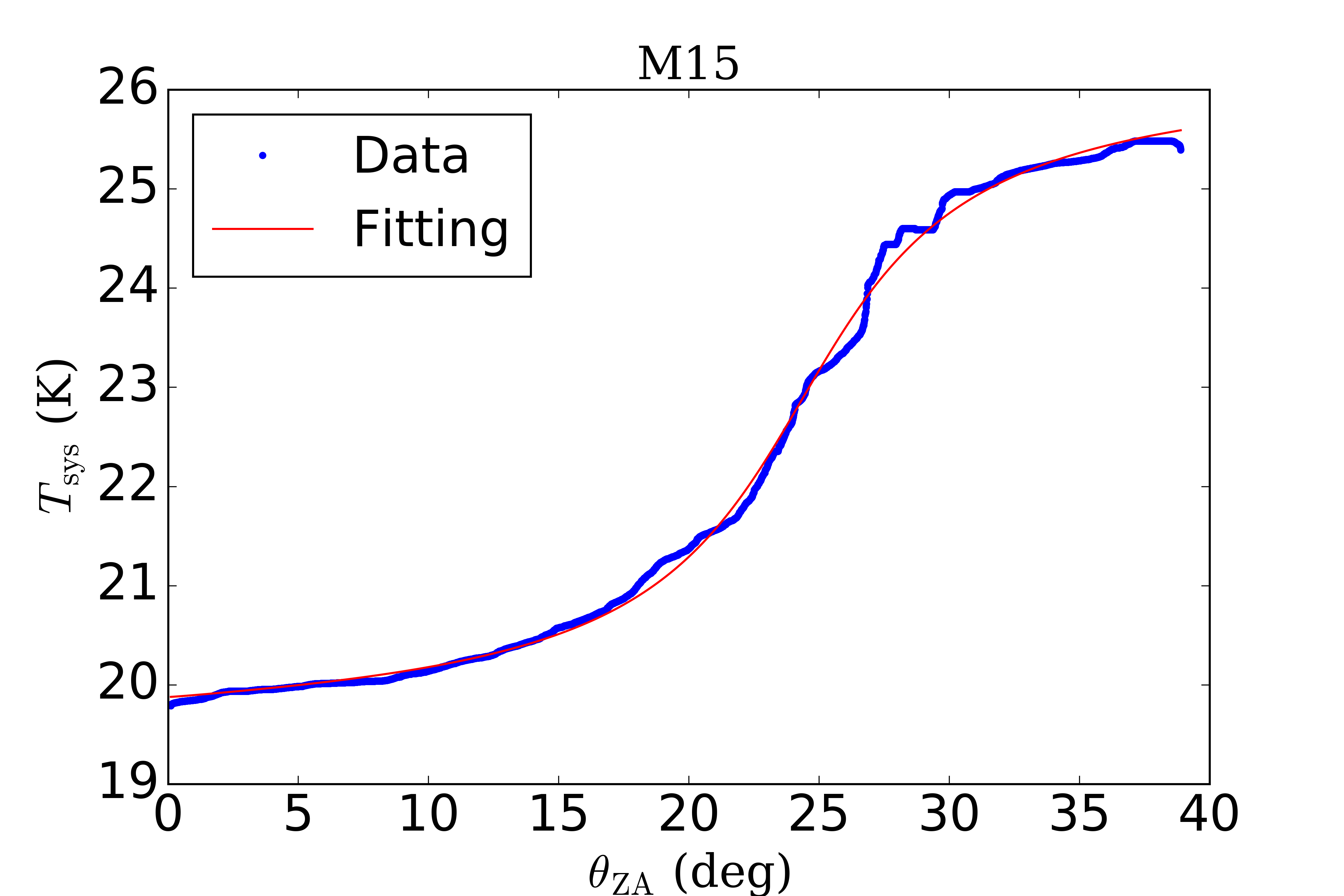}
  \includegraphics[width=0.45\textwidth]{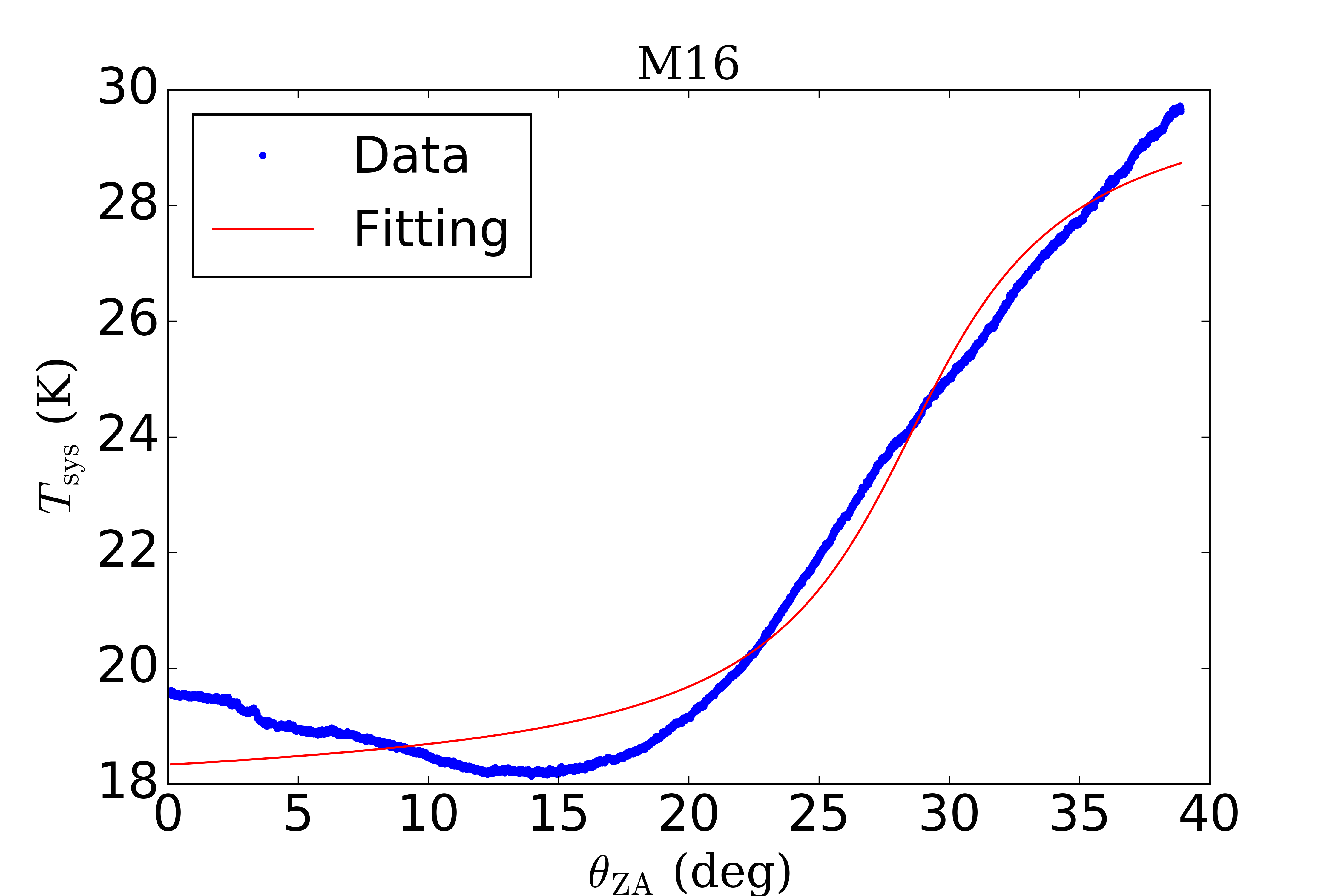}
  \includegraphics[width=0.45\textwidth]{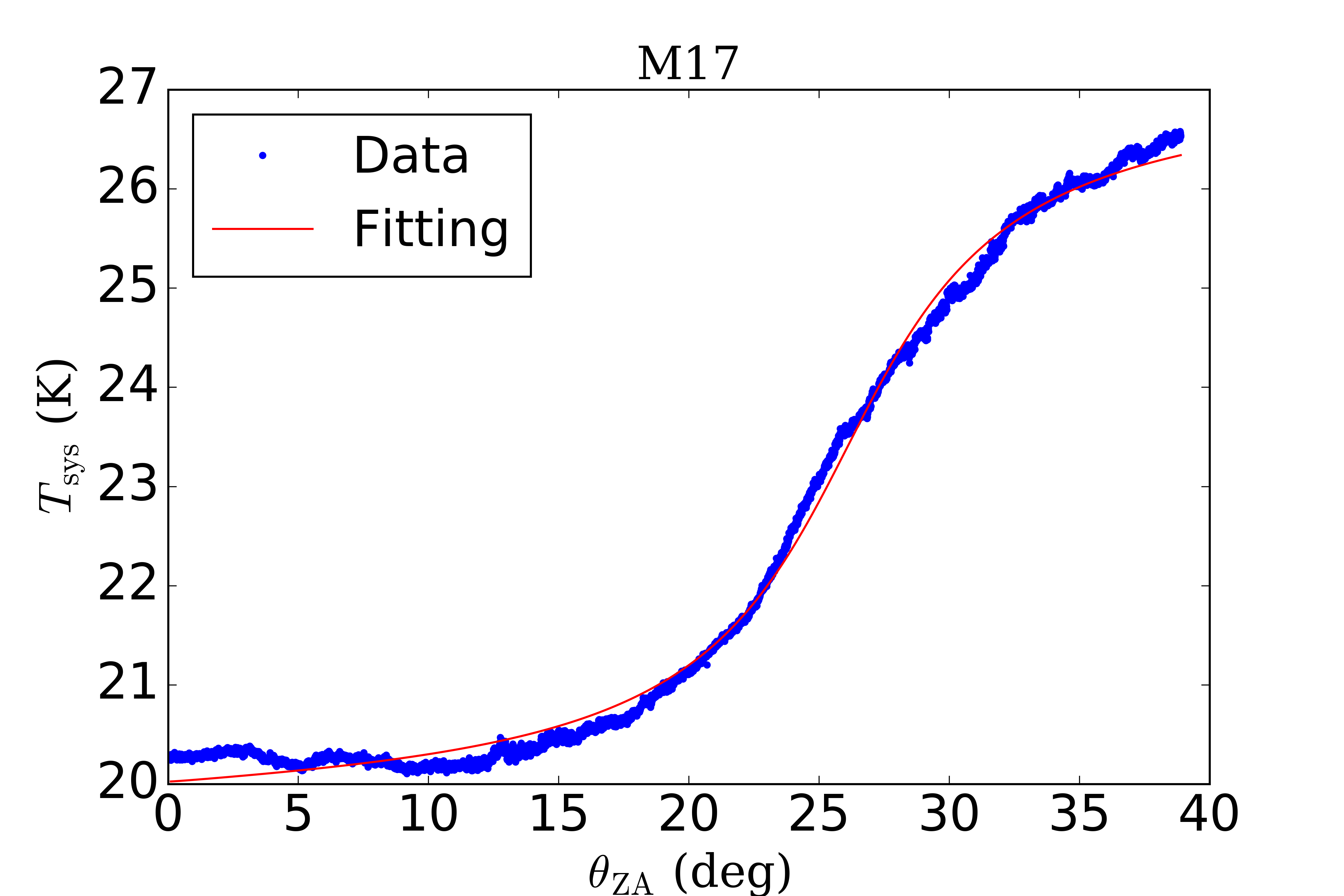}
  \includegraphics[width=0.45\textwidth]{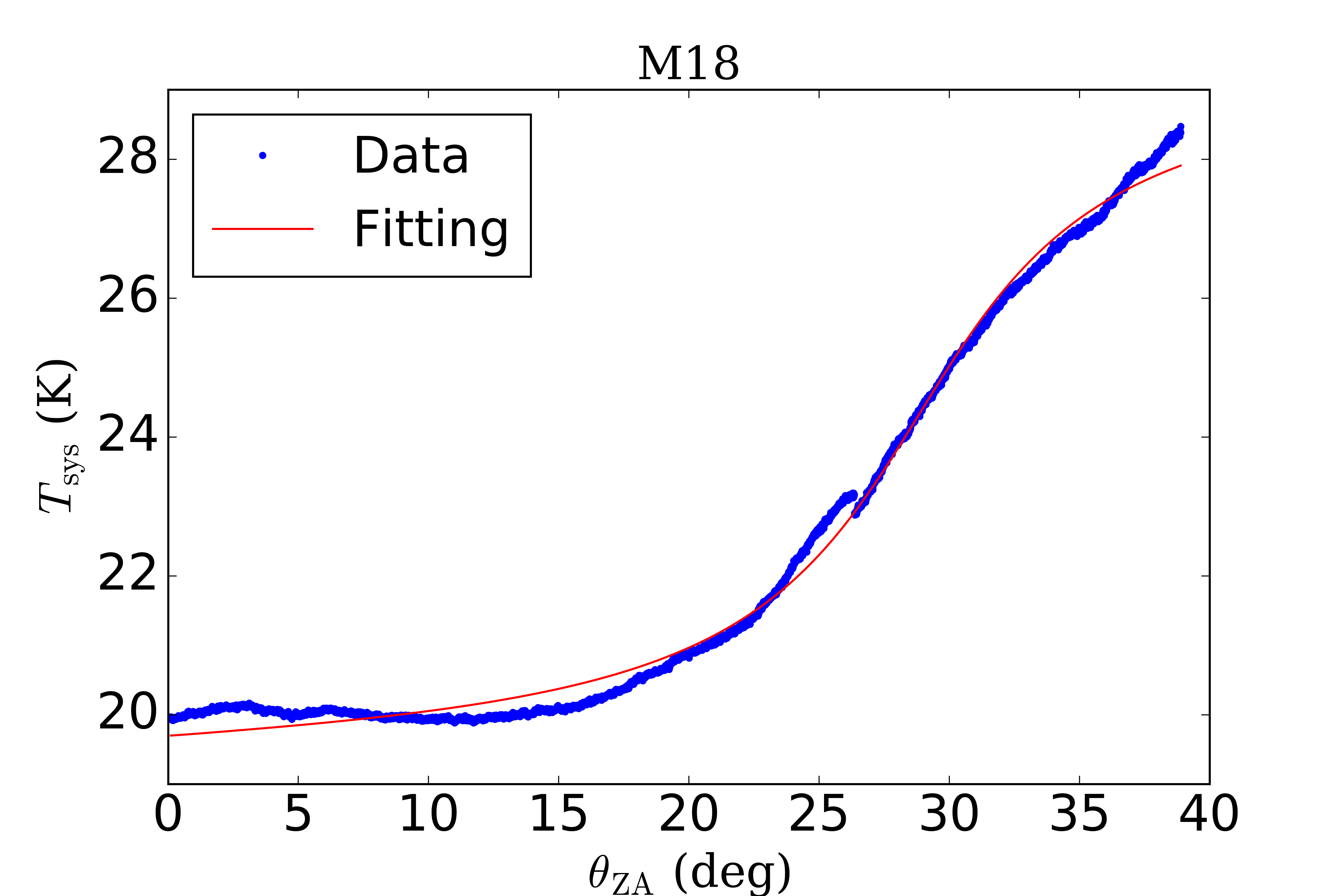}
  \caption{Measured $T\rm_{sys}$ curve as a function of zenith angle $\theta\rm_{ZA}$ at 1400 MHz for Beam 1, 2, 8 and 19.  The fitting result is represented with red solid line.}
\label{fig:tsysfit_restbeams}
\end{figure}

\end{document}